\documentclass[aps,pra,reprint,groupedaddress]{revtex4-1}

\usepackage{ae,amsmath,amssymb,amsfonts,bbm,amsthm}
\usepackage{braket}
\usepackage{graphicx}
\usepackage{calc}
\usepackage{color}
\usepackage{transparent}
\usepackage{dcolumn}
\usepackage{hyperref}
\usepackage{multirow}
\usepackage[caption=false]{subfig}
\usepackage{adjustbox}
\usepackage{here}
\usepackage{array}
\usepackage[titletoc,title]{appendix}
\usepackage{placeins}
\usepackage{enumitem}
\newcolumntype{C}[1]{>{\centering\arraybackslash}p{#1}}

\newcommand{\Mod}[1]{\ (\text{mod}\ #1)}

\usepackage{units}

\begin{document}

\title{Ultrafast Fault-Tolerant Long-Distance Quantum Communication\\ with Static Linear Optics}

\author{Fabian Ewert}
\email[]{ewertf@uni-mainz.de}
\affiliation{Institute of Physics, Johannes Gutenberg-Universität Mainz, Staudingerweg 7, 55128 Mainz, Germany}
\author{Peter van Loock}
\email[]{loock@uni-mainz.de}
\affiliation{Institute of Physics, Johannes Gutenberg-Universität Mainz, Staudingerweg 7, 55128 Mainz, Germany}

\begin{abstract}
We present an in-depth analysis regarding the error resistance and optimization of our all-optical Bell measurement and ultrafast long-distance quantum communication scheme proposed in [arXiv:1503.06777]. In order to promote our previous proposal from loss- to fault-tolerance, we introduce a general and compact formalism that can also be applied to other related schemes (including non-all-optical ones such as [PRL 112, 250501]). With the help of this new representation we show that our communication protocol does not only counteract the inevitable photon loss during channel transmission, but is also able to resist common experimental errors such as Pauli-type errors (bit- and phase-flips) and detector inefficiencies (losses and dark counts). Furthermore, we demonstrate that on the physical level of photonic qubits the choice of the standard linear optical Bell measurement with its limited efficiency is optimal for our setting in the sense that, apart from their potential use in state preparation, more advanced Bell measurements yield only a small decrease in resource consumption. We devise two state generation schemes that provide the required ancillary encoded Bell states (quasi-)on-demand at every station. The schemes are either based on nonlinear optics or on linear optics with multiplexing and exhibit resource costs that scale linearly or less than quadratic with the number of photons per encoded qubit, respectively. Finally, we show that it is possible to operate our communication scheme with on-off detectors instead of employing photon-number-resolving detectors.
\end{abstract}

\maketitle

\section{Introduction}
\label{sec:motivation}

Two conceptual, theoretical breakthroughs were made 15 years ago with regards to the implementation of quantum computation and communication. Knill, Laflamme, and Milburn (KLM) showed that, quite surprisingly at that time, linear optical elements such as beam splitters and phase shifters are sufficient for universal quantum computation \cite{KLM}. Duan, Lukin, Cirac, and Zoller (DLCZ) demonstrated that also quantum communication over large distances can be, in principle, achieved by means of linear optics together with atomic-ensemble quantum memories \cite{DLCZ}.

A crucial element in either scheme is quantum error detection (QED). In the DLCZ quantum repeater, entangled states are distributed over sufficiently small channel segments and then connected by entanglement swapping, where the primary source of errors is loss, especially during the transmission of the photonic states. Such channel losses can be immediately detected during the initial entanglement distribution step, while memory losses are suppressed later during the entanglement swapping procedure (or at the latest through a final postselection step). This way long-distance quantum communication (LDQC) becomes possible, as opposed to using direct quantum communication without QED and without memories where the transmission rate would drop exponentially with distance.

Could it be useful to replace QED by quantum error correction (QEC) for LDQC? First of all, the original quantum repeater idea \cite{Briegel98,Duer99} emphasized the fact that, in contrast to scalable quantum computation, scalable quantum communication is possible with the help of entanglement purification, corresponding to a form of QED against both channel losses and local gate errors, with no need for many levels of additional complex QEC codes. This huge simplification, however, comes at a price: entanglement purification is probabilistic and so are typically the entanglement distribution over lossy channel segments (including the local state preparations) and the subsequent entanglement swappings (at least when restricted to linear optics)
%%%%%%%%%%%%%%FOOTNOTE
\footnote{Alternative encodings beyond discrete qubit states such as continuous-variable Gaussian states allow for unconditional state preparation and deterministic entanglement swapping with linear optics, however, QED and entanglement purification are difficult in this case. Yet another promising alternative could be the use of both discrete and continuous variables that combines the advantages and avoids the disadvantages of the individual approaches \cite{HQR,HybridSwap}.}.
%%%%%%%%%%%%%%FOOTNOTE
As a consequence, this type of nested quantum repeater
is extremely slow, relying on two-way classical communication and long-lived quantum memories. In fact, to scale up their scheme to larger distances, DLCZ need sufficiently good atomic-ensemble memories at every repeater station in order to store the initial entangled states until other entanglement distributions and connections succeed. While certain variations of the original DLCZ scheme lead to a significant improvement of the repeater rates \cite{SangouardRMP}, conceptually, this type of quantum repeater is fundamentally limited by the long waiting times for the heralding signals at every repeater node.

In recent years, various proposals have been made to employ QEC codes for LDQC. Since these codes suppress errors deterministically, long waiting times and two-way classical communication (and hence the use of quantum memories) can be, in principle, completely avoided. While one class of schemes focused on the correction of operational errors with channel losses still suppressed through heralded distribution and purification of entanglement \cite{Jiang09,Munro10,Bernardes12,Bratzik14}, another class did include QEC against transmission losses making high-rate loss-tolerant \cite{Munro12,ATL,pant2016rate} or even fully fault-tolerant \cite{Fowler10,LLPRL,namiki16,muralidharan2016} LDQC possible. These latter schemes are limited only by the speed of the local gate operations and thus, they approach rates as obtainable in classical communication. Our scheme also belongs to this class and allows for ultrafast LDQC, but unlike Refs.~\cite{Munro12,Fowler10,LLPRL} it does so in an all-optical fashion without the use of difficult local quantum gates (implementable via local nonlinear matter-light interactions \cite{Munro12,LLPRL}).

For this purpose, by employing a certain version of loss-tolerant parity-codes \cite{RalphGilchrist05,Munro12,LLPRL}, we suggest to send encoded qubit states directly, which are then subject to a Bell measurement (BM) together with locally prepared, encoded Bell states after every few kilometers (see Fig.~\ref{fig:comm_scheme}). These local state teleportations allow for a non-destructive loss-error syndrome detection and a qubit state recovery in one step. The use of QEC by teleportation \cite{Knill} along the channel is conceptually similar to the protocol of Ref.~\cite{LLPRL}. However, in our scheme, every teleportation is performed with optical (encoded) Bell states and linear optical elements
%%%%%%%%%%%%%%%FOOTNOTE
\footnote{In terms of resources, our scheme is actually most close to that of Ref.~\cite{ATL} who employ nonlocally distributed loss-resistent cluster states and standard (non-logical) linear-optics BMs for entanglement connection. However, in order to suppress the effect of losses, fast feedforward operations are required at every repeater station to separate successful from failed BM events (depending only on the local BM results and independent of the classical information about neighboring BM outcomes like in multiplexed standard repeaters \cite{Collins}).}.
%%%%%%%%%%%%%%FOOTNOTE
It turns out that the encoding has two positive effects: the larger the code is, the more efficient the ideal BM (despite the linear-optics constraint \cite{CalsamigliaNL}) {\it and} the higher the amount of tolerable photon loss become (despite the increasing photon number of the codewords). In contrast to the all-optical scheme of Ref.~\cite{ATL}, our logical BMs are conceptually different and work entirely without feedforward. This does not only reduce the local operation times, but also makes on-chip integration along an optical fiber channel more feasible, as optical switching in this case is very sensitive to loss \cite{QTonChip14,LossMultiplObrien14}. We emphasize that our scheme, as originally developed in Ref.~\cite{Ewert2015arxivPRL} with mainly the effect of transmission loss under investigation, is the only all-photonic forward error-correcting quantum communication scheme so far (the scheme of Ref.~\citep{ATL} relies upon error correction on entangled states distributed among neighboring stations).

Here we show that our scheme is also fault-tolerant against various kinds of error sources beyond transmission loss. In addition to an extensive investigation of our schemes' performance in the presence of loss, including a comparison to the recent benchmarks of repeaterless quantum communication \cite{TGW,PLOB} which even very small instances of our code (with only eight photons per logical qubit) can beat, we develop a toolbox to analyze other, more general errors. This very general toolbox, which is not restricted to our all-optical approach but can be used for every QPC-based one-way communication scheme (e.g. the one presented in \cite{LLPRL}), allows us to demonstrate that our communication scheme is fault-tolerant with respect to Pauli-type errors and detector inefficiencies (such as lossy detectors and dark counts), i.e., the most common operational errors in optical quantum information processing schemes. Since we are also concerned with minimizing the experimental requirements of our scheme, we explore the use of on-off-detectors instead of photon-number-resolving detectors in our BM. We find that this replacement is indeed possible, provided the code size is adjusted accordingly. Furthermore, we investigate the use of more advanced BM schemes on the single-photon level \cite{Grice,Zaidi,Ewert} and find that, although more complicated BM schemes can allow for the use of smaller error correction codes, the resource cost reduction associated with this code size decrease is most probably not worth the additional experimental effort of the more advanced BM schemes. However, when considering the experimental costs for generating the required encoded Bell states at every repeater station, the use of more advanced BM schemes can yield significant advantages. We present two different schemes to prepare the encoded Bell states. One is based on nonlinear optics, utilizing coherent photon conversion \cite{langford11}, and exhibits a linear resource cost scaling with respect to the number of photons per logical qubit. The second scheme is based solely on single-photon sources, linear optical elements, and the use of multiplexing and feedforward. Advanced BM schemes allow to reduce the quadratic cost scaling of this state generation scheme to a much-closer-to-linear one, which in turn reduces the number of required single-photon sources by several orders of magnitude. Our logical BM may also be employed for gate teleportation in the context of quantum computation. In Ref.~\cite{Ewert2015arxivPRL}, we already discussed that such a generalization is subtle and the intrinsic loss resistance of the BM scheme has to be sacrificed at the expense of quantum computational universality. Moreover, feedforward operations are needed, though to a smaller extent compared to schemes like KLM \cite{KLM}. In this work here, we exclusively focus on quantum communication.

This work is structured as follows. In Sec.~\ref{sec:comm scheme} we present the general idea of our communication scheme and give an overview of the errors we expect to occur within our setup. Section~\ref{sec:QPC code} focuses on the quantum error correction code we use - the quantum parity code (QPC) - and its behavior under loss and Pauli errors. In Sec.~\ref{sec:QPC BM} we investigate the central building block of our communication scheme: the all-optical BM on the encoded qubits which is required for the teleportation that performs the error correction step. We show that the BM efficiency approaches one with increasing code size and that the BM still has high success rates if (some but not too many) photons are lost. In order to treat errors other than loss, in Sec.~\ref{sec:error analysis} we develop a toolbox to analyze how errors on the single-photon level propagate through the different encoding levels and lead to errors on the encoded qubits. This toolbox is applied to various error models in Sec.~\ref{sec: results}. There we show that our error correction scheme does not only take care of the inevitable photon loss during transmission, but also gives resistance to depolarizing errors and faulty photon detectors. Additionally, we demonstrate that the requirement of photon-number-resolving detectors can be dropped and on-off detectors can be used, provided the code size is chosen accordingly. Finally, we conclude with a summary of our results.

In the Appendices we provide proofs of several formulas given in the main text, especially the representation of encoded Bell states in terms of two-photon Bell states on which our error analysis relies. Furthermore, two appendices are dedicated to the two state generation schemes described above. We give a detailed description of how coherent photon conversion or linear optics with multiplexing can be utilized to obtain the encoded Bell states at every repeater station. We also investigate the resource cost scaling of the two schemes. Finally, we give a detailed view on how dark counts affect the capabilities of the standard linear-optical BM on the single-photon level. 

\section{The communication scheme}
\label{sec:comm scheme}

\begin{figure}[t]
	\includegraphics[scale=1]{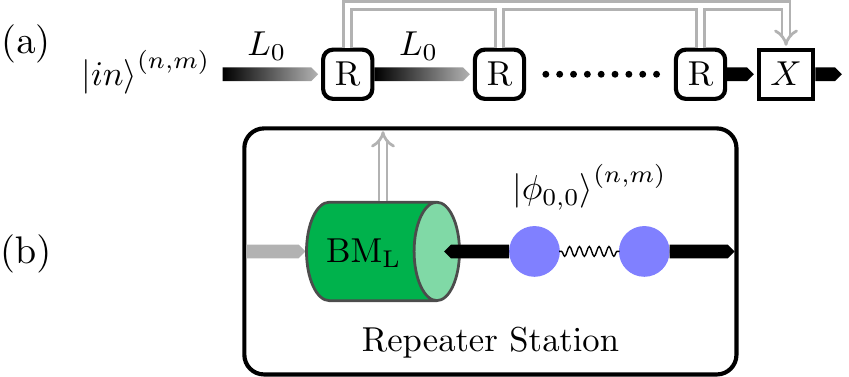}
	\caption{One-way communication scheme: (a) To send a quantum state $\ket{in}^{(n,m)}$ over a long distance, repeater stations (R) at shorter distances $L_0$ are used to recover an incoming qubit from accumulated losses (fading arrows). A classical signal (double line) defines a single Pauli correction $X$ at the receiver (not necessarily a bit flip). (b) Each repeater station is equipped with an encoded Bell state which in our first, simplest model is not corrupted by loss. A highly efficient, loss-resistant, logical Bell Measurement (BM$_{\text{L}}$) acting on the incoming signal and one half of the Bell state is performed. The other half of the Bell state is sent to the next station along with a classical signal containing the result of the BM. \label{fig:comm_scheme}}
\end{figure}

To send quantum information in the form of an arbitrary (flying, i.e. optical) qubit state over a total distance $L_{\rm tot}$ we propose to employ a quantum error correction code that is specifically designed to correct photon loss \cite{LLPRL}. We have found the quantum parity code (QPC, see Sec.~\ref{sec:QPC code}) to be an excellent candidate for this. The quantum state stored in this code is sent directly through the communication channel (without prior distribution of a long-distance entangled state). To counteract the loss of photons from the transmission, repeater stations are placed after every channel segment of length $L_0$. At every station an encoded Bell state $\ket{\phi_{0,0}}^{(n,m)}$ is available on demand (i.e., created and consumed locally), and a logical BM is performed on one half of the Bell state together with the incoming encoded qubit (see Fig.~\ref{fig:comm_scheme}). Whenever the BM succeeds the qubit state (up to an heralded Pauli error) is teleported to the other half of the repeater station's Bell state which is then sent onward to the next station. Since the Bell state was created locally, the outgoing qubit state does not carry any of the errors the incoming qubit state might have had. Therefore, the teleportation acts like an error correction scheme \cite{Knill}. The heralded Pauli error that the BM might induce can be kept and need not be corrected immediately. As long as the information on which errors occurred at the stations is kept (i.e., the knowledge of the Pauli frame), a final Pauli correction at the receiver is sufficient and any additional, experimentally demanding feed-forward procedures are avoided. It is crucial to our scheme that the BM has both a high success probability and a tolerance to errors in the incoming qubit. We found that the QPC code allows for an easy-to-implement all-optical BM that fulfills both of these requirements (see Secs.~\ref{sec:QPC BM}~-~\ref{sec: results}).

Regarding the errors that occur in our scheme we make the usual assumption that the error operators act independently on the individual physical qubits. The main error source to be dealt with is photon loss due to the channel transmission. Every photon that is sent over a channel segment of length $L_0$ suffers from potential loss according to a transmission coefficient of $\eta_t=\exp(-L_0/L_{\rm att})$ (with attenuation length $L_{\rm att} = \unit[22]{km}$). Those photons that are part of the ancillary Bell state in a repeater station, consumed locally in a BM, are not subject to the transmission loss channel. However, a variety of other errors may also be present that do not occur during the transmission but locally within the repeater station (potentially acting on both the incoming qubit and the locally created Bell states). The errors we are concerned with can roughly be classified into three groups: (i) The loss-type errors, which include the already mentioned transmission loss but also losses through misalignment of optical equipment or errors in the state generation that lead to missing photons in the ancillary Bell states. In contrast to transmission loss, the two latter types of errors are independent of the repeater spacing $L_0$. (ii) The Pauli-type errors that do not affect the total number of photons in an encoded qubit, but become manifest in the form of bit-flips or phase-flips on the single-photon (physical single-qubit) level, which is a common error model in all quantum information processing protocols. (iii) Detector errors, especially lossy detectors, which can also be seen as part of the first class of errors, and dark counts.

To investigate the performance of our communication scheme in the presence of errors it is useful to shift the point of view from the actual error channels to the outcome probabilities of the BM in each repeater station. These can be calculated by merging all error types that occur on the optical modes participating in the BM. Particularly, this means that the errors on the ancillary Bell states can be split in two halves. Those photons that are consumed locally in the BM and the errors related with these, are assigned to the BM in this repeater station, whereas the errors on the other half of the Bell state which is sent onward to the next station are assigned to the BM there. This splitting is possible because of the above assumption that all errors occur only on the single-photon level and are independent of each other. Since the various loss channels and Pauli-type errors, governed by a depolarizing error channel, commute, a single error channel for every group of errors suffices to model all errors that occur during state generation, transmission and BM.

\section{The quantum parity code}
\label{sec:QPC code}

In our proposal we use the quantum parity code QPC($n,m$) to protect quantum information from loss and other sources of errors. Several versions of this code have been used before \cite{RalphGilchrist05,Munro12,LLPRL}, but we found the one which is also used in Ref.~\cite{LLPRL} to be the most suitable for our purposes. It encodes a logical qubit into $n m$ physical qubits. The code can be understood as having three different levels of encoding. On the lowest level, which we call the physical level, we have standard dual-rail (two-mode) qubits. These are typically realized by two orthogonal polarization modes of photons $\{\ket{0} = \ket{H},\ket{1} = \ket{V}\}$, but also other realizations like spatial or temporal modes are possible. On the second level of encoding, the block level, $m$ physical qubits are collected to represent a block-qubit $\{\ket{0}^{(m)} = \ket{H}^{\otimes m}, \ket{1}^{(m)} = \ket{V}^{\otimes m}\}$. This repetition part of the code is crucial for the loss robustness as we will see later. The highest encoding level is the logical level. Here $n$ block-qubits are used to construct the logical qubits as $\ket{\pm}^{(n,m)} = \left(\ket{0}^{(m)} \pm \ket{1}^{(m)} \right)^{\otimes n} / \sqrt{2^n} = (\ket{\pm}^{(m)})^{\otimes n}$. The codewords are then naturally obtained by $\ket{\pm}^* = \left(\ket{0}^* \pm \ket{1}^* \right)/ \sqrt{2}$, where the $*$ denotes the encoding level (blank for physical, $(m)$ for block and $(n,m)$ for logical). 

In all three encoding levels the four Bell states are defined as
\begin{align}
	\ket{\phi_{k,l}}^* = \tfrac{1}{\sqrt{2}}\left(\ket{0,k}^* + (-1)^l \ket{1,1-k}^*\right),  \label{eq:bell_states}
\end{align}
with $k,l\in\{0,1\}$. It can be shown (see Appendix~\ref{sec:AppA}) that the Bell states of the higher encoding levels can be represented in terms of Bell states of the lower encoding levels
\begin{align}
	\ket{\phi_{k,l}}^{(m)} &\cong \frac{1}{\sqrt{2^{m-1}}}\sum_{\vec{r} \in A_{l,m}} \bigotimes_{i=1}^m \ket{\phi_{k,r_i}},\label{eq:phys2block}\\
	\ket{\phi_{k,l}}^{(n,m)} &\cong \frac{1}{\sqrt{2^{n-1}}}\sum_{\vec{s} \in A_{k,n}} \bigotimes_{i=1}^n \ket{\phi_{s_i,l}}^{(m)},\label{eq:block2logic}
\end{align}
where the index set is defined as $A_{l,m} = \Set{\vec{r}\in\{0,1\}^m | \sum_{i=1}^m r_i = l \Mod 2 }$.

This representation of encoded Bell states in terms of lower level Bell states is the basis on which the efficiency of the logical Bell measurement rests (see section \ref{sec:QPC BM}).

Note that the above representations \eqref{eq:phys2block},\eqref{eq:block2logic} are only true after an appropriate reordering of the modes (indicated by $\cong$). Quite naturally, the photons of two logical qubits in a Bell state are paired with their equivalent: photon $i$ in block $j$ of the logical qubit $A$ forms a Bell state with photon $i$ in block $j$ of the logical qubit $B$. In the remainder of this paper this reordering will be omitted in the notation (with the exception of the next example).

As an example we take a closer look at QPC(2,2). The codewords on the three encoding levels are $\ket{0} = \ket{H}$, $\ket{1}=\ket{V}$ on the physical level, $\ket{0}^{(2)} = \ket{HH}$, $\ket{1}^{(2)} = \ket{VV}$ on the block level and
\begin{align}
	\ket{0}^{(2,2)} &= \tfrac{1}{\sqrt{2}}\left(\ket{HHHH}+\ket{VVVV}\right)\\
	\ket{1}^{(2,2)} &= \tfrac{1}{\sqrt{2}}\left(\ket{HHVV}+\ket{VVHH}\right).
\end{align}
on the logical level. To illustrate the Bell state representations and the mode reordering consider the block level Bell state
\begin{align}
	\ket{\phi_{1,1}}^{(2)} = \tfrac{1}{\sqrt{2}} \left(\ket{HHVV} - \ket{VVHH}\right).
\end{align}
The representation \eqref{eq:phys2block} yields
\begin{align}
	\ket{\phi_{1,1}}^{(2)} &\cong \tfrac{1}{\sqrt{2}} \left(\ket{\phi_{1,0}} \ket{\phi_{1,1}} + \ket{\phi_{1,1}} \ket{\phi_{1,0}} \right) \\
	&= \tfrac{1}{\sqrt{2}} \left(\ket{HVHV} - \ket{VHVH}\right)\nonumber.
\end{align}
These two forms only coincide after the modes 2 and 3 have been interchanged. Similarly, modes 3 and 4 have to be interchanged with modes 5 and 6 (i.e. block 2 with block 3) to get
\begin{align}
	\ket{\phi_{1,1}}^{(2,2)} &= \tfrac{1}{\sqrt{2}} \left(\ket{\phi_{0,1}}^{(2)} \ket{\phi_{1,1}}^{(2)} + \ket{\phi_{1,1}}^{(2)} \ket{\phi_{0,1}}^{(2)} \right)
\end{align}
on the logical level.

As mentioned above, the main purpose of the QPC-code is to protect the quantum information from photon loss and other error sources. Assume a (potentially unknown) quantum state $\ket{\psi} = \alpha \ket{0}^{(2,2)} + \beta \ket{1}^{(2,2)}$ is sent over a lossy channel and a photon, say the one of the first physical qubit, is lost. $\ket{\psi}$ is then projected onto the mixed state
\begin{align}
	\ket{\psi}\bra{\psi} \rightarrow \tfrac{1}{2}[ & \left(\alpha \ket{0HHH} + \beta \ket{0HVV}\right) \times {\rm h.c.} \nonumber\\
	&+ \left(\alpha \ket{0VVV} + \beta \ket{0VHH}\right) \times {\rm h.c.}].
\end{align}
Here we used the common notation $0$ to describe the vacuum state in mode 1. It should not be confused with the state of the physical qubit $\ket{0}=\ket{H}$ used outside of this example. The mixed state above consists of states from two error spaces, which can be identified by the photon in mode 2 (the remaining photon of block 1). In both of the error spaces the quantum information, i.e. the superposition with the complex amplitudes $\alpha$ and $\beta$ is still available. Therefore the original state $\ket{\psi}$ can be restored by error correction (the details of this will be outlined in section \ref{sec:QPC BM}). If, however, a second photon is lost as well, the quantum information is lost: either the second photon of block 1 is lost and $\ket{\psi}$ becomes
\begin{align}
	\tfrac{1}{2}[ & \left(\alpha \ket{00HH} + \beta \ket{00VV}\right) \times {\rm h.c.} \nonumber\\
	&+ \left(\alpha \ket{00VV} + \beta \ket{00HH}\right) \times {\rm h.c.}],
\end{align}
which means the error spaces can no longer be discriminated (and a random bit-flip error remains), or one of the photons of block 2, say photon 3, is lost and one gets
\begin{align}
	\tfrac{1}{2}( & |\alpha|^2 \ket{0H0H}\times {\rm h.c.} + |\beta|^2 \ket{0H0V}\times {\rm h.c.}\nonumber \\
	&+ |\alpha|^2 \ket{0V0V}\times {\rm h.c.} + |\beta|^2 \ket{0V0H}\times {\rm h.c.})
\end{align}
and the superposition is lost.
This generalizes to higher QPC$(n,m)$-codes in the following way. In all of the $n$ blocks at least one of the $m$ photons must remain to identify the correct error space. Additionally in at least one of the blocks all $m$ photons must remain to preserve the superposition with the amplitudes $\alpha$ and $\beta$. 

Most errors other than loss can be described with the help of Pauli errors, i.e. bit-flips and phase-flips, on the physical level. To correct these, larger QPC codes are necessary. If, for example, a Pauli-X-error occurs on the first photon $\ket{\psi}$ becomes
\begin{align}
	&\tfrac{\alpha}{\sqrt{2}} (\ket{VHHH}+\ket{HVVV}) + \tfrac{\beta}{\sqrt{2}} (\ket{VHVV}+\ket{HVHH}).
\end{align}
However, a state of the same error space is obtained from an X-error on the second photon,
\begin{align}
	&\tfrac{\alpha}{\sqrt{2}} (\ket{HVHH}+\ket{VHVV})
	+\tfrac{\beta}{\sqrt{2}} (\ket{HVVV}+\ket{VHHH}),
\end{align}
but the amplitudes $\alpha$ and $\beta$ are flipped. Thus the initial state cannot be recovered reliably. Similarly, phase flips on photons 1 and 3 lead to the same error space, but to different amplitudes: $(\alpha,\beta) \leftrightarrow (\alpha,-\beta)$. The smallest QPC code that can reliably correct a single Pauli-error is QPC(3,3).

As a final note on the properties of the QPC codes we take a look at Pauli-errors on the logical level. They occur frequently as a result of the teleportation but can easily be corrected by appropriate combinations of Pauli gates on the physical level:
\begin{align}
	X^{(n,m)} &= X^{\otimes m} \otimes \mathbf{1}^{\otimes (n-1)m} \\ Z^{(n,m)} &= \left(Z \otimes \mathbf{1}^{\otimes m-1}\right)^{\otimes n}
\end{align}

\section{Bell Measurements on QPC}
\label{sec:QPC BM}

The main component of our communication scheme is the local teleportation at every repeater station that acts as an error correction. To perform this teleportation a BM, i.e. a projection onto the four Bell states \eqref{eq:bell_states}, however, on the logical level, is required. The Bell state representations \eqref{eq:phys2block} and \eqref{eq:block2logic} show that a logical BM can be done via $n m$ parallel BMs on the physical level: every pattern of results on the physical level is either unique to one of the logical Bell states, or it indicates that some kind of error must have occurred.

\begin{figure}[t]
	\includegraphics[width=0.95\columnwidth]{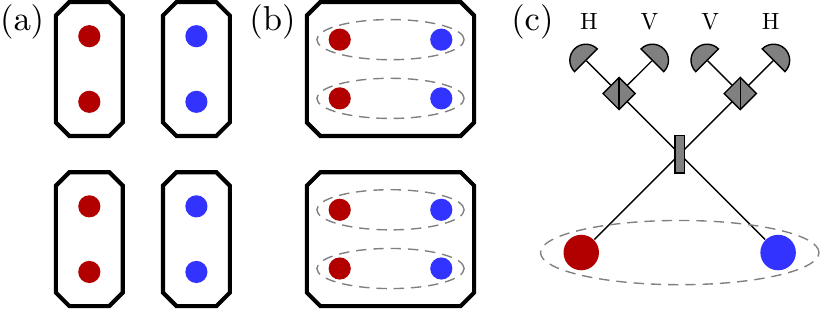}	
	\caption{Block structure and Bell measurement (BM): (a) The block structure for two QPC(2,2)-qubits. The polarization qubits on the left (red) belong to the incoming signal and are thus subject to channel errors, while those on the right (blue) are part of the encoded Bell state provided in each repeater station. (b) In a Bell state in QPC-encoding the qubits are joined blockwise.  The dashed ellipses highlight physical-level qubit pairs that are combined at the BM. (c) Optical BM setup on the physical level adapted to polarization encoding.\label{fig:BM_setup}}
\end{figure}

Let us first take a look at the efficiency of the BM in the case that no errors occur. It has been known for quite some time that the maximal efficiency of a BM on dual-rail qubits with standard linear optical tools (beam splitters, phase shifters and photon detectors) is limited to one half \cite{CalsamigliaNL}. The BMs on the physical level of our scheme are subject to this limit, but thanks to the form of the QPC the logical BMs are not, and can reach efficiencies arbitrarily close to one. The standard BM on polarization-encoded qubits, depicted in Fig.~\ref{fig:BM_setup}~(c), consists of a 50:50 beam splitter, followed by two polarizing beam splitters that reflect vertically polarized photons, and 4 photon detectors. If a $\ket{\phi_{0,l}}$-state enters the BM, both photons will be directed to a single detector, independent of $l$. The states $\ket{\phi_{1,0}}$ and $\ket{\phi_{1,1}}$ lead to clicks in two detectors: one for vertically and one for horizontally polarized photons. If the clicks are on the same side of the setup (defined by the output modes of the beam splitter), it was a $\ket{\phi_{1,0}}$-state, whereas for $\ket{\phi_{1,1}}$ the photons end up on opposing sides. This shows that the physical BM unambiguously identifies the index $k$ of the Bell state $\ket{\phi_{k,l}}$ but the index $l$ is identified if and only if $k=1$. Later, in Sec.~\ref{subsec:advBM}, we shall also consider advanced BM schemes that sometimes can also distinguish the $\ket{\phi_{0,l}}$-states and their influence on our communication protocol.

On the block level the BM for the state $\ket{\phi_{k,l}}^{(m)}$ will be successful if $k$ is identified correctly in at least one of the physical BMs, since its value is the same in all of them (see Eq.~\eqref{eq:phys2block}), and if the values $r_i$ are identified correctly in every physical BM. The first condition will always be met, but the second one only if $k=1$. Thus, the block BM has the same efficiency of one half as the physical BM. On the logical level $\ket{\phi_{k,l}}^{(n,m)}$, however, the identification of $l$ only has to be successful in one block, while the identification of $s_i$ must succeed in every block (see Eq.~\eqref{eq:block2logic}). The latter will always be accomplished and the first condition will be met almost every time: The only case where it is not fulfilled is $s_i=0$ $\forall i$. The Bell state representation \eqref{eq:block2logic} shows that this case can only occur in the logical Bell states $\ket{\phi_{1,l}}^{(n,m)}$ and in these only in one of the $2^{n-1}$ terms of the superposition. Thus, the efficiency of the logical BM in the case of no errors is
\begin{align}
	p_{\text{BM, ideal}} = \frac{2+2\left(1-2^{-(n-1)}\right)}{4} = 1-2^{-n}.
\end{align}
This approaches one with an increasing number of blocks $n$.

As an example we again choose the code QPC$(2,2)$. The block Bell states can be written as
\begin{align}
	\ket{\phi_{k,0}}^{(2)} &= \tfrac{1}{\sqrt{2}} \left(\ket{\phi_{k,0}}\ket{\phi_{k,0}}+\ket{\phi_{k,1}}\ket{\phi_{k,1}}\right),\label{eq:qpc22block0}\\
	\ket{\phi_{k,1}}^{(2)} &= \tfrac{1}{\sqrt{2}} \left(\ket{\phi_{k,0}}\ket{\phi_{k,1}}+\ket{\phi_{k,1}}\ket{\phi_{k,0}}\right).\label{eq:qpc22block1}
\end{align}
For $k=0$ all physical Bell states involved are $\ket{\phi_{0,l}}$-states, which cannot be identified unambiguously. In contrast,  for $k=1$ all physical states are $\ket{\phi_{1,l}}$-states and will always be identified correctly. Therefore the block level BM exhibits exactly the same behavior as the standard physical BM. On the logical level, however, we get the following:
\begin{align}
	\ket{\phi_{0,l}}^{(2,2)} &= \tfrac{1}{\sqrt{2}}\! \left(\ket{\phi_{0,l}}^{(2)}\! \ket{\phi_{0,l}}^{(2)} \!+\! \ket{\phi_{1,l}}^{(2)}\! \ket{\phi_{1,l}}^{(2)} \right),\label{eq:qpc22log0}\\
	\ket{\phi_{1,l}}^{(2,2)} &= \tfrac{1}{\sqrt{2}}\! \left(\ket{\phi_{0,l}}^{(2)}\! \ket{\phi_{1,l}}^{(2)} \!+\! \ket{\phi_{1,l}}^{(2)}\! \ket{\phi_{0,l}}^{(2)} \right).\label{eq:qpc22log1}
\end{align}
For the states $\ket{\phi_{1,l}}^{(2,2)}$ there is at least one block state $\ket{\phi_{1,l}}^{(2)}$ in every summand of the superposition, which is enough to identify $l$ and therefore the Bell state. Thus, the fully encoded logical $\ket{\phi_{1,l}}^{(2,2)}$-states can still be perfectly distinguished, just like $\ket{\phi_{1,l}}$ and $\ket{\phi_{1,l}}^{(2)}$. However, for the states $\ket{\phi_{0,l}}^{(2,2)}$ the index $l$ can only be identified in the second summand. Thus, the states $\ket{\phi_{0,l}}^{(2,2)}$ will be identifiable in half of all cases. This yields an overall efficiency of the logical BM of $75\%$.

As already mentioned before, the size $m$ of the blocks does not influence the success rate of the logical BM, if no errors are present. However, as pointed out in Sec.~\ref{sec:QPC code}, it is crucial to preserve a good BM efficiency also in the presence of losses and other errors. Let us assume the incoming photon in the first mode of the first block was lost during transmission. The physical BM this photon was headed to will then detect only a single photon and report this loss. It is important to note at this point that the photon detectors must be able to resolve between a single and two photons. Otherwise, the loss of a photon cannot be distinguished from an intact $\ket{\phi_{0,l}}$-state. In Sec.~\ref{subsec: onoff} we shall also discuss the possibility of using on-off detectors. Of course, no value of $k$ and $l$ can be found in a physical BM, if a photon was lost. On the block level, however, the index $k$ can still be obtained from the second physical BM in this block (see Eqs.~\eqref{eq:qpc22block0} and \eqref{eq:qpc22block1}). Only the value of $l$ is unidentifiable here. On the logical level the $50\%$ chance of correctly identifying the states $\ket{\phi_{0,l}}^{(2,2)}$ remains unchanged, since in the only summand that can be useful for the correct identification of $l$ there is still a second term $\ket{\phi_{1,l}}^{(2)}$. On the other hand, for the $\ket{\phi_{1,l}}^{(2,2)}$-state, the second term of the superposition becomes useless, since the remaining state $\ket{\phi_{0,l}}^{(2)}$ does not allow for an identification of $l$, and thus the total success rate drops to $50\%$. The same arguments hold for a single loss of any other photon in the logical state.
When a second photon is also lost during the transmission we find that the logical BM of the QPC$(2,2)$-code will fail, because either the information about $k$ is inaccessible in one of the blocks or none of the blocks provides information about $l$ (similar to the discussion in Sec.~\ref{sec:QPC code}). Only if two photons are lost that were intended to take part in the same physical BM, the logical BM can still succeed. But since only one of these photons is transmitted in the first place, the second loss must be due to non-unit-efficiency detectors (or any other type of local loss mechanism). It is possible to calculate the BM success rates for any code sizes $n$ and $m$ and an arbitrary number of losses and calculate transmission rates from these. In fact, this was our original approach which is outlined in the supplemental material of [prl]. However, for the purpose of covering more general types of errors in the context of our communication scheme, another point of view is more useful. It is described in the next section.

\section{Error analysis}
\label{sec:error analysis}

In this section we describe a general method to calculate the secure key rates of our communication scheme when a certain error model is investigated. It will be applied to various error models in Sec.~\ref{sec: results}. The idea of this method is to first calculate the outcome probabilities of the physical BMs from the error model, and then use the Bell state representations \eqref{eq:phys2block} and \eqref{eq:block2logic} to calculate the outcome probabilities of the BMs in the higher encoding levels. Effectively, we explore how the errors that happen on the actual photons propagate through the various levels of encoding. Finally, the outcome probabilities on the logical level can be used to calculate the transmission rate as well as the quantum bit error rate, which together yield the secure key rate.

\subsection{Propagation of errors}
\label{subsec:prop of errors}

The outcome probability refers to the probability to have a certain result in the BM when the entire error channel is applied to the original, undisturbed Bell state $\ket{\phi_{k,l}}^*$. This undisturbed Bell state is a theoretical construct and might not have existed at any point in time, because we assume the ancillary Bell states at every repeater station, each corresponding to one half of this undisturbed Bell state, to be generated on demand. The undisturbed Bell states merely provide a useful basis for the logical two-qubit space of the incoming qubit and one half of the ancillary Bell state. This approach allows us to take into account the entire error channel in one step (see also Sec.~\ref{sec:comm scheme}).

The outcome probabilities on the three encoding levels are stored in three matrices: $P$ (physical), $B$ (block), and $L$ (logical). In every matrix the column defines the input state, while the row refers to the BM result. Quite naturally, the sum of all entries of a certain column must be one. Since the four Bell states are distinguished by the two indices $(k,l)$ (not to be confused with the code parameters $(n,m)$) and ambiguous results for both indices are possible, initially there are in total nine possible BM outcomes on every level: $(0,0),(0,1),(1,0),(1,1),(0,?),(1,?),(?,?),(?,0),(?,1)$. However, of these results, the last two, where the index $l$ is identified, but $k$ is ambiguous, can never occur in the standard BM and, due to the form of the QPC, are also impossible on the higher encoding levels. For almost every error model some of the results are inaccessible, and we suppress them when displaying the matrix elements $P_{u,v}$, $B_{u,v}$ and $L_{u,v}$ in tables. Nevertheless, the index $u$ will always range from 1 to 7 and correspond to the position in the above list of BM outcomes.

As an example let us consider the case, where the only error present is photon loss due to the channel transmission. The error channel for a single physical qubit is then given by
\begin{align}
	\rho \rightarrow \eta_t \rho + (1-\eta_t)\ket{vac}\bra{vac},\quad \eta_t = e^{-\frac{L_0}{L_{\rm att}}},
\end{align}
where $L_{\rm att}$ is the attenuation length of the optical fiber (we assume $L_{\rm att}= \unit[22]{km}$ in our calculations) and $\eta_t$ is the probability of the photon not to be lost during transmission.

The probability of losing the photon $(1-\eta_t)$ is independent of the actual Bell state entering the BM and the loss is detected by the BM, since the detectors can distinguish between one and two photons. If one of the Bell states $\ket{\phi_{0,l}}$ enters the BM, both photons are directed to the same detector and no information about $l$ can be found. On the other hand, a $\ket{\phi_{1,l}}$-state will always be identified correctly, provided the photon is not lost. This gives the outcome probabilities $P_{u,v}^{\text{(only loss)}}$ on the physical level that are displayed in Table~ \ref{tab:Ponlyloss}.

\begin{table}[b]
\begin{ruledtabular}
\newcommand{\wida}{1.3cm}
\begin{tabular}{cC{\wida}C{\wida}C{\wida}C{\wida}}
$P_{u,v}^{(\text{only loss})}$ & $\ket{\phi_{0,0}}$ & $\ket{\phi_{0,1}}$ & $\ket{\phi_{1,0}}$ & $\ket{\phi_{1,1}}$ \\\hline\\[-1em]
$(1,0)$ & $0$ & $0$ & $\eta_t$ & $0$ \\
$(1,1)$ & $0$ & $0$ & $0$ & $\eta_t$ \\
$(0,?)$ & $\eta_t$ & $\eta_t$ & $0$ & $0$ \\
$(?,?)$ & $1-\eta_t$ & $1-\eta_t$ & $1-\eta_t$ & $1-\eta_t$
\end{tabular}
\end{ruledtabular}
\caption{Outcome probabilities on the physical level if only photon loss is present. The matrices $P$ (as well as the matrices $B$ and $L$) are defined as $7 \times 4$ matrix. However, in many occasions, entire rows have vanishing entries corresponding to BM outcomes that are inaccessible in the given error model. These ``zero-rows'' (in this particular instance the rows $1,2,$ and $6$ corresponding to outcomes $(0,0)$, $(0,1)$, and $(1,?)$) are not displayed in the tables throughout this work.}
\label{tab:Ponlyloss}
\end{table}

To obtain the outcome probabilities $B_{u,v}$ on the block level from those on the physical level, we use the Bell state representation \eqref{eq:phys2block} and an interpretation scheme that is chosen depending on the corresponding error model. This interpretation scheme determines, how combinations of BM results on the physical level should be interpreted on the block level. For example, if in one of the physical BMs a loss was detected, we are unable to identify the second index $l$ of that block. The interpretation scheme is given in the form of the expressions $f_u(\gamma)$ and is incorporated in the following formulas via $\delta_{f_u(\gamma)}$, which should be understood as a restriction on an index set. Here, $u$ is the row index of the outcome probability matrix $B$ and $\gamma \in \mathbb{N}_0^7$ specifies how often every BM result occurrs on the physical level: $\gamma_1=$number of results (0,0), $\gamma_2=$ number of results (0,1), and so on. As a general restriction we have $|\gamma|=\gamma_1+...+\gamma_7=m$, where $m$ corresponds to the size of a block, and so, the total number of physical BMs per block. In most cases, the expressions $f_u(\gamma)$ are given as follows:
\begin{align*}
	f_1(\gamma):&\quad \gamma_1+\gamma_2=m \wedge \gamma_2 \text{ is even} & (0,0)\\
	f_2(\gamma):&\quad \gamma_1+\gamma_2=m \wedge \gamma_2 \text{ is odd} & (0,1)\\
	f_3(\gamma):&\quad \gamma_3+\gamma_4=m \wedge \gamma_4 \text{ is even} & (1,0)\\
	f_4(\gamma):&\quad \gamma_3+\gamma_4=m \wedge \gamma_4 \text{ is odd} & (1,1)\\
	f_5(\gamma):&\quad \gamma_1+\gamma_2+\gamma_5 > \gamma_3+\gamma_4+\gamma_6 &\\
	&\quad \wedge \gamma_1+\gamma_2<m & (0,?)\\
	f_6(\gamma):&\quad \gamma_1+\gamma_2+\gamma_5 < \gamma_3+\gamma_4+\gamma_6 &\\
	&\quad \wedge \gamma_3+\gamma_4<m & (1,?)\\
	f_7(\gamma):&\quad \gamma_1+\gamma_2+\gamma_5 = \gamma_3+\gamma_4+\gamma_6 & (?,?)
\end{align*}
The index $k$ should be the same in every physical BM, so a majority voting decides ($\gamma_1+\gamma_2+\gamma_5$ vs. $\gamma_3+\gamma_4+\gamma_6$). The identification of the second index $l$ can only be successful, if no photons are lost. Moreover, we found that better communication rates will be achieved with the additional requirement that all physical BMs in a block must give the same value of $k$ in order to accept this block for the identification of $l$. (For the standard BM scheme as depicted in Fig.~\ref{fig:BM_setup}, this is always satisfied, since the index $l$ can only be identified if $k=1$. However, when using advanced BM schemes in a scenario, where also depolarizing errors are present, the extra requirement becomes important. For details see Sec.~\ref{subsec:advBM}.) For some error channels, the standard expressions $f_u(\gamma)$ are inappropriate, especially when considering on-off detectors, where losses can no longer be identified. We will give the appropriate expressions in Sec.~\ref{subsec: onoff}.

The outcome probabilities $B_{u,1}$ corresponding to an undisturbed $\ket{\phi_{0,0}}^{(m)}$-state on the block level can now be given as
\begin{align}
B_{u,1} &= \frac{1}{2^{m-1}}\sum_{\substack{r=0\\r \text{ even}}}^m  \sum_{\substack{\alpha,\beta\in \mathbbm{N}_0^{7}\\|\alpha|=m-r\\|\beta|=r}} \binom{m}{\alpha,\beta} P_{\cdot,1}^{\alpha} P_{\cdot,2}^{\beta}\ \delta_{f_u(\alpha+\beta)}.\nonumber\\[-1.8em]\label{eq:Bu1}
\end{align}
Here we use the common multi-index notation $x^{\alpha} = x_1^{\alpha_1}x_2^{\alpha_2}...$ together with the multinomial coefficient $\binom{m}{\alpha,\beta} = \binom{m}{\alpha_1,...,\alpha_7,\beta_1,...,\beta_7} = \frac{m!}{\alpha_1!...\beta_7!}$
and define the vectors $P_{\cdot,v}$ as the column vectors of the matrix $P_{u,v}$.
The sum index $r$ gives the number of incoming (physical) $\ket{\phi_{0,1}}$ states in this block. According to the Bell state representation \eqref{eq:phys2block} it must be even. The index vectors $\alpha$ and $\beta$ describe the number of outcomes and have the same form as $\gamma$. However, they still distinguish between incoming $\ket{\phi_{0,0}}$ and $\ket{\phi_{0,1}}$ states. The BM itself has no access to this distinction, which is represented by the fact that the expressions $f_u$ depend only on the combined numbers $\gamma = \alpha + \beta$. The combinatorical expression in Eq.~\eqref{eq:Bu1} can be simplified (see Appendix~\ref{sec:AppB}) to
\begin{align}
B_{u,1} = \frac{1}{2^{m}}\sum_{\substack{\gamma \in \mathbbm{N}_0^{7}\\|\gamma|=m}}& \binom{m}{\gamma}\ \delta_{f_u(\gamma)}\label{eq:Bu1a}\\
&\  \times \left[\left(P_{\cdot,1}\!+\!P_{\cdot,2}\right)^{\gamma} + \left(P_{\cdot,1}\!-\!P_{\cdot,2}\right)^{\gamma}\right],\nonumber
\end{align}
which has far less restrictions and is easier to compute. Furthermore, for most error models some of the summands $(P_{w,1}+P_{w,2})$ and/or $(P_{w,1}-P_{w,2})$ can be evaluated to become zero, thus eliminating the corresponding summation indices $\gamma_w$.

\begin{table*}[t]
\begin{ruledtabular}
\newcommand{\widb}{3.94cm}
\begin{tabular}{cC{\widb}C{\widb}C{\widb}C{\widb}}
$B_{u,v}^{(\text{only loss})}$ & $\ket{\phi_{0,0}}^{(m)}$ & $\ket{\phi_{0,1}}^{(m)}$ & $\ket{\phi_{1,0}}^{(m)}$ & $\ket{\phi_{1,1}}^{(m)}$ \\\hline\\[-1em]
$(1,0)$ & $0$ & $0$ & $\eta_t^m$ & $0$ \\
$(1,1)$ & $0$ & $0$ & $0$ & $\eta_t^m$ \\
$(0,?)$ & $1-(1-\eta_t)^m$ & $1-(1-\eta_t)^m$ & $0$ & $0$ \\
$(1,?)$ & $0$ & $0$ & $1-(1-\eta_t)^m-\eta_t^m$ & $1-(1-\eta_t)^m-\eta_t^m$ \\
$(?,?)$ & $(1-\eta_t)^m$ & $(1-\eta_t)^m$ & $(1-\eta_t)^m$ & $(1-\eta_t)^m$ \\\hline\hline\\[-1em]
$L_{u,v}^{(\text{only loss})}$ & $\ket{\phi_{0,0}}^{(n,m)}$ & $\ket{\phi_{0,1}}^{(n,m)}$ & $\ket{\phi_{1,0}}^{(n,m)}$ & $\ket{\phi_{1,1}}^{(n,m)}$ \\\hline\\[-1em]
$(0,0)$ & \multicolumn{2}{c}{\multirow{2}{*}{$\{[1-(1-\eta_t)^m]^n - [1-(1-\eta_t)^m-\tfrac{\eta_t^m}{2}]^n - [\tfrac{\eta_t^m}{2}]^n\}\mathbbm{1}_{2\times 2}$}} & \multicolumn{2}{c}{\multirow{2}{*}{$0_{2\times 2}$}} \\
$(0,1)$ & \multicolumn{2}{l}{} & \multicolumn{2}{l}{} \\
$(1,0)$ & \multicolumn{2}{c}{\multirow{2}{*}{$0_{2\times 2}$}} & \multicolumn{2}{c}{\multirow{2}{*}{$\{[1-(1-\eta_t)^m]^n - [1-(1-\eta_t)^m-\tfrac{\eta_t^m}{2}]^n + [\tfrac{\eta_t^m}{2}]^n\}\mathbbm{1}_{2\times 2}$}} \\
$(1,1)$ & \multicolumn{2}{l}{} & \multicolumn{2}{l}{} \\
\end{tabular}
\end{ruledtabular}
\caption{Outcome probabilities on the block level (top) and on the logical level (bottom) if only photon loss is present.}
\label{tab:BLonlyloss}
\end{table*}

To obtain the outcome probabilities for the other three block Bell states $\ket{\phi_{k,l}}^{(m)}$, only minor changes must be made in Eqs.~\eqref{eq:Bu1} and \eqref{eq:Bu1a}. If the second index $l$ is 1 ($B_{u,2}$ and $B_{u,4}$), the sum index $r$ must be odd instead of even in Eq.~\eqref{eq:Bu1}. In Eq.~\eqref{eq:Bu1a} this changes the $+$ in the middle of the square brackets to $-$. On the other hand, for $k=1$ ($B_{u,3}$ and $B_{u,4}$), the vectors $P_{\cdot,1}$ and $P_{\cdot,2}$ are replaced by $P_{\cdot,3}$ and $P_{\cdot,4}$, respectively.

Let us now return to our example that includes only transmission loss. Since the matrix $P^{(\text{only loss})}$ has non-zero entries only in the rows 3,4,5, and 7, we immediately obtain $\gamma_1=\gamma_2=\gamma_6=0$. This tells us that the conditions $f_1$ and $f_2$ cannot be met. Additionally, for $B_{u,1}^{(\text{only loss})}$ and $B_{u,2}^{(\text{only loss})}$ only the first and second columns of $P^{(\text{only loss})}$ must be considered, where we have $\gamma_3=\gamma_4=0$. This yields, for example,
\begin{align}
	B_{7,1}^{(\text{only loss})} &= \frac{1}{2^m} \sum_{\substack{\gamma_1,\gamma_2\\\gamma_1+\gamma_2=m}} \binom{m}{\gamma_1,\gamma_2} \delta_{\gamma_2=0}\nonumber\\
	&\qquad\qquad\qquad \times [2(1-\eta_t)]^{\gamma_1} [2\eta_t]^{\gamma_2}\nonumber\\
	&=(1-\eta_t)^m.
\end{align}
As expected, this is the probability of losing all $m$ photons in the block, which is the only case, where no information can be obtained. Similar calculations give the other entries of $B^{(\text{only loss})}$, which are displayed in Table~\ref{tab:BLonlyloss}

The step from the block level to the logical level is very similar to the procedure as shown above. Yet, this time we must use the Bell state representation \eqref{eq:block2logic} and choose a different set of interpretation rules $g_u(\lambda)$, where $\lambda$ plays the same role as $\gamma$ did before:
\begin{align*}
	g_1(\lambda):&\quad \lambda_7=0 \wedge \lambda_3 + \lambda_4 +\lambda_6 \text{ is even} &\\
	&\quad \wedge \lambda_1+\lambda_3 > \lambda_2+\lambda_4 & (0,0)\\
	g_2(\lambda):&\quad \lambda_7=0 \wedge \lambda_3 + \lambda_4 +\lambda_6 \text{ is even} &\\
	&\quad \wedge \lambda_1+\lambda_3 < \lambda_2+\lambda_4 & (0,1)\\
	g_3(\lambda):&\quad \lambda_7=0 \wedge \lambda_3 + \lambda_4 +\lambda_6 \text{ is odd} &\\
	&\quad \wedge \lambda_1+\lambda_3 > \lambda_2+\lambda_4 & (1,0)\\
	g_4(\lambda):&\quad \lambda_7=0 \wedge \lambda_3 + \lambda_4 +\lambda_6 \text{ is odd} &\\
	&\quad \wedge \lambda_1+\lambda_3 < \lambda_2+\lambda_4 & (1,1)\\
	g_5(\lambda):&\quad \lambda_7=0 \wedge \lambda_3 + \lambda_4 +\lambda_6 \text{ is even} &\\
	&\quad \wedge \lambda_1+\lambda_3 = \lambda_2+\lambda_4 & (0,?)\\
	g_6(\lambda):&\quad \lambda_7=0 \wedge \lambda_3 + \lambda_4 +\lambda_6 \text{ is odd} &\\
	&\quad \wedge \lambda_1+\lambda_3 = \lambda_2+\lambda_4 & (1,?)\\
	g_7(\lambda):&\quad \lambda_7>0 & (?,?)
\end{align*}
In order to identify the first index $k$ we must know it in every block, thus $\lambda_7=0$. The identification of the second index $l$, however, is done via majority voting in this step ($\lambda_1+\lambda_3$ vs. $\lambda_2+\lambda_4$). Analogously to Eqs.~\eqref{eq:Bu1} and \eqref{eq:Bu1a}, we get the outcome probabilities $L_{u,1}$ corresponding to an undisturbed $\ket{\phi_{0,0}}^{(n,m)}$-state on the logical level:

\begin{align}
L_{u,1} &= \frac{1}{2^{n-1}}\sum_{\substack{s=0\\s \text{ even}}}^n \sum_{\substack{\mu,\nu\in \mathbbm{N}_0^{7}\\|\mu|=n-s\\|\nu|=s}} \binom{n}{\mu,\nu} B_{\cdot,1}^{\mu} B_{\cdot,3}^{\nu}\ \delta_{g_u(\mu+\nu)}.\nonumber\\[-1.8em]\label{eq:Lu1}
\end{align}

Here $s$ is the number of incoming $\ket{\phi_{1,0}}^{(m)}$ block states. Note that, because of the different Bell state representation, we now have to use the columns 1 and 3 of the matrix $B$, whereas before we had to use columns 1 and 2 of matrix $P$. The same simplification as before (see Appendix~\ref{sec:AppB}) yields

\begin{align}
L_{u,1} = \frac{1}{2^{n}}\sum_{\substack{\lambda \in \mathbbm{N}_0^{7}\\|\lambda|=n}}& \binom{n}{\lambda}\ \delta_{g_u(\lambda)} \label{eq:Lu1a}\\
&\  \times \left[\left(B_{\cdot,1}\!+\!B_{\cdot,3}\right)^{\lambda} + \left(B_{\cdot,1}\!-\!B_{\cdot,3}\right)^{\lambda}\right].\nonumber
\end{align}

The required modifications to obtain the outcome probabilities for the other three Bell states $\ket{\phi_{k,l}}^{(n,m)}$ are: for $k=1$ ($L_{u,3}$ and $L_{u,4}$) the index $s$ in Eq.~\eqref{eq:Lu1} must be odd, which turns the $+$ in the square brackets in Eq.~\eqref{eq:Lu1a} into a $-$; for $l=1$ ($L_{u,2}$ and $L_{u,4}$) the vectors $B_{\cdot,1}$ and $B_{\cdot,3}$ are replaced by $B_{\cdot,2}$ and $B_{\cdot,4}$, respectively.

Of the seven possible BM outcomes on the logical level only the first four are relevant: those that indicate a successful BM. Since there is no additional code on top of the QPC that could potentially make use of the partial information of outcomes 5 and 6, we display only the four relevant matrix rows of $L$ in Table~\ref{tab:BLonlyloss}. The calculation of the outcome probabilities $L_{u,v}^{(\text{only loss})}$ in our exemplary transmission-loss-only scenario can be done in perfect analogy to that of $B_{u,v}^{(\text{only loss})}$. These results are also part of Table~\ref{tab:BLonlyloss}.

Note that the resulting $4 \times 4$ array as shown in Table~\ref{tab:BLonlyloss} is diagonal with all diagonal entries non-zero. This means that whenever one of the four logical Bell states is obtained as an outcome of the BM, this BM result is error-free, just as required for an unambiguous BM on the logical level. As opposed to the BMs on the physical level, however, all four Bell states can sometimes be identified. Nonetheless, in a realistic communication scenario, there will be other sources of error beyond loss and the logical BMs do become faulty. This effect and its influence on the transmission rates will be discussed next.

\subsection{Transmission rates}
\label{subsec:trans rates}

The outcome probabilities $L_{u,v}$ on the logical level allow for an investigation of the performance of the logical BM. The probability for the BM to identify a Bell state correctly (assuming equal a priori probabilities) is given by
\begin{align}
	L_{\rm id} = \tfrac{1}{4}\left(L_{1,1}+L_{2,2}+L_{3,3}+L_{4,4}\right),
\end{align}
while the probabilities to induce an unheralded Pauli error are given by
\begin{align}
	L_{\rm X} &= \tfrac{1}{4}\left(L_{3,1}+L_{4,2}+L_{1,3}+L_{2,4}\right),\\
	L_{\rm Y} &= \tfrac{1}{4}\left(L_{4,1}+L_{3,2}+L_{2,3}+L_{1,4}\right),\\
	L_{\rm Z} &= \tfrac{1}{4}\left(L_{2,1}+L_{1,2}+L_{4,3}+L_{3,4}\right),
\end{align}
respectively. These unheralded Pauli errors are not to be confused with the heralded ``errors'' that the results of the BM induce as part of the teleportation protocol. The latter can, of course, be corrected, either immediately or later and altogether in a final step of the communication protocol (see Sec.~\ref{sec:comm scheme}) and so they do not influence the efficiency of the communication scheme. The former, however, describe unheralded errors that reduce the efficiency of the BM and thus that of the communication scheme.

When analyzing the communication scheme, we are first interested in the raw transmission rate. This is related to the probability that all the BMs at the intermediate repeater stations herald a successful teleportation. It is given by
\begin{align}
	p_{\rm trans} = (L_{\rm id} + L_{\rm X} + L_{\rm Y} + L_{\rm Z})^N,
\end{align}
where $N=\tfrac{L_{\rm tot}}{L_0}$ is the number of repeater stations taking part in the transmission (including the one at the receiver). This probability only represents the number of accepted transmissions. It contains no information whether the transmitted logical qubit has undergone an unheralded Pauli error. Apart from qubit state fidelities, a common figure of merit to account for these unheralded errors is the secure key rate that can be achieved when using the communication scheme for quantum key distribution (QKD). In order to be comparable to other communication schemes, we choose the secure key rate of the BB84 protocol \cite{BB84}, which is also the figure of merit for the communication scheme in Ref. \cite{LLPRL}. Since the BB84 protocol utilizes measurements in the $X$- and in the $Z$-basis, we define the associated quantum bit error rates (QBERs) as the rate of obtaining an $X$-flipped ($Z$-flipped) result at the receiver divided by the raw transmission rate $p_{\rm trans}$. An $X$-flipped ($Z$-flipped) result is obtained, when the total number of hidden $X$ and $Y$ errors ($Z$ and $Y$ errors) occurring in the BMs of the transmission is odd. We therefore get
\begin{align}
	Q_{\rm X/Z} &= \frac{1}{p_{\rm trans}} \sum_{\substack{a_{\rm id},a_{\rm X},a_{\rm Y},a_{\rm Z} \\ a_{\rm id}+a_{\rm X}+a_{\rm Y}+a_{\rm Z} =N}} \binom{N}{a_{\rm id},a_{\rm X},a_{\rm Y},a_{\rm Z}}  \nonumber \\
	&\qquad \qquad \qquad \qquad \times  L_{\rm id}^{a_{\rm id}}L_{\rm X}^{a_{\rm X}}L_{\rm Y}^{a_{\rm Y}}L_{\rm Z}^{a_{\rm Z}} \delta_{a_{\rm X/Z} + a_{\rm Y} \text{ odd}} \nonumber\displaybreak\\
	&= \frac{1}{2}\left[1-\frac{(L_{\rm id} \mp L_{\rm X} - L_{\rm Y} \pm L_{\rm Z})^N}{(L_{\rm id} + L_{\rm X} + L_{\rm Y} + L_{\rm Z})^N}\right].
\end{align}
The secure key rate obtainable by BB84 is then given by
\begin{align}
	R &= \tfrac{1}{t_0}\max\left[p_{\rm trans} \left(1-2 h(Q) \right),0\right],
\end{align}
where $t_0$ is the elementary time needed at every repeater station until the incoming signal qubit has been processed and a fresh encoded Bell state is ready for teleporting and error-correcting the next qubit. It corresponds to the standard time unit in our proposal. The combined QBER is given by $Q=(Q_{\rm X} + Q_{\rm Z})/2$ and $h(Q)$ is the binary entropy function
\begin{align}
	h(Q) &= -Q \log_2(Q) - (1-Q) \log_2(1-Q).
\end{align}
It should be noted at this point that the given secure key rates are only lower bounds to the actually achievable key rates. This is due to the fact that a lot of averaging happens at several points of the analysis. In a recent work \cite{namiki16}, Namiki et al. showed for a very similar setup (the one described in Ref. \cite{LLPRL}) that by keeping track of all the information obtained from the BMs on the physical level, better estimations on the secure key rate are possible
%%%%%%%%%%%%%%FOOTNOTE
\footnote{The approach used in \cite{namiki16} that concentrates on utilizing the results of all physical Bell measurements and obtaining the secure key rate from a Monte-Carlo simulation, is in principle also applicable to our scheme. However, Namiki et al. apply their theoretical results only on QPC-codes of relatively small size ($n,m\leq 4$) where our scheme still suffers quite heavily from the non-unit Bell measurement efficiency of linear optics. In addition to the requirement of larger code sizes, the reduced symmetry in the optical Bell measurement compared to a standard BM via a CNOT gate makes the numerical implementation of the results of \cite{namiki16} rather expensive in terms of computer resources. We are therefore content with using the coarse grained approach presented in the main text and in Ref. \cite{LLPRL}.}.
%%%%%%%%%%%%%%FOOTNOTE

When looking at our transmission-loss-only scenario, we find that the success probability of the raw transmission $p_{\rm trans}^{\text{(only loss)}}$ is given by
\begin{align}
	p_{\rm trans}^{\text{(only loss)}} &= \left\{\left[1\!-\!(1\!-\!\eta_t)^m\right]^n\!-\!\left[1\!-\!(1\!-\!\eta_t)^m\!-\!\frac{\eta_t^m}{2}\right]^n\right\}^N.
\end{align}
Furthermore, we see that transmission loss alone cannot induce Pauli errors ($L_{\rm X}=L_{\rm Y}=L_{\rm Z}=0$) and that, consequently, the quantum bit error rate is $Q^{\text{(only loss)}}=0$. Therefore, the secure key rate obtained for this simple error model is just the raw transmission probability divided by the elementary time $t_0$, $R^{\text{(only loss)}}=\tfrac{1}{t_0} p_{\rm trans}^{\text{(only loss)}}$ (see the rate analysis of the loss-only scenario in Ref. [prl]: the qubit transmission rates calculated there are equivalent to secure key rates). A further discussion of this rate is part of Sec.~\ref{subsec:loss}.
\section{Results}
\label{sec: results}

In this section we apply the results of Sec.~\ref{sec:error analysis} to various error models. We start with the case of only loss that has been partly discussed already, but here we shall also include the effect of detector loss. We then focus on Pauli-type errors, especially the depolarizing error channel, and on the effect of dark counts in the photon detectors of the BM. This is followed by a discussion on whether an improvement of the BM efficiency on the physical level is worth the additional experimental effort. Finally, the possibility of using on-off detectors instead of PNRDs will be explored.

\subsection{Loss only}
\label{subsec:loss}

The case where transmission loss is the only error source present in our setup has already been described in Sec.~\ref{sec:error analysis}. However, losses may also occur within the repeater stations, especially within the photon detectors. Lossy detectors with efficiency $\eta_d$ are usually modeled by a beam splitter with transmission coefficient $\eta_d$ in front of a perfect photon detector with the second input mode of the beam splitter being vacuum. This model does not only reproduce the detector efficiency $\eta_d$, which is defined as the probability to detect a single photon when a single photon enters the detector, but also gives the probabilities for the measurement results when multiple photons enter the detector, e.g. for two incoming photons, the probability of detecting exactly two photons is $\eta_d^2$. For every successful (or partly successful) BM on the physical level, two photons (of which one stems from the propagating incoming signal and the other one comes from the local ancilla Bell state) must be detected by the four detectors of the BM. We therefore obtain the probability that the transmitted photon is not lost on the way and both photons are detected $\eta = \eta_d^2\exp(-\frac{L_0}{L_{\rm att}})$. This takes the place of $\eta_t$ in Sec.~\ref{sec:error analysis}, yielding the secret key rate
\begin{align}
	R^{\text{(loss)}}\!&=\! \frac{1}{t_0}\!\left\{\left[1\!-\!(1\!-\!\eta)^m\right]^n \!-\!\left[1\!-\!(1\!-\!\eta)^m\!-\!\frac{\eta^m}{2}\right]^n\right\}^{N}. \label{eq:Rloss}
\end{align}
The case where only loss is present (either due to transmission or lossy detectors) is one of the rare cases, for which we found a compact analytic formula for the secure key rate, i.e. Eq.~\eqref{eq:Rloss}. It can be understood quite nicely by recalling the correction properties of the QPCs (see Sec.~\ref{sec:QPC code}): for the loss to be correctable, at least one intact photon pair must be left in every block and one block must remain completely intact. The first term in Eq.~\eqref{eq:Rloss}, $\left[1-(1-\eta)^m\right]^n$, is the probability that in every one of the $n$ blocks less than all $m$ photon pairs are subject to (either transmission or detector) loss. The second term, $\left[1-(1-\eta)^m-\tfrac{\eta^m}{2}\right]^n$, can be best understood by first ignoring the factor $\tfrac{1}{2}$. In that case, it gives the probability that in every block at least one but not all photon pairs are left. These terms must be discarded due to the second condition of at least one intact block. The factor $\tfrac{1}{2}$ now corresponds to the BM efficiency at the block level: of all the blocks where all photon pairs are left, only half are useful, because the other half gives $\ket{\phi_{0l}}^{(m)}$ states, where the index $l$ cannot be identified.

\begin{figure}[t]
\hspace{2mm}\includegraphics[width=0.25\textwidth]{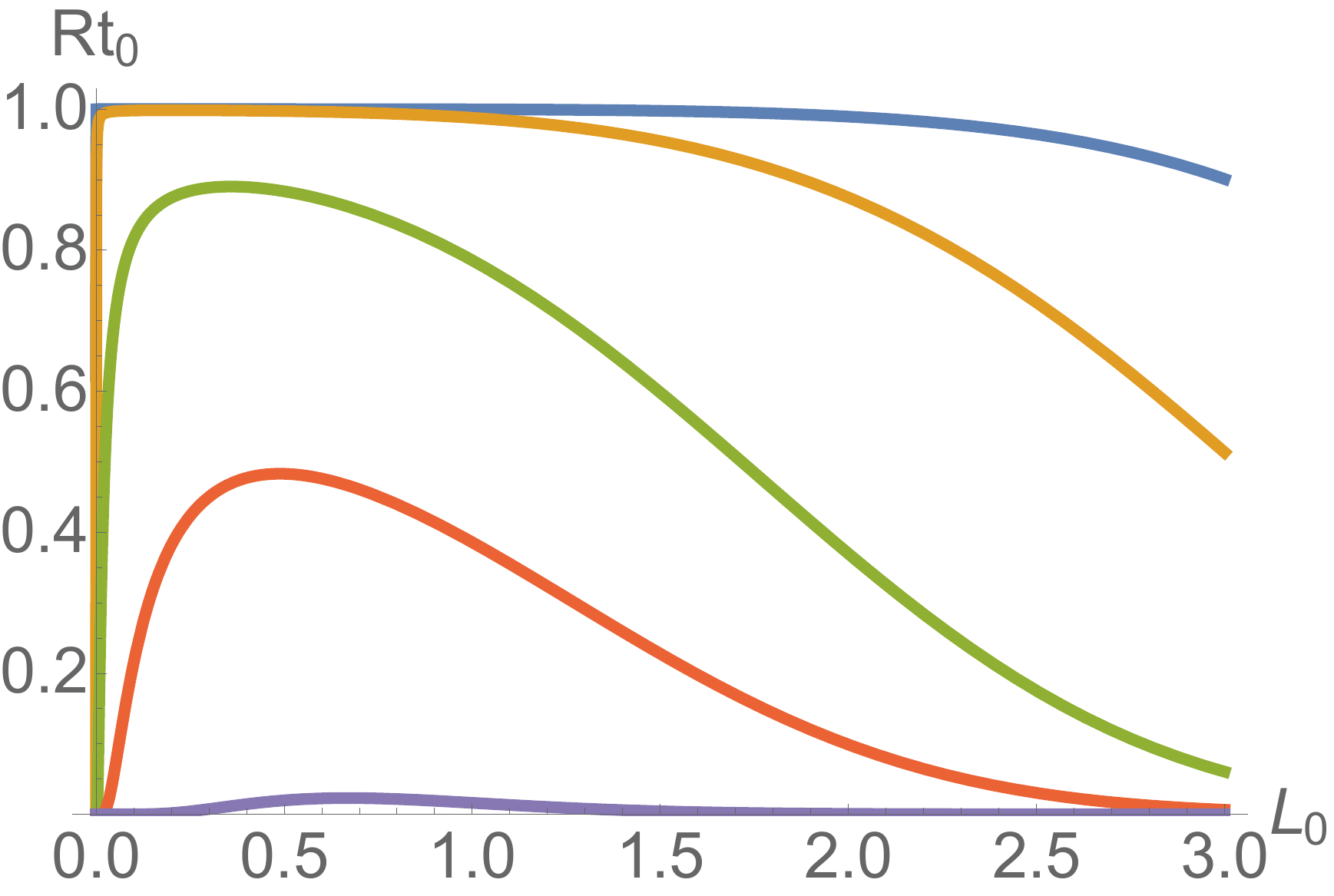} \hfill \includegraphics[width=0.215\textwidth]{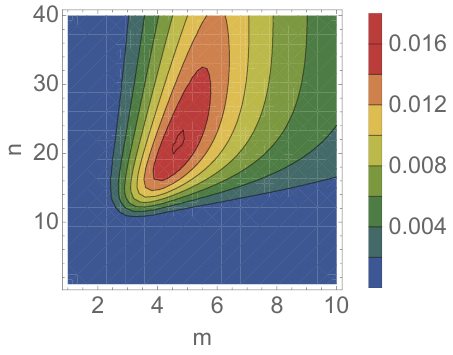}
\caption{Left: Total success probability $R t_0$ vs. repeater spacing $L_0$ in km for a communication distance of $L_{\rm tot}=\unit[1\,000]{km}$ and various encodings (from bottom to top: (10,3),(13,4),(16,4),(23,5),(35,6)). Right: Inverse of the cost function $C(\unit[1\,000]{km})$ as a function of the code parameters $n,m$. At every point the optimal repeater spacing $L_0$ is chosen. The most cost efficient code is $(23,5)$ with a repeater spacing of $L_0\approx \unit[2.4]{km}$ yielding $R t_0 = 76.18\%$.}\label{fig:plotspsucc}
\end{figure}

Next, we will discuss some properties of the secure key rate in Eq.~\eqref{eq:Rloss} as a function of the loss $\eta$ and hence as a function of the repeater spacing $L_0$. In the limit of very small loss ($\eta\rightarrow 1$), corresponding to very good detectors and small repeater spacing, the term in the curly braces in Eq.~\eqref{eq:Rloss} goes to $1-2^{-n}$, which is just the success probability of the logical BM when no loss occurs. However, the small repeater spacing $L_0$ in the exponent $N=\tfrac{L_{\rm tot}}{L_0}$, corresponding to very bad detectors or large repeater spacing (typically we may assume reasonable $\eta_d$ and so this means large repeater spacing), pushes the rate to zero. On the other hand, for very large loss ($\eta\rightarrow 0$) we find that the terms in the two square brackets in Eq.~\eqref{eq:Rloss} go to zero and hence the secure key rate decays with increasing repeater spacing $L_0$. In between the two extreme cases of very small and very large repeater spacings (or, more generally, loss) the secure key rate has a single maximum. Even for large total distances $L_{\rm tot}$ it is possible to find appropriate values of $n$, $m$, and $L_0$ such that high secure key rates ($R^{\text{(loss)}} t_0 \lessapprox 1$) can be achieved. In most cases the repeater spacing that, for given $L_{\rm tot}$, $n$ and $m$, maximizes the secure key rate is in the range of hundreds of meters up to a few kilometers. Some examples are shown in Fig.~\ref{fig:plotspsucc}.

In addition to the search of code parameters that yield high secure key rates, we are also interested in optimizing these parameters with respect to the experimental costs \cite{LLPRL}. To do so, in our case, we assume the cost of a setup to be proportional to the total number of photons used in the setup. At every repeater station a logical Bell state has to be supplied, corresponding to a total number of photons of $(2 n m)\tfrac{L_{\rm tot}}{L_0}$. In order to find an optimal code for a given distance that the communication scheme shall cover we therefore use the cost function
\begin{align}
	C = \frac{n m}{R L_0}. \label{eq:cost}
\end{align}
For example, for a total distance of \unit[1\,000]{km} and perfect detectors ($\eta_d=1$) the inverse $1/C$, which corresponds to the repeater rate per photons used, is shown in Fig.~\ref{fig:plotspsucc}. We find that $(n,m)=(23,5)$ and $L_0=\unit[2.4]{km}$ are most cost efficient and give a secure key rate $R^{\text{(loss)}} = 0.76 t_0^{-1}$. In a more demanding scenario of a total distance of \unit[10\,000]{km} and $\eta_d=0.97$ the optimal choices $(n,m)=(50,7)$ and $L_0=\unit[1.6]{km}$ yield $R^{\text{(loss)}} = 0.77 t_0^{-1}$. As a general trend we see that the number of blocks $n$ should be chosen significantly larger than the block size $m$. Additionally, we found that both lossier detectors and larger total distances require not only the use of larger codes, but also a reduced repeater spacing.

It should be kept in mind that this cost analysis relies on the assumption that the cost of generating the ancillary QPC($n,m$) Bell states is linear in the number of photons in the final state. In Appendix~\ref{sec:AppC} we present a scheme to generate the required QPC-Bell states with the help of coherent photon conversion, a nonlinear technique first presented in Ref.~\cite{langford11} which exactly achieves this linear scaling. However, it is also possible to restrict state generation to the methods of linear optics. In this case a large overhead of single photon sources and the use of feedforward operations is necessary for the so-called multiplexing. In Appendix~\ref{sec:AppD} we give an extensive description of this linear optical state generation (see also Refs.~\cite{ATL,pant2016rate,Li2015}) and also show that the resource cost in this case scales approximately as $(n m)^{\log_2(8/3)} \approx (n m)^{1.415}$. This scaling is, of course, worse than the linear one achieved by the coherent-photon-conversion-based state generation scheme and it will change what code sizes are most cost efficient. In general, a scaling larger than linear will yield smaller QPC sizes as an optimal choice. Here and in the following we restrict the discussion of the optimal code sizes for the various error models to the linearly scaling cost function.

When comparing our scheme to other recent quantum communication schemes it is important to also take the elementary time $t_0$ into consideration. Since our logical Bell states are assumed to be available on demand, $t_0$ corresponds only to the duration of the linear-optics processing with photon detection. Compared to those times required in a matter-based scheme with $t_0 \sim 1\mu s$ (even assuming future enhanced ion-cavity coupling strengths \cite{LLPRL}) or an all-optical scheme including feedforward \cite{ATL} with $t_0 \sim 10 ns$ (provided all circuits can be integrated \cite{Prevedel}), corresponding to rates $R\sim \unit[]{MHz}$ or $R\sim \unit[0.1]{GHz}$, respectively, our static linear optical scheme allows, in principle, for $\unit[]{GHz}-$rates and beyond
%%%%%%%%%%%%%%%FOOTNOTE
\footnote{Provided that the signal repetition rates and the detector bandwidths are sufficiently high.}. 
%%%%%%%%%%%%%%FOOTNOTE
\subsection{Depolarizing Errors}
\label{subsec:depol}

\begin{table}[b]
\begin{ruledtabular}
\begin{tabular}{ccccc}
$P_{u,v}^{(\text{depol})}$ & $\ket{\phi_{0,0}}$ & $\ket{\phi_{0,1}}$ & $\ket{\phi_{1,0}}$ & $\ket{\phi_{1,1}}$ \\\hline\\[-1em]
$(1,0)$ & $\tfrac{\epsilon}{2}\eta$ & $\tfrac{\epsilon}{2}\eta$ & $\left(1-\tfrac{3 \epsilon}{2}\right)\eta$ & $\tfrac{\epsilon}{2}\eta$ \\
$(1,1)$ & $\tfrac{\epsilon}{2}\eta$ & $\tfrac{\epsilon}{2}\eta$ & $\tfrac{\epsilon}{2}\eta$ & $\left(1-\tfrac{3 \epsilon}{2}\right)\eta$ \\
$(0,?)$ & $\left(1-\epsilon\right)\eta$ & $\left(1-\epsilon\right)\eta$ & $\epsilon \eta$ & $\epsilon \eta$ \\
$(?,?)$ & $1-\eta$ & $1-\eta$ & $1-\eta$ & $1-\eta$
\end{tabular}
\end{ruledtabular}
\caption{Outcome probabilities on the physical level if photon loss and depolarizing errors are present.}
\label{tab:Pdepol}
\end{table}

Another type of errors that is frequently observed in optical experiments is the class of photon-number-preserving errors. In the case of dual-rail encoding on the physical qubits (polarization encoding is a form of dual-rail encoding), these photon-number-preserving errors can be represented in terms of Pauli errors. We therefore investigate the influence that Pauli errors on the physical level have on the transmission and secure key rates. As a prototype of Pauli-type errors, we choose the depolarizing error channel
\begin{align}
	\rho \rightarrow (1-2\epsilon) \rho + \frac{\epsilon}{2}\sum_{i=0}^3 \sigma_i \rho \sigma_i,
\end{align}
where $\sigma_i$ are the four Pauli operators $\{\mathbbm{1},X,Y,Z\}$. Since the four Pauli operators are weighted equally, the depolarizing error channel is the most generic Pauli-type error channel and gives a quantum error correction code the broadest challenge. The outcome probabilities $P_{u,v}^{(\text{depol})}$ on the physical level are given in Table~\ref{tab:Pdepol}. Of course, the transmission loss due to fiber attenuation is also present, and as in Sec.~\ref{subsec:loss} lossy detectors can be incorporated by choosing $\eta$ accordingly. To calculate the outcome probabilities on the higher encoding levels [$B_{u,v}^{(\text{depol})}$ and $L_{u,v}^{(\text{depol})}$] we again use the index restrictions $f_u(\gamma)$ and $g_u(\lambda)$ given in Sec.~\ref{subsec:prop of errors}.

\begin{figure}[t]
\includegraphics[width=0.235\textwidth]{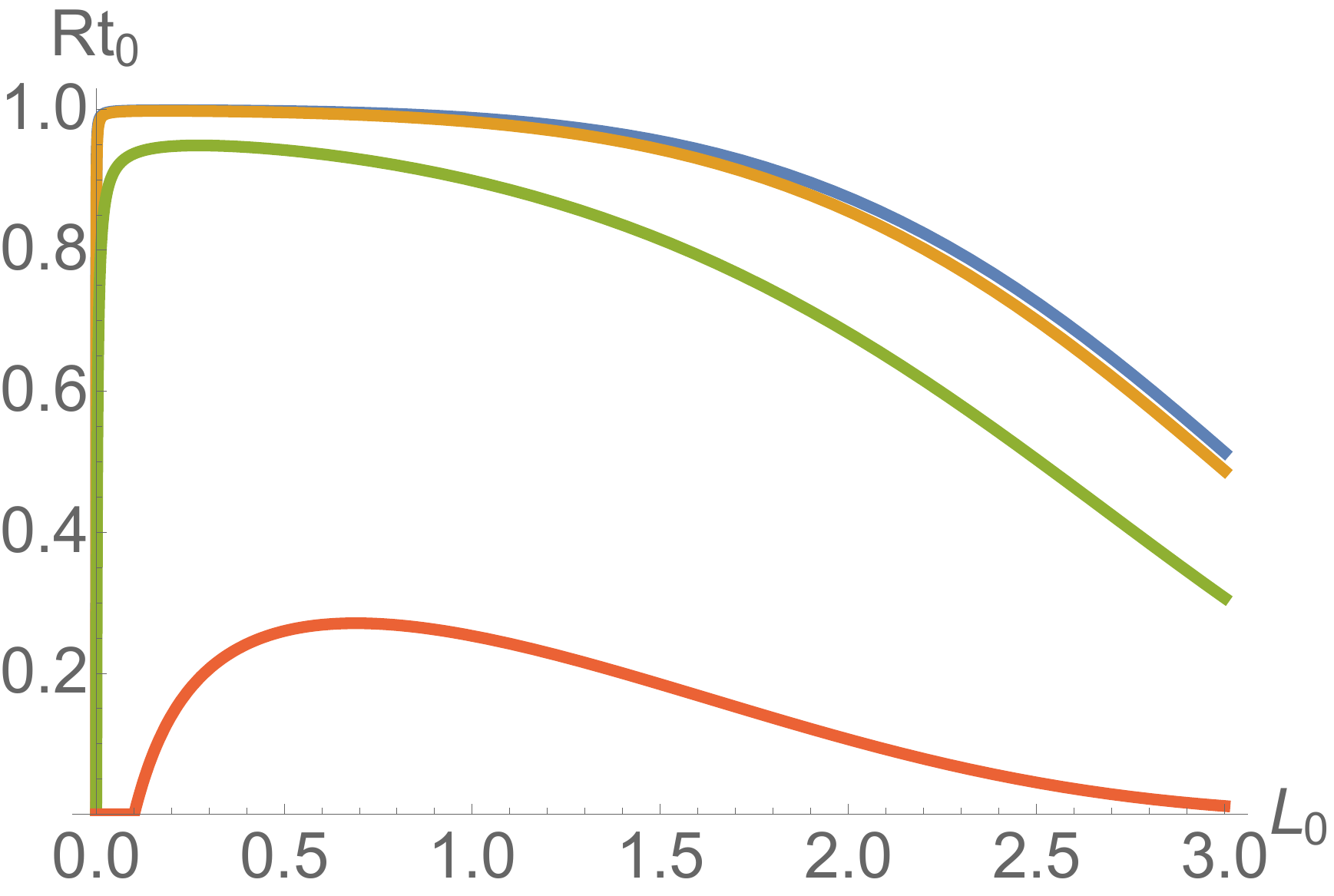} \hfill \includegraphics[width=0.235\textwidth]{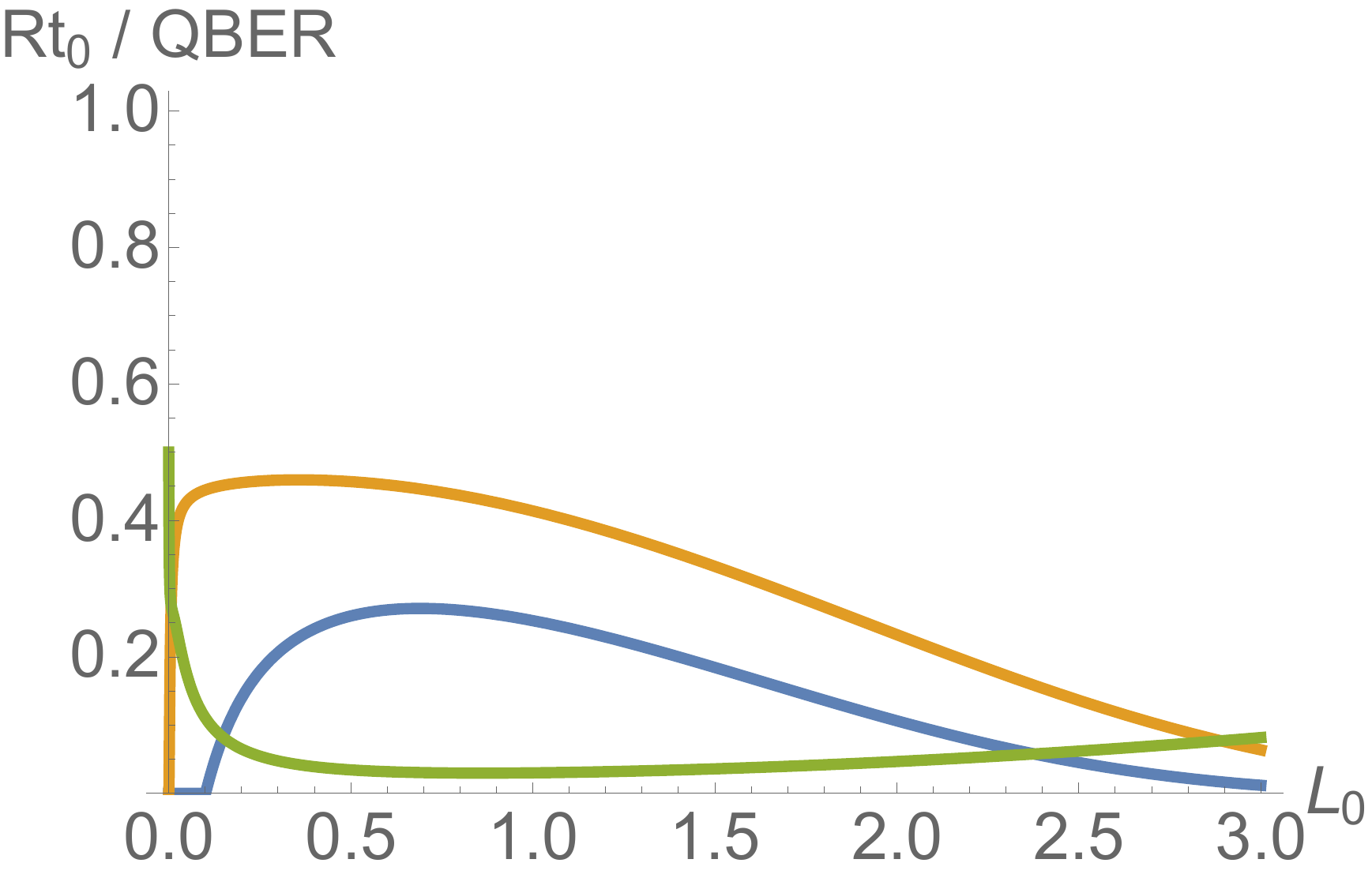}
\caption{Influence of depolarizing errors on the secure key rate ($L_{\rm tot}=\unit[1\,000]{km}$ and rates as $R t_0$ in both plots) -- Left: Secure key rate of QPC(23,5) (which is optimal in terms of cost for $\epsilon=0$) as a function of the repeater spacing. From top to bottom: $\epsilon=0,10^{-4},10^{-3},5\cdot10^{-3}$. Right: Raw transmission rate (orange, top), secure key rate (blue, mid) and QBER (green, bottom) of QPC(23,5) for $\epsilon=5\cdot10^{-3}$.  For very small repeater spacings the QBER exceeds $11\%$ and thus the secure key rate vanishes. This is an extreme case, since $(23,5)$ is not the optimal parameter choice anymore. For the optimal (i.e., most cost efficient) code (28,6) the QBER is much smaller and therefore the difference between raw transmission rate and secure key rate is smaller as well. In any case, the main effect of the depolarizing error is a reduction of the raw transmission rate.}\label{fig:depol}
\end{figure}

Unfortunately, both the transmission rate and the secure key rate cannot be represented in a compact analytic form like in Eq.~\eqref{eq:Rloss}. Nevertheless, various conclusions can be drawn from the numerical results. As expected, Pauli errors on the physical level can lead to logical Pauli errors. The QBER no longer vanishes in this case and therefore the secure key rate is lower than the raw transmission rate. Although there are parameter ranges, where the QBER is above $11\%$, which is the point where the secure key rate drops to zero, in most cases the main effect of the depolarizing errors is a decrease in the raw transmission rate (see Fig.~\ref{fig:depol}).

In general, errors  below $\epsilon = 10^{-4}$ hardly have an effect on the secure key rate. Errors of the order  of $\epsilon \approx 10^{-3}$ reduce the secure key rates, but the optimal code parameters ($n$, $m$, $L_0$) are almost identical to those for $\epsilon=0$. When considering even larger errors ($\epsilon \approx 10^{-2}$) the optimal code size grows and the repeater spacing has to be reduced. For example, $L_{\rm tot}=\unit[1\,000]{km}$, $\eta_d=1$ and $\epsilon=10^{-2}$ lead to optimal (i.e., most cost efficient) code parameters $(n,m)=(50,8)$ and $L_0=\unit[2.0]{km}$ and a secure key rate $R^{(\text{depol})} = 0.69 t_0^{-1}$. It appears to be a common theme that the secure key rate at the optimal parameter choices becomes $R \approx 0.7 t_0^{-1}$, even though rates much closer to one are possible with smaller repeater spacings (resulting in higher costs).

\subsection{Dark Counts}
\label{subsec:darkcounts}

In Sec.~\ref{subsec:loss} we already discussed that photon detectors usually have a limited efficiency and do not detect every photon that enters. This is modeled by a beam splitter with transmission coefficient $\eta_d$ in front of a perfect detector. Another error that often occurs in photon detectors is a dark count. Usually a dark count is defined as a click of the detector, when no photon was present. However, when considering photon-number-resolving detectors, the situation becomes more complicated, because it is then generally possible to detect more photons than those that were actually present. To model a detector with dark counts we extend the idea for the lossy detector. Again, a beam splitter with transmission coefficient $\eta_d$ is put in front of a perfect detector, but instead of leaving the second input mode of the beam splitter empty (i.e., using vacuum), we use a thermal state
\begin{align}
	\rho_{\rm th}(\bar{n}) = \sum_{j=0}^\infty \frac{\bar{n}^j}{(\bar{n}+1)^{j+1}} \ket{j}\bra{j},
\end{align}
where $\bar{n}$ is the average photon number of the state.

This model can be used to derive the probability $p_{\mu \rightarrow \nu}^{(\rm dc)}$ of detecting $\nu$ photons, when $\mu$ photons entered the detector. The calculation of these probabilities is rather lengthy and can be found in Appendix~\ref{sec:AppE}. Usually the dark count probability is defined as the probability to detect at least one photon, although no photon entered the detector. In this model, it is given by
\begin{align}
	p_{\rm dc} = 1-p_{0 \rightarrow 0}^{\rm (dc)} = \frac{\bar{n}(1-\eta_d)}{1+\bar{n}(1-\eta_d)} \approx \bar{n}(1-\eta_d), \label{eq:pdc}
\end{align}
where the approximation holds for high efficiency detectors and small average photon numbers.

Of course, only a small set of the probabilities $p_{\mu \rightarrow \nu}^{(\rm dc)}$ is actually needed for our purposes. First, since there is no other source of additional photons in our communication scheme, the number of photons entering a detector is limited to a maximum of two. Second, the detectors are not required to resolve the photon number for more than two photons: if a detector recognizes more than two photons and the other detectors of the physical BM find none, it is most likely that the incoming state was a $\ket{\phi_{0l}}$ state [recall that in the ideal case these states lead to the click pattern (2,0,0,0) or permutations thereof]. On the other hand, if more than one detector recognizes multiple photons, the BM should be regarded as a failure. Effectively, the detectors must only be able to distinguish the cases $0,1,$ and $\geq2$ photons.

With the help of the probabilities $p_{\mu \rightarrow \nu}^{(\rm dc)}$ it is possible to derive the outcome probabilities $P_{u,v}^{(\text{dc})}$ for the physical BMs. It should be kept in mind that in addition to the detector inefficiencies, the transmission loss is also still present. It is, for example, possible to lose a photon during the transmission, but still detect two photons in the corresponding physical-level BM, because a dark count occurred. Whether the state is identified correctly or not in this case depends on which detector was affected.

When analyzing the outcome probabilities $P_{u,v}^{(\text{dc})}$, we find that in the leading order (with respect to the average photon number $\bar{n}$) the probability of identifying an incoming $\ket{\phi_{0,0}}$-state as a $\ket{\phi_{1,0}}$-state is given by
\begin{align}
	\epsilon^{\rm (dc)} = p_{-1} \bar{n} (1-\eta_d),
\end{align}
which is just the product of the probabilities of losing exactly one of the two original photons (either during transmission or within the detector) $p_{-1}=(1-\eta_t)\eta_d +\eta_t 2(1-\eta_d)\eta_d$, having a photon in the thermal state (approximately $\bar{n}$ for small average photon numbers), and reflecting that photon at the beam splitter in front of the detector $(1-\eta_d)$. As we have seen in Eq.~\eqref{eq:pdc} the term $\bar{n} (1-\eta_d)$ corresponds to the probability of a dark count. When looking at the other false identifications, we always find the same probability $\epsilon^{\rm (dc)}$ (in the leading order), see Table~\ref{tab:pdc} in Appendix~\ref{sec:AppE}. Therefore, the effect of the dark counts bears some resemblance to a depolarizing error channel with $\epsilon = 2 \epsilon^{\rm (dc)}$. However, it is also possible that a dark count occurs in a detector that clearly signals a failure, e.g. when there is a single click in both horizontal detectors (see section \ref{sec:QPC BM}). This occurs with the same probability $\epsilon^{\rm (dc)}$ and does not have a counterpart in the depolarizing error channel. Consequently, the leading-order terms of the correct identifications as well do not fit into the depolarizing error channel interpretation. Additionally, dark counts can, of course, also occur without prior loss of another photon. Although click patterns with a total of three clicks can contain information on the original Bell state, we have decided to interpret these events as failures of the physical-level BM. For details on the interpretation of click patterns see Appendix~\ref{sec:AppE}. To summarize, the effect of the dark counts is more complex than just introducing depolarizing errors. Nevertheless, the numerical analysis shows that the QPC-based BM is also robust against the effects of dark-count-afflicted detectors, as long as $\epsilon^{\rm(dc)}$ is small enough and the detector inefficiency $1-\eta_d$ is not too big. For example, for $\eta_d=0.97$ and $\bar{n}=0.03$ the optimal parameters $(n,m)=(38,6)$ and $L_0=\unit[1.9]{km}$ yield a repeater rate of $R^{\rm (dc)} = 0.69 t_0^{-1}$ and correspond to $\epsilon^{\rm (dc)} \approx 0.7 \cdot 10^{-4}$. In this parameter range the comparison to a transmission with depolarizing errors with $\epsilon=1.4\cdot10^{-4}$, no dark counts but detector losses of $\eta_d=0.97$ is rather adequate as we obtain the optimal parameters $(n,m)=(37,6)$ and $L_0=\unit[2.1]{km}$ yielding $R^{\rm(depol)} = 0.70 t_0^{-1}$.

\subsection{Advanced BM Schemes}
\label{subsec:advBM}

After considering the main error types that occur during the communication scheme (photon loss due to channel transmission and operational errors such as Pauli-type errors and detector inefficiencies), we now concentrate on the actual BM scheme. Compared to other communication schemes, especially that in Ref.~\cite{LLPRL}, our scheme requires more photons to encode a single qubit. This is mainly due to the limitation of the physical BM to an efficiency of $\tfrac{1}{2}$. It is therefore sensible to ask whether more advanced optical BM schemes on the physical level could improve the performance of our communication scheme. Particularly, more effective BMs which, of course, come with a higher experimental cost, could reduce the required size of the QPCs for faithful long-distance communication.

\begin{table}[b]
\begin{ruledtabular}
\begin{tabular}{ccccc}
$P_{u,v}^{(\text{adv})}$ & $\ket{\phi_{0,0}}$ & $\ket{\phi_{0,1}}$ & $\ket{\phi_{1,0}}$ & $\ket{\phi_{1,1}}$ \\\hline\\[-1em]
$(0,0)$ & $p_{\rm adv} \eta_t$ & $0$ & $0$ & $0$ \\
$(0,1)$ & $0$ & $p_{\rm adv} \eta_t$ & $0$ & $0$ \\
$(1,0)$ & $0$ & $0$ & $\eta_t$ & $0$ \\
$(1,1)$ & $0$ & $0$ & $0$ & $\eta_t$ \\
$(0,?)$ & $\left(1-p_{\rm adv}\right)\eta_t$ & $\left(1-p_{\rm adv}\right)\eta_t$ & $0$ & $0$ \\
$(?,?)$ & $1-\eta_t$ & $1-\eta_t$ & $1-\eta_t$ & $1-\eta_t$
\end{tabular}
\end{ruledtabular}
\caption{Outcome probabilities on the physical level if an advanced BM scheme is used that has a probability of $p_{\rm adv}$ to correctly identify the states $\ket{\phi_{0,l}}$. Additionally, transmission loss is considered.}
\label{tab:Padv}
\end{table}

In recent years, various suggestions have been made to realize a static, all-optical BM that can beat the well-known notorious $\tfrac{1}{2}$-limit of Ref.~\cite{CalsamigliaNL}. All these schemes use additional resources to achieve higher efficiencies, namely either squeezing \cite{Zaidi} or additional photons that can be entangled \cite{Grice} or not \cite{Ewert}. A common property of these schemes, which is also a necessary condition to be useful for their use in the QPC-based BM, is that the unambiguous identification of the states $\ket{\phi_{1,l}}$ is always successful when no errors occurred. At the ame time the probability of correctly identifying the states $\ket{\phi_{0,l}}$ is increased. Additionally, this increase is independent of $l$ for most schemes \cite{Grice,Ewert}, which simplifies the further analysis.

Since there is no need to restrict our analysis to a particular one of the above-mentioned advanced BM schemes, we introduce a parameter $p_{\rm adv}$, which is the probability to correctly identify the states $\ket{\phi_{0,l}}$. For the standard linear optics Bell measurement we have $p_{\rm adv}=0$, whereas for an ideal BM we have  $p_{\rm adv}=1$. The total efficiency of the BM is given by $p_{\rm BM} = \tfrac12 (1+p_{\rm adv})$. The outcome probabilities $P_{u,v}^{\text{(adv)}}$ are given in Table~\ref{tab:Padv}. Here we consider transmission loss, but lossy detectors are not taken into account since, depending on the corresponding advanced scheme, a varying number of photons are detected. The same obviously holds for dark counts. Table~\ref{tab:Padv} does also not consider depolarizing errors, but including these is very straightforward.

Applying the error propagation analysis presented in Sec.~\ref{sec:error analysis} to  $P_{u,v}^{\text{(adv)}}$  as depicted in Table~\ref{tab:Padv} with the usual propagation rules $f_u(\gamma)$ and $g_u(\lambda)$ yields the following secure key rate:
\begin{align}
	R^{\text{(adv)}}&=\frac{1}{t_0}\bigg\{\left[1-(1-\eta_t)^m\right]^n \label{eq:Radv}\\
	 &\qquad\quad-\left[1-(1-\eta_t)^m-\frac{(1+p_{\rm adv}^m)\eta_t^m}{2}\right]^n\bigg\}^{N}. \nonumber
\end{align}
This is very similar to the secure key rate of Eq.~\eqref{eq:Rloss}. The only difference (besides $\eta_d\equiv1$) is the factor $(1+p_{\rm adv}^m)$ in the last summand. From this term alone we already see that the impact of an increased physical BM efficiency will be rather small unless we have a very good BM with $p_{\rm adv}$ close to one, because the term $(1+p_{\rm adv}^m)$ grows rather slow for smaller values of $p_{\rm adv}$.

\begin{figure}[t]
\includegraphics[width=0.235\textwidth]{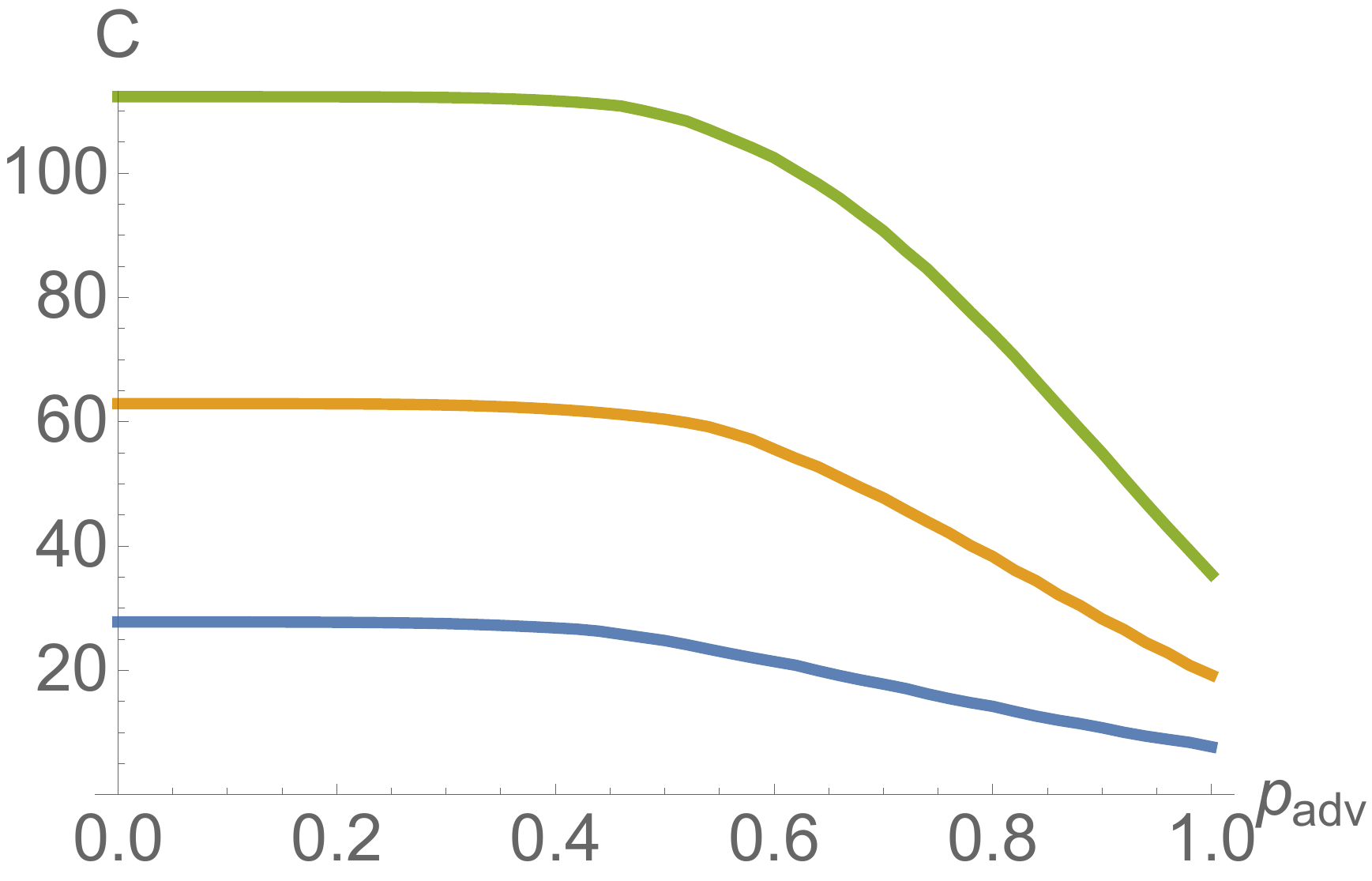} \hfill \includegraphics[width=0.235\textwidth]{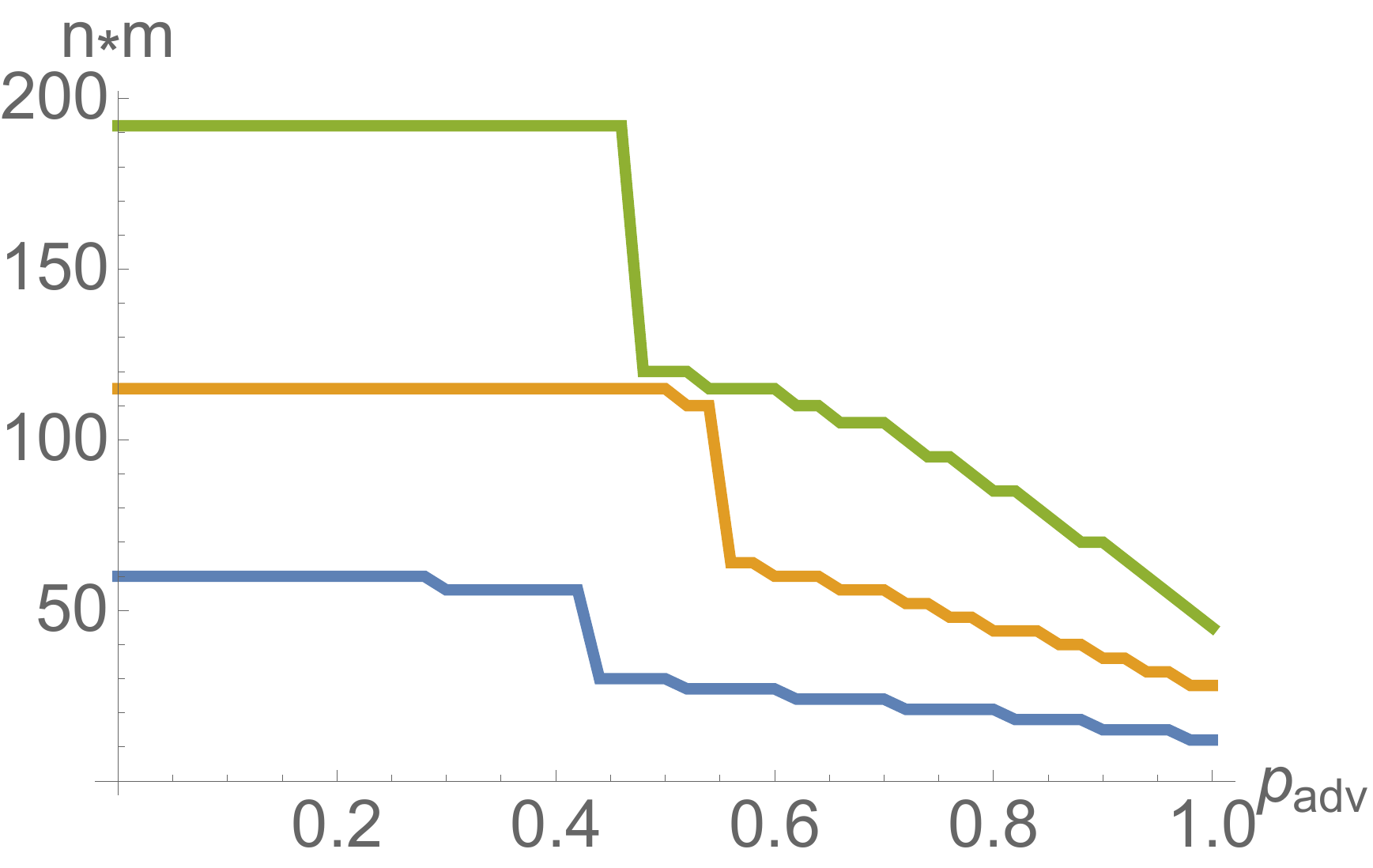}
\caption{Cost function $C$ (left) and optimal code size $n\cdot m$ (right) as a function of the probability $p_{\rm adv}$ to correctly identify a $\ket{\phi_{0,l}}$-state. From bottom to top the total communication distance is $L_{\rm tot}=\unit[100]{km}$, $\unit[1\,000]{km}$ and $\unit[10\,000]{km}$. The absolute values of the cost function $C$ should not be compared between the variuos communication distances, because for an actual comparison a factor $L$ has to be multiplied. However, when doing so, the shape of the the cost function becomes unrecognizable for the smaller $L$ in this representation. Independent of $L$ the optimal code remains the same for small values of $p_{\rm adv}$. At $p_{\rm adv} \approx 0.5$ the size of the optimal code starts to drop and so does the cost. The large drop in the code size, which also marks the start of the decrease of the cost function occurs, when the more efficient BM allows to reduce the block size $m$.}\label{fig:padv}
\end{figure}

When considering depolarizing errors in a setting with advanced BM schemes it is now possible that all physical BMs of a block herald success, i.e. no photon of the block was lost and every BM gives an unambiguous result, but the physical BMs give differing results for the index $k$. As already mentioned in the discussion of the error propagation rules $f_u(\gamma)$ in Sec.~\ref{subsec:prop of errors}, better communication rates are obtained when the result for the index $l$ that such a block gives is disregarded. This is due to the fact that differing values in $k$ herald a bit-flip in at least one of the physical qubits. This bit-flip can either be an $X$- or a $Y$-error, which are equally probable in the depolarizing error channel. However, the $Y$-error also includes a phase-flip, which changes the result for $l$. Therefore, in a block where the index $k$ is not measured unanimously in the physical BMs, the resulting value for $l$ is known to be completely random.

Fig.~\ref{fig:padv} shows the optimized cost function $C$ [see Eq.~\eqref{eq:cost}] and the optimal code size $n\cdot m$ as a function of the BM parameter $p_{\rm adv}$ for various communication distances $L_{\rm tot}$. Since the advanced BM scheme has to be applied to every pair of physical qubits, it is reasonable to assume the additional cost to be proportional to the total number of photons used ($n m L_{\rm tot} / L_0$). This is the same coefficient that is used in the cost function $C$. The difference in $C$ for varying values of $p_{\rm adv}$ therefore indicates the allowed overhead cost to achieve the higher BM efficiency. We found that the shape of the graph $C(p_{\rm adv})$ is almost independent of the communication distance $L_{\rm tot}$. Especially, the fraction $\tfrac{C(p_{\rm adv}=0)}{C(p_{\rm adv}=1)}$ is almost constant at a value of about 3. For the advanced BM schemes based only on linear optics \cite{Grice, Ewert}, the experimental requirements of a single BM grow very quickly when approaching BM efficiencies close to 1. Thus, it appears that utilizing them to reduce the code size is not the most resource effective method. Most notably, even the use of nonlinearites to achieve unit BM efficiency should be questioned (again under the assumption of a linear scaling for the cost of generating QPC encoded states). In other words, our linear-optics communication scheme based on standard linear-optics BMs is almost as good, producing almost as good rates at almost the same cost, as those schemes based on local CNOT operations and light-matter interactions such as that of Ref.~\cite{LLPRL}.

Nonetheless, the result that the use of more advanced optical BM schemes, such as the $\tfrac{3}{4}$-efficiency scheme of Ref.~\cite{Ewert}, increases the cost effectiveness of our communication scheme only marginally, is rather surprising, since for the all-optical communication scheme presented by Azuma et al. in Ref.~\cite{ATL}, the more detailed analysis performed in Ref.~\cite{pant2016rate} showed that an increase of the BM efficiency to $\tfrac{3}{4}$ (both for the linear-optics state preparation and for the online repeater operations) allows for a substantial reduction of the required resources. This highlights that the quantum parity code used in our communication scheme is very well adapted to the limited capability of the standard linear optical BM. However, when our state preparations are also restricted to linear optics \cite{ATL,pant2016rate,Li2015}, an improved physical BM on dual-rail qubits does significantly reduce the cost of resources (in terms of single-photon sources) also in our case as well (see Appendix \ref{sec:AppD}).

\subsection{On-Off Detectors}
\label{subsec: onoff}

We have seen above that the relatively cheap standard optical BM yields good transmission rates at costs comparable to those of schemes using matter qubits for processing. Now we even go one step further and eliminate the last experimentally expensive component from our BM -- the photon-number-resolving detectors that can resolve $0,1$, and $\geq2$ photons -- and replace them by simple on-off detectors. As was already stated, the PNRDs are required to distinguish a $\ket{\phi_{0,l}}$-state from any Bell state that has lost a photon. At first, it seems that this excludes the possibility to use on-off detectors for a loss resistant QPC-based BM, but we found that with appropriate changes to the interpretation rules $f_u(\gamma)$ the QPC-based BM can still be used for our communication scheme without increasing the required cost too much.

When $\ket{\phi_{0,l}}$-states can no longer be distinguished from loss, rows $5$ and $7$ in $P_{u,v}$ have to be joined. The resulting matrix $P_{u,v}^{(\text{on-off})}$ is shown in Table~\ref{tab:Ponoff} for the case that both transmission loss and depolarizing errors are present. To be precise, Table~\ref{tab:Ponoff} is correct only for perfect photon detectors $(\eta_d=1)$. In any other case, the usual choice $\eta = \eta_d^2 \eta_t$ introduces a small error, since the on-off detectors are still able to detect whether both photons have been lost. These cases, however, have only a marginal effect, since they change the result of the logical BM only, if in one block all $2m$ photons are lost. This is extremely unlikely and, in fact, we found that the corresponding change of the secure key rate is below the numerical precision in almost every case. Therefore, we neglect this subtlety in our presentation here. (It has been taken into account in the actual calculations though.)

To obtain the outcome probabilities on the block level $B_{u,v}^{(\text{on-off})}$ a new set of interpretation rules $\tilde{f}_u(\gamma)$ is required. Since a distinction between $k=0$ and $k=\ ?$ on the physical level is no longer possible, the majority voting ($\gamma_1+\gamma_2+\gamma_5$ vs. $\gamma_3+\gamma_4+\gamma_6$) in the standard form of $f_u(\gamma)$ is not appropriate here. This can be seen very clearly in the case, where there are no depolarizing errors ($\epsilon=0$). In case of a $\ket{\phi_{0,l}}^{(m)}$ block state, which consists only of $\ket{\phi_{0,r_i}}$ states on the physical level, the BM will always give $m$ results $(0,?)$. Therefore, even a single $k=1$ result indicates that the incoming state must have been a  $\ket{\phi_{1,l}}^{(m)}$ block state. However, the standard interpretation rules $f_u(\gamma)$ demand more than $m/2$ results $k=1$ on the physical level for the state to be identified as a $\ket{\phi_{1,l}}^{(m)}$ state. In this case of $\epsilon=0$ the appropriate interpretation rules are therefore:
\begin{align*}
	\tilde{f}_3(\gamma):&\quad \gamma_3+\gamma_4=m \wedge \gamma_4 \text{ is even} & (1,0)\\
	\tilde{f}_4(\gamma):&\quad \gamma_3+\gamma_4=m \wedge \gamma_4 \text{ is odd} & (1,1)\\
	\tilde{f}_5(\gamma):&\quad \gamma_3+\gamma_4=0 & (0,?)\\
	\tilde{f}_6(\gamma):&\quad 0<\gamma_3+\gamma_4<m & (1,?)
\end{align*}
For $u\in\{1,2,7\}$ no acceptable combinations of measurement results on the physical level exist. For the further computation of $L_{u,v}^{(\text{on-off})}$ no changes to the interpretation rules $g_u(\lambda)$ are required. Following the error propagation formalism presented in Sec.~\ref{sec:error analysis} we find the raw transmission rate
\begin{align}
	p_{\rm trans}^{\text{(on-off)}} = \left[1-\left(1-\frac{\eta^m}{2}\right)^n\right]^N.
\end{align}
This rate is higher than that obtained with PNRDs, because the check whether at least one photon is left in every block is no longer possible and only those events are filtered out where no block allowed for an identification of the index $l$. This increased transmission rate comes, of course, at the cost that the lack of loss-detection can lead to undetected Pauli-errors. In fact, we found that logical bit-flips ($X$) are induced, whereas logical phase-flips ($Z$ and consequently $Y$) are still impossible. The QBER for the induced bit-flips is then given by
\begin{align}
	Q_X^{\text{(on-off)}} = \frac{1}{2}\left[1-\frac{p_{\rm trans}^{\text{(only loss)}}}{p_{\rm trans}^{\text{(on-off)}}}\right]. \label{eq:qxonoff}
\end{align}
At first glance, the appearance of $p_{\rm trans}^{\text{(only loss)}}$  here, i.e. the transmission rate associated with the use of PNRDs, is surprising, however, it can be understood quite nicely. The QBER $Q_{X}$ is defined as the fraction of the probability for successful transmission with a hidden bit-flip and the probability for any successful transmission. When using on-off detectors, the latter is given by $p_{\rm trans}^{\text{(on-off)}}$. The remaining numerator in Eq.~\eqref{eq:qxonoff} ($p_{\rm trans}^{\text{(on-off)}} - p_{\rm trans}^{\text{(only loss)}})$ is the probability that a state enters the BM that would be filtered out as a failure by the PNRDs, but is accepted when using on-off detectors (because every input state that PNRDs accept will also be accepted by on-off detectors). These are exactly those states, where in at least one block all photons were lost. In that case the on-off-detector-based BM will always assign the block the value $k=0$. Thus, in half of all cases it induces a logical bit-flip error.

\begin{table}[b]
\begin{ruledtabular}
\begin{tabular}{ccccc}
$P_{u,v}^{(\text{on-off})}$ & $\ket{\phi_{0,0}}$ & $\ket{\phi_{0,1}}$ & $\ket{\phi_{1,0}}$ & $\ket{\phi_{1,1}}$ \\\hline\\[-1em]
$(1,0)$ & $\tfrac{\epsilon}{2}\eta$ & $\tfrac{\epsilon}{2}\eta$ & $\left(1-\tfrac{3 \epsilon}{2}\right)\eta$ & $\tfrac{\epsilon}{2}\eta$ \\
$(1,1)$ & $\tfrac{\epsilon}{2}\eta$ & $\tfrac{\epsilon}{2}\eta$ & $\tfrac{\epsilon}{2}\eta$ & $\left(1-\tfrac{3 \epsilon}{2}\right)\eta$ \\
$(0,?)$ & $1-\epsilon \eta$ & $1-\epsilon\eta$ & $1-(1-\epsilon)\eta$ & $1-(1-\epsilon)\eta$
\end{tabular}
\end{ruledtabular}
\caption{Outcome probabilities on the physical level if on-off detectors are used and depolarizing errors are present in addition to the inevitable photon loss.}
\label{tab:Ponoff}
\end{table}

Numerical results show that the resulting secure key rate $R^{\text{(on-off)}}$ is, as expected, smaller than that associated with the use of PNRDs. However, the difference is rather small. When optimizing the code parameters with the help of the cost function $C$ and assuming that only transmission loss is present, i.e. perfect detectors with $\eta_d=1$, we even found that the required code size is reduced. Since the reduced code size comes along with a reduced repeater spacing, the total cost is nevertheless slightly higher. When also including lossy detectors ($\eta_d<1$) the advantage in terms of a code size reduction for the on-off detectors is lost, and the codes must be chosen larger. However, the secure key rate and the cost $C$ are still comparable to the case of PNRDs. To be more precise, the cost $C$ is increased by about $25\%$ (see Table~\ref{tab:onoffcodes}).

\begin{table}[b]
\begin{ruledtabular}
\begin{tabular}{cccccccc}
$\epsilon$ & $\eta_d$ & det. type & $\kappa$ & $(n,m)$ & $L_0$[km] & $R t_0$ & $C$[a.u.] \\\hline
$0$ & $1$ & PNRD && $(23,5)$ & $2.4$ & 0.76 & 62.9\\
$0$ & $1$ & on-off && $(21,5)$ & $1.9$ & 0.72 & 76.3\\
$0$ & $0.97$ & PNRD && $(37,6)$ & $2.1$ & 0.73 & 144.6\\
$0$ & $0.97$ & on-off && $(45,7)$ & $2,1$ & 0.78 & 193.5\\
$10^{-3}$ & $1$ & PNRD && $(22,5)$ & $2.0$ & 0.65 & 84.0\\
$10^{-3}$ & $1$ & on-off & 2 & $(30,8)$ & $1.9$ & 0.63 & 199.4\\
$10^{-3}$ & $1$ & on-off* & 3 & $(26,8)$ & $1.4$ & 0.76 & 194.7\\
$10^{-3}$ & $0.97$ & PNRD && $(36,6)$ & $1.8$ & 0.60 & 198.8\\
$10^{-3}$ & $0.97$ & on-off & 3 & $(67,11)$ & $1.3$ & 0.77 & 735.4\\
$10^{-3}$ & $0.97$ & on-off* & 3 & $(58,10)$ & $1.3$ & 0.71 & 631.3
\end{tabular}
\end{ruledtabular}
\caption{Optimal codes and their rates and cost when combining different error sources (depolarizing errors and detector loss) and different detector types and interpretation schemes. The on-off schemes marked with a star accept basically all input states, as there are no ties in the majority voting for $k$ on each block. The parameter $\kappa$ is an additional decision parameter explained in the text.}
\label{tab:onoffcodes}
\end{table}

Let us now turn to the case, where also depolarizing errors are present. Here the interpretation rules $f_u(\gamma)$ have to be changed again: If, for example, the physical BMs of a block give one result $(1,0)$ and $m-1$ results $(0,?)$, two scenarios are possible: Either the original block state was a $\ket{\phi_{0,l}}^{(m)}$ state and in one of the photons a bit-flip occurred or the original state was $\ket{\phi_{1,l}}^{(m)}$ and $m-1$ photons where lost (of course, other scenarios including multiple bit-flips are also possible, but far less probable). The BM should identify the incoming state according to the most probable scenario. Which one this is depends on the parameters of the setting ($m,\eta_d,L_0,\epsilon$). We propose to enforce this decision by introducing a boundary parameter $\kappa$. It is associated to the number of $k=1$ results in the block. If more (less) than $\kappa$ physical BMs in the block give $k=1$ the block BM gives $(k,l)=(1,?)$ [$(k,l)=(0,?)$]. If exactly $\kappa$ results with $k=1$ are obtained, the block (and therefore the entire logical state) is marked as ambiguous (we refer to these events as tie). As already stated in Sec.~\ref{subsec:advBM}, the blocks with a non-unanimous voting for $k$ should not be considered for identifying $l$. The new interpretation rules are now given as:
\begin{align*}
	\bar{f}_3(\gamma):&\quad \gamma_3+\gamma_4=m \wedge \gamma_4 \text{ is even} & (1,0)\\
	\bar{f}_4(\gamma):&\quad \gamma_3+\gamma_4=m \wedge \gamma_4 \text{ is odd} & (1,1)\\
	\bar{f}_5(\gamma):&\quad \gamma_3+\gamma_4 < \kappa & (0,?)\\
	\bar{f}_6(\gamma):&\quad \kappa < \gamma_3+\gamma_4 < m & (1,?)\\
	\bar{f}_7(\gamma):&\quad \gamma_3+\gamma_4 = \kappa & (?,?)
\end{align*}
Again, no acceptable combinations of measurement results on the physical level exist for $u\in\{1,2\}$ and the interpretation rules for the final step from block to logical level $g_u(\lambda)$ remain unchanged.

\begin{figure}[t]
\includegraphics[width=0.235\textwidth]{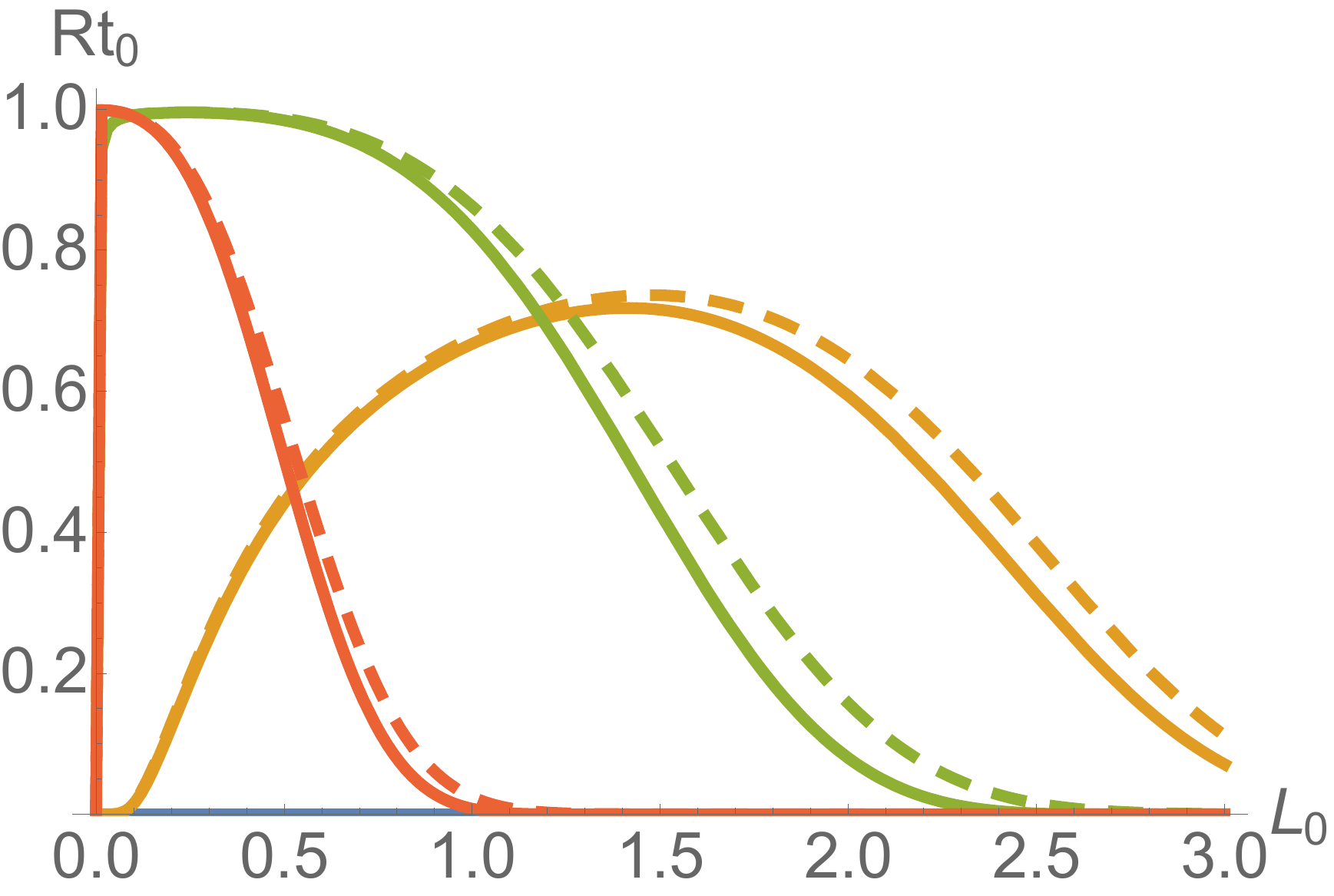} \hfill \includegraphics[width=0.235\textwidth]{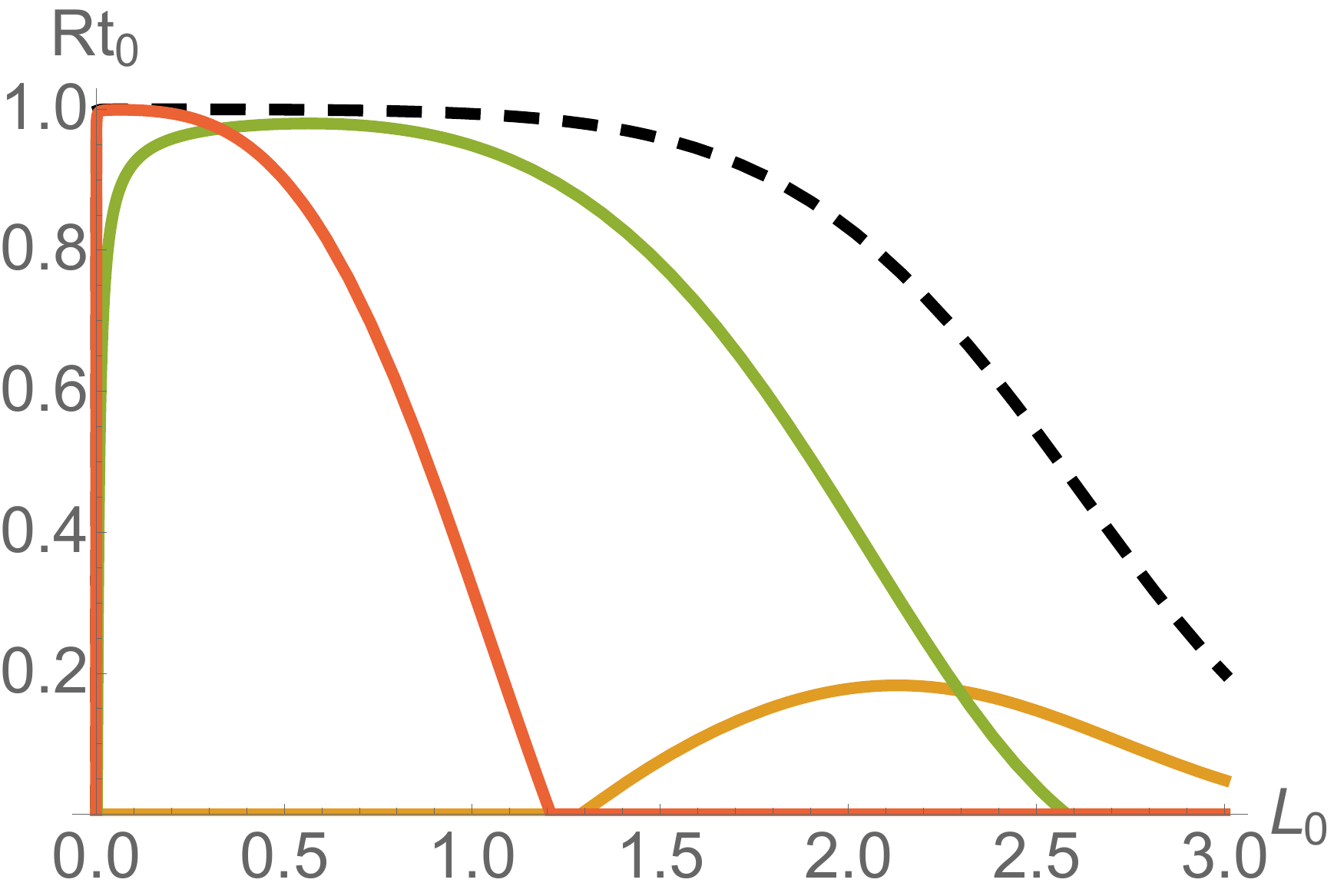}
\caption{The interpretation rule determines the secure key rate of a repeater when using on-off detectors.\\
Left: Secure key rate and raw transmission rate (dashed) of QPC$(30,8)$ for $L_{\rm tot}=\unit[1\,000]{km}$, $\epsilon=10^{-3}$, $\eta_d=1$, and various choices of $\kappa$: blue (bottom): $\kappa=1$, orange (right): $\kappa=2$, green (mid): $\kappa=3$, red (left): $\kappa=4$.\\
Right: The same experimental parameters $L_{\rm tot}, \epsilon, \eta_d$ and the same code (30,8) as Left, but the alternative interpretation rule is applied, where all combinations of physical BM results are accepted for the block level (marked with a * in Table~\ref{tab:onoffcodes}). The raw transmission rate (dashed) becomes independent of $\kappa$ and is significantly larger, but also a large QBER is induced [see especially the case of $\kappa=2$ (orange)].}
\label{fig:kappa}
\end{figure}

As before, the inclusion of depolarizing errors into the setup makes it impossible to give a compact analytic form for the raw transmission rate or the secure key rate. However, our numerical results suggest that with the above interpretation rules and appropriate code parameter choices, faithful and efficient transmission is still possible, although at an increased cost. Especially the combination of depolarizing errors and detector losses requires much larger codes compared to the case with PNRDs (see table~\ref{tab:onoffcodes}).

It should be noted that the secure key rate depends strongly on the choice of $\kappa$ (see Fig.~\ref{fig:kappa}). In fact, in some cases it is even better to slightly change the above interpretation rules: Instead of discarding those cases with exactly $\kappa$ results of $k=1$, these can also be incorporated into one of the two acceptance sets, e.g. by assigning to them the result $(k,l)=(1,?)$ (this interpretation rule is marked with a star in Table~\ref{tab:onoffcodes}). Depending on the situation this may yield considerably higher secure key rates and/or lower costs. However, this comes at the price of inducing a high QBERe. By accepting all blocks the raw transmission rate is increased, but this also induces a higher rate of bit-flip errors. Especially, if the gain in the secure key rate is comparatively small, we would rather not include a lot of events with (hidden) bit-flips into the set of accepted transmissions. Fortunately, the decision as to whether to include the tie events or not, as well as the choice of $\kappa$, have no influence on the actual optical setup, but are merely a software adjustment that can be performed at any time. Furthermore, the communication scheme does herald results, but does not react to it. Thus, all possible choices of $\kappa$ can be investigated in the same attempt.

As a further note, it is also possible to include dark counts into the analysis of on-off detectors. The model presented in Sec.~\ref{subsec:darkcounts} remains the same. However, due to the fact that dark counts can induce click combinations in the physical BM that are identified as an error with certainty, e.g. clicks in both vertical detectors, the introduction of a voting parameter $\kappa$ becomes more complicated, since now three possible outcome classes must be taken into account. We therefore refrain from a detailed description, especially, since no unexpected interaction between the two error sources occurs. Basically, to suppress both errors, the code size has to be increased further, just as it was the case for each error source alone.

To summarize this section, the use of on-off detectors instead of PNRDs introduces a number of subtleties, especially regarding the interpretation of the physical BM results. Nevertheless, these complications can be handled and it is possible to replace PNRDs and still obtain high secure key rates with codes of a larger but still similar size to before and therefore comparable resource costs.

\section{Summary and Discussion}

In the previous sections we have analyzed a static linear-optics BM on the quantum parity codes with regards to various error sources. We found that the encoded BM does not only have an efficiency extremely close to one but is also resistant against the inevitable photon loss occurring in any optical communication scheme. Furthermore, we have seen that communication distances of thousands of kilometers can also be reached when detector losses $(1-\eta_d)$ in the range of a few percent in every detector and depolarizing errors $\epsilon \lesssim 10^{-2}$ on all single-photon qubits are present. Investigating the effect of dark counts in the detectors showed that they induce errors similar to depolarizing errors of the strength $\epsilon \approx 2(1-\eta_t)\bar{n}(1-\eta_d)$, but can also lead to ambiguous click patterns that reduce the secure key rate. Nevertheless, by adjusting the code size $(n,m)$ and the repeater spacing $L_0$ accordingly, faithful communication is still possible, as long as the dark count probability $\bar{n}(1-\eta_d)$ does not grow too big.

\begin{figure}[t]
\centering
\includegraphics[width=0.35\textwidth]{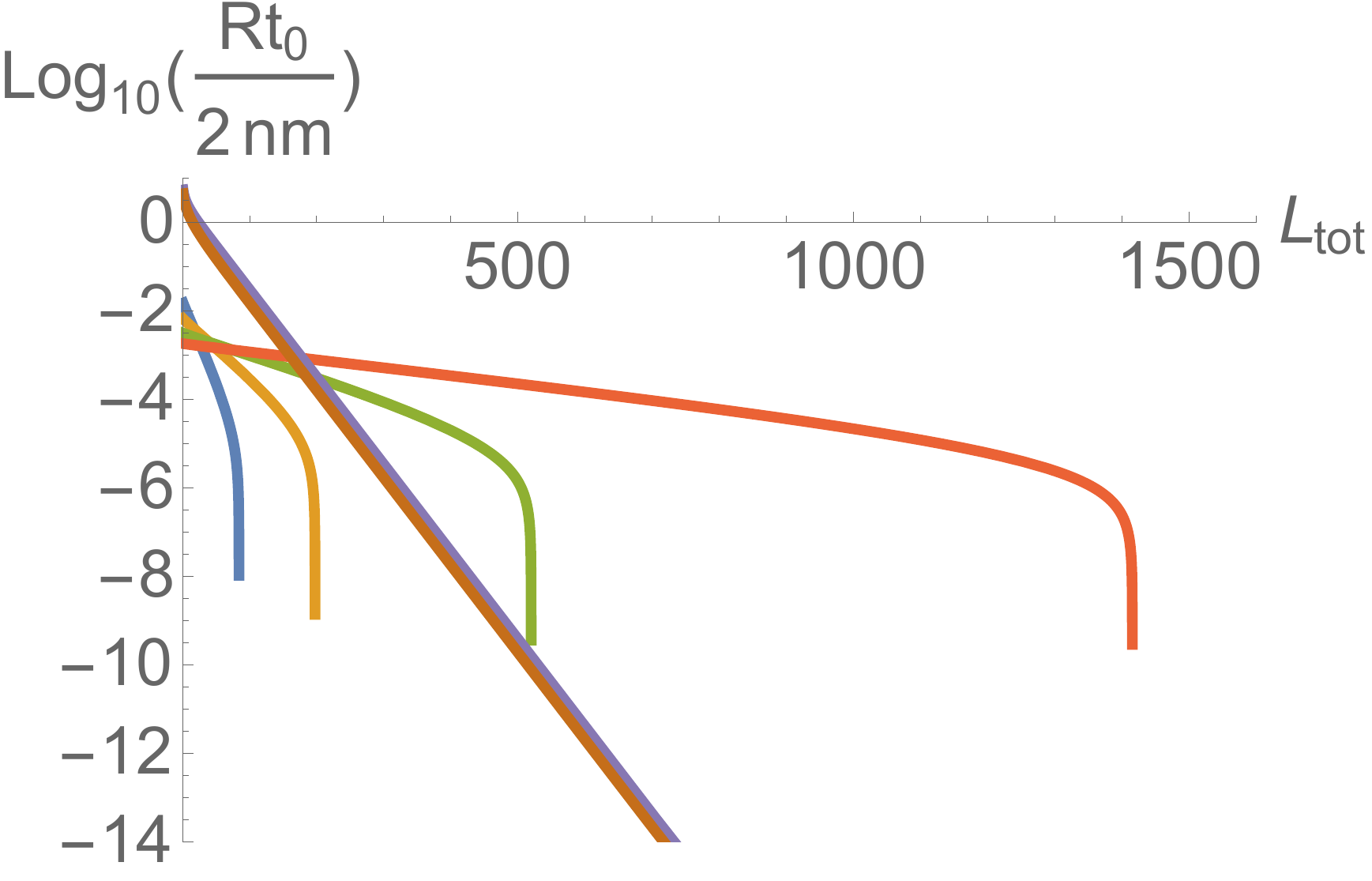}
\caption{Depending on the code size, the all-optical QPC-based communication scheme can easily beat the TGW bound (dark blue) and the PLOB bound (dark red). Shown is the secure key rate per number of modes used $\tfrac{R t_0}{2 n m}$ for $L_0=\unit[1.5]{km}$, $\epsilon=5 \cdot 10^{-3}$, $\eta_d=0.97$, $2\epsilon^{\rm (dc)}=5.2 \cdot 10^{-4}$, and various code sizes: blue: $(10,3)$, orange: $(18,4)$, green: $(31,5)$, red: $(45,6)$ (from left to right). The codes were chosen by fixing $m$ and then optimizing $n$ such that the highest communication distance could be reached.}
\label{fig:tgw}
\end{figure}

As a clear indicator that our communication scheme does indeed yield advantages over the direct communication of quantum states without any repeater structure we have compared the secure key rate per modes used $\tfrac{R t_0}{2 n m}$ with the TGW bound (Takeoka Guha Wilde, \cite{TGW}) and its refinement, the PLOB bound (Pirandola Laurenza Ottaviani Banchi, \cite{PLOB}), in Fig.~\ref{fig:tgw}. There it becomes obvious that our communication scheme can comfortably beat these bounds provided the code size is chosen accordingly. In Fig.~\ref{fig:tgw} we have chosen a constant repeater spacing of $L_0=\unit[1.5]{km}$ and a demanding scenario of detector losses $1-\eta_d=3\%$, depolarizing errors of $\epsilon=5\cdot 10^{-3}$, and dark counts with an average photon number of $\bar{n}=0.1$ in the thermal state (see Sec.~\ref{subsec:darkcounts}) which leads to a depolarizing-type error of $2\epsilon^{\rm (dc)}=5.2 \cdot 10^{-4}$. The codes have been chosen by fixing the block size $m$ and optimizing the number of blocks $n$ with respect to the maximal total communication distance which is marked by the sharp drop. If, on the other hand, we keep the experimental parameters $L_0, \eta_d, \epsilon$, and $\bar{n}$ fixed and optimize the code size  with the help of the cost function $C$ while also fixing the total communication distance at $\unit[1\,000]{km}$, we obtain the code QPC(71,8). In the style of Fig.~\ref{fig:tgw} this code has its drop at about $\unit[4\,000]{km}$. This indicates that the maximally achievable distance with a given code is much higher than the one where it is chosen as the most cost efficient one. We note that in an idealized repeater scenario where no errors are present besides those originating from channel losses, a QPC as small as $(23,5)$ represents the most cost-efficient choice for a total communication distance of $L_{\rm tot} =\unit[1\,000]{km}$. We also found that the smallest code that can beat the TGW bound in the idealized setting of only transmission loss is actually much smaller: QPC(6,2) suffices to beat the TGW bound with our all-optical scheme. This can even be further reduced to QPC(4,2) when using the advanced BM scheme of Ref.~\cite{Ewert} for the physical BMs in the repeater stations. Thus, in order to beat the benchmarks of repeaterless quantum communication, encoded states are needed in our scheme which are clearly in sight of experimental realizations.

In addition to the effect of the error types described above, we have investigated the effect of using different BM schemes on the single-photon level. When replacing the standard linear-optics BM on polarization qubits which is limited to an efficiency of $\tfrac{1}{2}$ by more advanced schemes, it was shown that in terms of cost effectiveness only BMs with an efficiency very close to one yield a considerable advantage. The optical schemes presented in \cite{Zaidi,Ewert} that yield efficiencies of about $75\%$ only have a marginal effect on the repeater rates and the resource consumption. When comparing the all-optical communication scheme using the standard BM on the single-photon level to one using deterministic BMs, e.g. that of Ref.~\cite{LLPRL} which uses atomic processing qubits, we found that, as long as the deterministic BM has more than thrice the experimental cost of the standard BM, our scheme is more resource efficient. From these observations we conclude that the quantum parity code is extremely well adapted to the experimental restrictions of static linear optics. This is further bolstered by the fact that the all-optical (and static) BM allows for much smaller processing times at every repeater station and thus for a higher repetition rate than atomic-based protocols.

Besides improving the cost effectiveness of our scheme with the use of more efficient but also more complex BM schemes, we also investigated the possibility of a further experimental simplification, namely replacing the photon-number-resolving detectors by on-off detectors. We found that, although this approach introduces a variety of subtle challenges, especially regarding the interpretation of BM results, the use of on-off detectors is definitely possible. While in the presence of multiple large operational errors (depolarizing errors and detector losses) the code size must be increased quite a bit, and the practicability of on-off detectors is questionable in this regime, in the case of smaller but not necessarily vanishing errors, the difference in both code size and repeater spacing between PNRDs and on-off detectors is quite small. Therefore, on-off detectors should provide a much easier-to-implement, but similarly effective alternative to PNRDs in this regime.

\section{Conclusion}

We have presented a high-efficiency, loss-and-error-resistant linear-optics Bell measurement on the quantum parity codes together with its application to quantum communication in the form of a static linear-optical one-way communication scheme reaching transmission rates in the GHz-regime over thousands of kilometers. In irder to analyze both the Bell measurement efficiency and the secure key rate of the communication scheme, we established a general toolbox that traces the effect of errors on the single-photon level through the different encoding levels of the quantum parity code to determine the induced error rates on the highest encoding level, i.e. on the logical qubits. With the help of this toolbox we demonstrated the resistance of our communication scheme to transmission loss, Pauli-type errors, and detector inefficiencies, i.e. lossy detectors and dark counts. Furthermore, we showed that the standard linear-optical Bell measurement with its limited efficiency of one half is perfectly sufficient for ultrafast fault-tolerant long-range quantum communication and that the code size reduction obtained by using more advanced Bell measurement schemes is most probably not worth the additional experimental effort if state preparations are excluded. In fact, it is even possible to further reduce experimental demands by replacing photon-number-resolving detectors by much simpler on-off detectors. With no need for matter qubits (neither as quantum memories nor as local quantum processors) or feedforward operations, our communication scheme is most suitable to be integrated along an optical fiber channel via chips that contain quantum sources \cite{Silverstone14,Silverstone2015,Spring13}, interferometers \cite{Peruzzo11}, and photon detectors \cite{Calkins13}.
\begin{acknowledgments}
We acknowledge support from Q.com (BMBF) and Hipercom (ERA-NET CHISTERA).
\end{acknowledgments}

\bibliography{repeaterbib}
\onecolumngrid
\appendix
\newpage
\section{Bell State Representations}
\label{sec:AppA}

Here we derive our representation of all encoded Bell states in terms of lower-encoding-level Bell states (Eqs. \eqref{eq:phys2block} and \eqref{eq:block2logic} in the main text),
\begin{align}
	\ket{\phi_{k,l}}^{(m)} &\cong \frac{1}{\sqrt{2^{m-1}}}\sum_{\vec{r} \in A_{l,m}} \bigotimes_{i=1}^m \ket{\phi_{k,r_i}},\label{eq:phys2blocka}\\
	\ket{\phi_{k,l}}^{(n,m)} &\cong \frac{1}{\sqrt{2^{n-1}}}\sum_{\vec{s} \in A_{k,n}} \bigotimes_{i=1}^n \ket{\phi_{s_i,l}}^{(m)},\label{eq:block2logica}
\end{align}
with $A_{l,m} = \Set{\vec{r}\in\{0,1\}^m | \sum_{i=1}^m r_i = l \Mod 2 }$.

Remember that QPC($n,m$) is constructed by $\ket{0}^{(m)} = \ket{0}^{\otimes m}$ ($\ket{1}^{(m)} = \ket{1}^{\otimes m}$) and $\ket{\pm}^{(n,m)} = \left(\ket{\pm}^{(m)}\right)^{\otimes n}$. We therefore obtain on the block level
\begin{align}
	\ket{\phi_{k,l}}^{(m)} &= \frac{1}{\sqrt{2}}\left[\ket{0}^{\otimes m} \ket{k}^{\otimes m} + (-1)^l \ket{1}^{\otimes m} \ket{1-k}^{\otimes m} \right]\nonumber\\
	& \cong \frac{1}{\sqrt{2}} \left[\ket{0,k}^{\otimes m} + (-1)^l \ket{1,1-k}^{\otimes m} \right]\nonumber\\
	& = \frac{1}{\sqrt{2^{m+1}}} \left[\left(\ket{\phi_{k,0}} + \ket{\phi_{k,1}}\right)^{\otimes m} + (-1)^l \left(\ket{\phi_{k,0}} - \ket{\phi_{k,1}}\right)^{\otimes m} \right]\nonumber\\
	&= \frac{1}{\sqrt{2^{m+1}}} \left[\sum_{\vec{r} \in \{0,1\}^m} \bigotimes_{i=1}^m \ket{\phi_{k,r_i}} + (-1)^l \sum_{\vec{r} \in \{0,1\}^m} \bigotimes_{i=1}^m (-1)^{r_i} \ket{\phi_{k,r_i}} \right]\nonumber\\
	&= \frac{1}{\sqrt{2^{m+1}}} \sum_{\vec{r} \in \{0,1\}^m} \left[1 + (-1)^{l + \sum_{j=1}^m r_j} \right] \bigotimes_{i=1}^m \ket{\phi_{k,r_i}}\nonumber\\
	&= \frac{1}{\sqrt{2^{m-1}}}\sum_{\vec{r} \in A_{l,m}} \bigotimes_{i=1}^m \ket{\phi_{k,r_i}}, \label{eq:phys2blockderivation}
\end{align}
where the symbol $\cong$ indicates the mode reordering mentioned in Sec.~\ref{sec:QPC code}. On the other hand, independent of the encoding level, the Bell states can always be written in the Pauli $X$-basis as
\begin{align*}
	\ket{\phi_{k,l}}^* &= \frac{1}{\sqrt{2}}\left[\ket{0,k}^* + (-1)^l \ket{1,1-k}^*\right]\\
	& = \frac{1}{2\sqrt{2}} \left[ \left(\ket{+}^*+\ket{-}^*\right)\left(\ket{+}^*+(-1)^k\ket{-}^*\right) + (-1)^l \left(\ket{+}^*-\ket{-}^*\right)\left(\ket{+}^*-(-1)^k\ket{-}^*\right) \right]\\
	& = \frac{1}{\sqrt{2}} \left[ \ket{(-1)^l +,+}^* + (-1)^k \ket{(-1)^l -,-}^* \right].
\end{align*}
On the logical level, we use this to obtain
\begin{align*}
	\ket{\phi_{k,l}}^{(n,m)} &= \frac{1}{\sqrt{2}} \left[ \left(\ket{(-1)^l+}^{(m)}\right)^{\otimes n}  \left(\ket{+}^{(m)}\right)^{\otimes n} +(-1)^k \left(\ket{(-1)^l-}^{(m)}\right)^{\otimes n}  \left(\ket{-}^{(m)}\right)^{\otimes n} \right]\\
	& \cong \frac{1}{\sqrt{2}} \left[ \left(\ket{(-1)^l+,+}^{(m)}\right)^{\otimes n} +(-1)^k \left(\ket{(-1)^l-,-}^{(m)}\right)^{\otimes n} \right]\\
	& = \frac{1}{\sqrt{2^{n+1}}} \left[\left(\ket{\phi_{0,l}}^{(m)} + \ket{\phi_{1,l}}^{(m)}\right)^{\otimes n} +(-1)^k \left(\ket{\phi_{0,l}}^{(m)} - \ket{\phi_{1,l}}^{(m)}\right)^{\otimes n} \right].
\end{align*}
At this point it becomes clear that, compared to the derivation on the block level \eqref{eq:phys2blockderivation}, the indices $k$ and $l$ have simply swapped their roles. This yields \eqref{eq:block2logica} immediately.
\newpage
\section{Error Propagation Formula Simplification}
\label{sec:AppB}
We present the simplification of the error propagation formulas \eqref{eq:Bu1} and \eqref{eq:Lu1} of the main text into \eqref{eq:Bu1a} and \eqref{eq:Lu1a}, respectively: 
\begin{align*}
B_{u,1} &= \frac{1}{2^{m-1}}\sum_{\substack{r=0\\r \text{ even}}}^m  \sum_{\substack{\alpha,\beta\in \mathbbm{N}_0^{7}\\|\alpha|=m-r\\|\beta|=r}} \binom{m}{\alpha,\beta} P_{\cdot,1}^{\alpha} P_{\cdot,2}^{\beta}\ \delta_{f_u(\alpha+\beta)}\\
&= \frac{1}{2^{m}}\sum_{r=0}^m \left[1+(-1)^r\right] \sum_{\substack{\gamma,\beta\in \mathbbm{N}_0^{7}\\|\gamma|=m\\|\beta|=r}} \binom{m}{\gamma-\beta,\beta} P_{\cdot,1}^{\gamma-\beta} P_{\cdot,2}^{\beta}\ \delta_{f_u(\gamma)}\\
&=\frac{1}{2^{m}}\sum_{r=0}^m \left[1+(-1)^r\right] \sum_{\tilde{\beta},\tilde{\gamma} \in \mathbbm{N}_0^{6}} \binom{m}{\tilde{\gamma}-\tilde{\beta},\tilde{\beta}} \binom{m-|\tilde{\gamma}|}{m-r-|\tilde{\gamma}-\tilde{\beta}|,r-|\tilde{\beta}|} \tilde{P}_{\cdot,1}^{\tilde{\gamma}-\tilde{\beta}}P_{7,1}^{m-r-|\tilde{\gamma}-\tilde{\beta}|} \tilde{P}_{\cdot,2}^{\tilde{\beta}}P_{7,2}^{r-|\tilde{\beta}|} \delta_{f_u(\tilde{\gamma},m-|\tilde{\gamma}|)}\\
&=\frac{1}{2^{m}} \sum_{\tilde{\beta},\tilde{\gamma} \in \mathbbm{N}_0^{6}} \binom{m}{\tilde{\gamma}-\tilde{\beta},\tilde{\beta}} \tilde{P}_{\cdot,1}^{\tilde{\gamma}-\tilde{\beta}} \tilde{P}_{\cdot,2}^{\tilde{\beta}}\delta_{f_u(\tilde{\gamma},m-|\tilde{\gamma}|)} \sum_{r=-|\tilde{\beta}|}^{m-|\tilde{\beta}|} \binom{m-|\tilde{\gamma}|}{r} \left[1+(-1)^{r+|\tilde{\beta}|}\right] P_{7,1}^{m-|\tilde{\gamma}|-r} P_{7,2}^{r}\\
&= \frac{1}{2^{m}} \sum_{\tilde{\beta},\tilde{\gamma} \in \mathbbm{N}_0^{6}} \binom{m}{\tilde{\gamma}-\tilde{\beta},\tilde{\beta}} \tilde{P}_{\cdot,1}^{\tilde{\gamma}-\tilde{\beta}} \tilde{P}_{\cdot,2}^{\tilde{\beta}}\delta_{f_u(\tilde{\gamma},m-|\tilde{\gamma}|)} \left[(P_{7,1}+P_{7,2})^{m-|\tilde{\gamma}|} + (-1)^{|\tilde{\beta}|} (P_{7,1}-P_{7,2})^{m-|\tilde{\gamma}|}\right]\\
&=\frac{1}{2^{m}} \sum_{\tilde{\gamma} \in \mathbbm{N}_0^{6}} \binom{m}{\tilde{\gamma}} \delta_{f_u(\tilde{\gamma},m-|\tilde{\gamma}|)} \left[(\tilde{P}_{\cdot,1}+\tilde{P}_{\cdot,2})^{\tilde{\gamma}} (P_{7,1}+P_{7,2})^{m-|\tilde{\gamma}|} + (\tilde{P}_{\cdot,1}-\tilde{P}_{\cdot,2})^{\tilde{\gamma}} (P_{7,1}-P_{7,2})^{m-|\tilde{\gamma}|}\right]\\
&= \frac{1}{2^{m}} \sum_{\substack{\gamma \in \mathbbm{N}_0^{7}\\ |\gamma|=m}} \binom{m}{\gamma} \delta_{f_u(\gamma)} \left[(P_{\cdot,1}+P_{\cdot,2})^{\gamma} + (P_{\cdot,1}-P_{\cdot,2})^{\gamma} \right].
\end{align*}
The main idea here is to order terms such that the sum over $r$ contains only the terms $P_{7,1}$ and $P_{7,2}$ and can then be executed explicitly. This is achieved by a multiple use of multinomial coefficient decompositions $\binom{m}{k_1,k_2,k_3} = \binom{m}{k_1,k_2}\binom{m-k_1-k_2}{k_3}$ and representing $P_{\cdot,1}$ as $(\tilde{P}_{\cdot,1},P_{7,1})$. The derivation of \eqref{eq:Lu1a} from \eqref{eq:Lu1} is done in perfect analogy.
\newpage

\section{Nonlinear Optics: Generating QPC($n,m$)-states with Coherent Photon Conversion}
\label{sec:AppC}
In order to generate both the encoded qubits and the ancillary encoded Bell states at the repeater stations we propose two different schemes. The first one, presented in this section is based on coherent photon conversion (CPC), as it was  proposed in Ref.~\cite{langford11}. Here a four-wave-mixing interaction,
\begin{align}
	H = \gamma a b^\dagger c^\dagger d + \gamma^* a^\dagger b c d^\dagger,
\end{align}
as realizable by a standard commercial, polarization-maintaining photonic crystal fiber (PCF), is pumped in one mode, e.g. that expressed by the annihilation operator $d$, with a bright classical beam to obtain effectively a three-wave-mixing Hamiltonian,
\begin{align}
	\tilde{H} = \tilde{\gamma} a b^\dagger c^\dagger + \tilde{\gamma}^* a^\dagger b c,
\end{align}
with a strong, tunable, nonlinear coupling $\tilde{\gamma}\propto \gamma E_d$ where $E_d$ is the (tunable) electric field amplitude of the pumping beam. The Hilbert space $\{\ket{1,0,0},\ket{0,1,1} \}$ is an eigenspace of this Hamiltonian $\tilde{H}$. Therefore, a state from this Hilbert space undergoes Rabi-like oscillations when evolving under $\tilde{H}$. Especially, for an appropriate combination of coupling strength and interaction time, namely $\tfrac{\tilde{\gamma} t}{\hbar} = \tfrac{\pi}{2}$, this can be used as a photon doppler, because
\begin{align*}
	e^{-i\frac{\pi}{2} \tilde{H}} \ket{100} = \ket{011}.
\end{align*}
As a result, the two photons are now in modes $b$ and $c$. If desired, a transformation into two modes of frequency $\omega_a$ is possible with the use of two more CPC elements and by weakly pumping the remaining mode ($c$ or $b$). For more details, see Fig.~1b of Ref.~\cite{langford11}.

Since the vacuum passes the CPC element unchanged, the photon doppler setup can be used to transform qubit states into Bell- or GHZ-type states:
\begin{align}
	e^{-i\frac{\pi}{2} \tilde{H}} \left( \alpha \ket{0} + \beta \ket{1} \right) \otimes \ket{00} = \ket{0} \otimes \left( \alpha \ket{00} + \beta \ket{11} \right).
\end{align}
With the help of polarizing beam splitters it is also possible to build up a scheme that can do the same for polarization-encoded qubits (see Fig.~2a) and b) in the Supplementary Material of Ref.~\cite{langford11}):
\begin{align*}
	\alpha\ket{H} + \beta \ket{V} \quad \adjustbox{raise={-2.5mm}}{\includegraphics[scale=0.7]{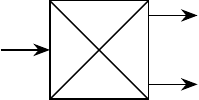}} \quad \alpha\ket{HH}+ \beta\ket{VV}
\end{align*}
When concatenating this photon doppling and using half wave plates oriented at $22.5^\circ$ to the optical axis (which realizes a Hadamard gate on polarization-encoded qubits), a scheme for generating arbitrary QPC($n,m$)-states is obtained:

\begin{figure}[H]
	\centering \includegraphics[scale=1]{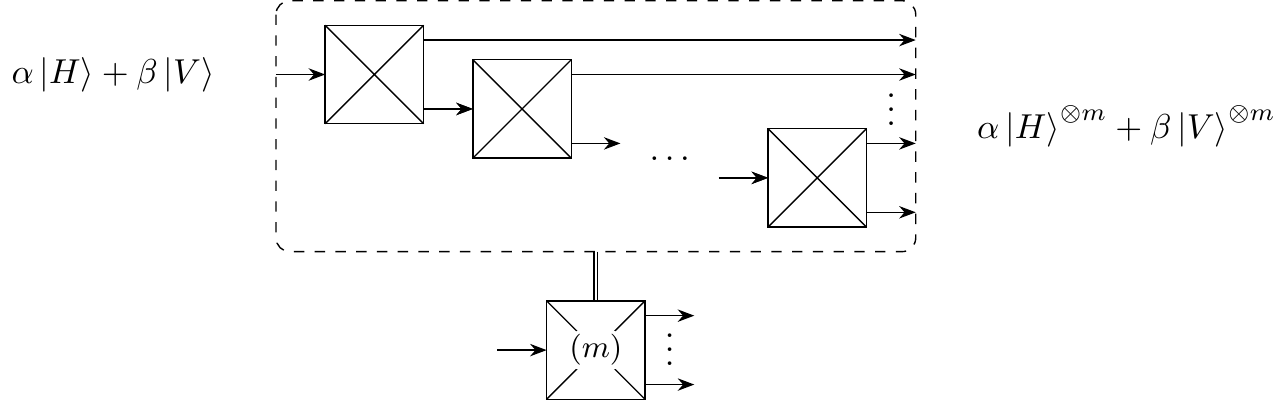}
	\caption{Concatenating $m-1$ photon dopplers gives a GHZ-type state with $m$ photons. \label{fig:m_doppler}}
\end{figure}
\begin{figure}[H]
	\centering \includegraphics[scale=1]{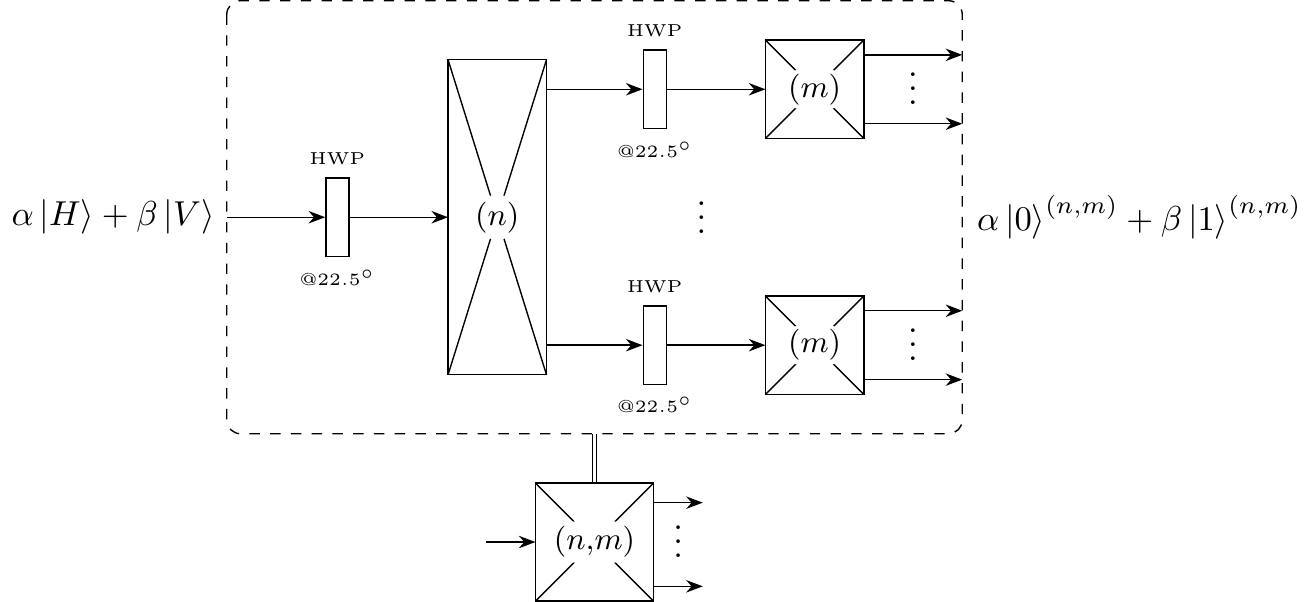}
	\caption{To generate an arbitrary QPC($n,m$) encoded state, an array of $n m -1$ photon dopplers and $n+1$ half-wave-plates, i.e. Hadamard gates, is used. \label{fig:QPC_Gen}}
\end{figure}
\begin{figure}[H]
	\centering \includegraphics[scale=1]{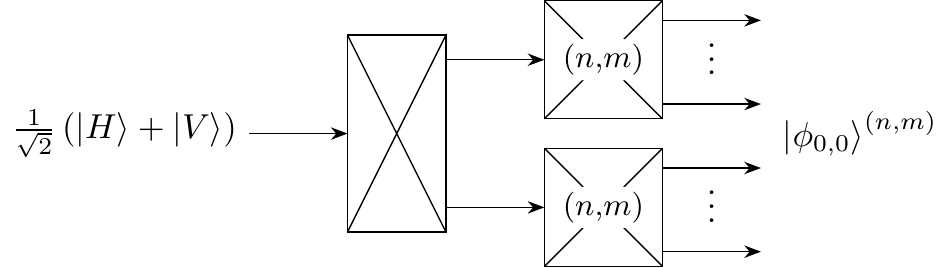}
	\caption{The ancillary Bell states at the repeater stations can be produced using two copies of the QPC($n,m$) generation scheme and one photon doppler to obtain the inital Bell state. This setup requires a total of $2 n m-1$ photon doppler modules. \label{fig:QPC_Bell_Gen}}
\end{figure}

A state generation scheme as depicted in Fig.~\ref{fig:m_doppler}~-~\ref{fig:QPC_Bell_Gen} gives exactly the (linear) cost scaling of $n m$ upon which the cost function $C$ as given in the main text relies, since for creating a QPC($n,m$)-encoded Bell state $2 n m-1$ copies of the photon doppler scheme are required.

It should be noted, however, that in this state generation scheme we depend on a strong nonlinear interaction, realizable with the techniques of coherent photon conversion. These nonlinearities could, in principle, also be used to obtain a unit-efficiency Bell measurement on the physical level. Employing these in our communication scheme instead of the standard linear optics BMs would allow the use of smaller quantum parity codes. However, the first main point of our proposal is to present a highly effective and loss-resistant Bell measurement on the quantum parity codes with the comparatively simple tools of linear optics.

Next, we take a look at the fault resistance of this state generation scheme. As was explained in Sec.~\ref{sec:comm scheme}, all errors that occur on individual physical-level qubits can be incorporated into the error channel of the transmission protocol and are taken care of there. However, errors that occur during the state generation often lead to non-local errors in the resulting state. We analyze three stages of the state generation.

The first stage is generating the initial one-photon state $\ket{\psi} = \alpha\ket{H}+\beta\ket{V}$. Current photon sources cannot produce pure states of this form on demand and instead give a mixed state with a vacuum portion $\rho = \eta_s \ket{\psi}\bra{\psi} + (1-\eta_s) \ket{0}\bra{0}$. The vacuum state passes the CPC-elements unchanged and, as a result, the outcome of the state generation scheme can then be written as $\eta_s \ket{\psi}^{(n,m)}\bra{\psi}^{(n,m)} + (1-\eta_s) \ket{0}^{\otimes 2 n m}\bra{0}^{\otimes 2 n m}$. For the case that such a multimode-vacuum emerges in a repeater station during a time interval $t_0$ instead of a QPC Bell state, the logical BM will (heraldedly) fail at this station and thus the transmission does not succeed. The transmission rate when including the corresponding probabilistic element from the imperfect photon sources becomes $R = \eta_s^{L_{\rm tot}/L_0} p^{L_{\rm tot}/L_0}/t_0$. Due to this exponential scaling, the value for $\eta_s$ must be extremely close to one to still obtain acceptable communication rates. For example, for a transmission distance of $\unit[1\,000]{km}$ and a repeater spacing of $\unit[2]{km}$ the vacuum probability $1-\eta_s$ must be smaller than $0.0014$ to allow for repeater rates $R>0.5/t_0$. Unfortunately, current photon sources do not reach this near-deterministic regime, but are more commonly at values of about $\eta_s\approx0.5$. Yet, by using multiple heralded photon sources and a little feedforward (so-called multiplexing) at the state preparation stage, it is possible to obtain a sufficient photon generation probability. For example, ten sources with $\eta_s=0.5$ yield at least one photon with probability $1-(1-\eta_s)^{10} \approx 0.9990$. Note that, in this case, it is not necessary to have $10$ instances of the nonlinear photon multiplying scheme presented in the above figures, because the feedforward operation takes place before that.

The second stage of the state generation scheme includes all steps to transform the single-photon state into an $n$-photon state  ($2n$ for the Bell state generation). Every photon is then the seed for one of the blocks of the QPC state. Should any of these photons be lost, the entire corresponding block is missing in the final state. This will lead to a (heralded) failure of the entire BM. Therefore the probability to lose any photons in this stage must be reduced as much as possible. Luckily, both the linear and the non-linear parts of the photon doppling can be performed near-deterministically \cite{langford11}. Should, nevertheless, the obtained success probability for the second stage of the state generation scheme be too small, multiplexing this stage as well would still be an option. Only very few instances would be required at every repeater station, as the base probability is already quite high. The important task of heralding the failure to initiate the feed-forward operation can be performed by generating an $n+1$-photon GHZ state instead of the $n$-photon GHZ state and measuring the additional photon in the $X$-basis. Due to the simple concatenation structure of the GHZ-state generation (see Fig.~\ref{fig:m_doppler}), the last photon is only present if all other $n$ photons are as well. The possible (heralded) phase-flip induced by the $X$-basis measurement can easily be corrected with the help of a $Z$-gate on any of the remaining modes. Using, for example, three multiplying instances (Fig.~\ref{fig:m_doppler}) with a total success rate of $0.9$ each, together with 10 photon sources of efficiency $\eta_s=0.5$ per instance, we obtain a probability of $\approx 0.9990$ for a successful generation of the desired GHZ state.

In the third stage of the generation scheme, the individual blocks are built up. Should a photon be lost in this stage, the resulting block can still be used for the identification of the Bell-state index $k$ in this block, as long as at least one physical BM succeeds. Only its ability to identify the index $l$ is gone. The remaining block is thus still useful, even though with the loss of one photon no more extra photons will be added to the state due to the linear concatenation scheme in Fig.~\ref{fig:m_doppler}. The number of lost photons can be further decreased by using a more tree-like concatenation structure. Then only those modes of the branches starting at the point of loss will be empty instead of all later modes. This decreased number of photon loss increases the chance for at least one physical BM to be successful. We expect the effect of losses in this stage on the transmission rate to be quite small, as long as the chance to lose photons particularly early within the third stage is not too high. 

\newpage

\section{Linear Optics: Generating QPC($n,m$)-states with Multiplexing and Feedforward}
\label{sec:AppD}
In this section we present an alternative scheme to generate the required QPC encoded Bell states at every repeater station. Instead of utilizing nonlinear effects like the coherent photon conversion used in Appendix~\ref{sec:AppC} we restrict ourselves to linear optical components: single-photon sources, both standard and polarizing beam splitters, and photon detectors. Additionally, we investigate the resource cost requirement of this scheme. As a measure for the cost we use the number of single-photon sources required to near-deterministically create a QPC($n,m$) Bell state. We will show that the number of photon sources scales approximately as $N_s = D (n m)^{\log_2(8/3)} \approx D (n m)^{1.415}$.

To generate the QPC encoded Bell states with a very high probability and this comparatively little number of photon sources we rely on two methods that go beyond the standard, passive, linear optical toolkit:
\begin{itemize}
	\item As much as we would like to generate the QPC states with a static linear optical setup, the success probability for this would be extremely small, and thus demand for an extremely large number of photon sources. To circumvent this we use a multiplexing scheme similar to that of Ref.~\cite{pant2016rate}. Our scheme is described in detail below. The general idea is to start with a large number of entangled states each conatining only a few photons and then joining them with the help of Bell measurements to increase the number of entangled photons. The probabilistic nature of both the generation of the small entangled states and the linear optical Bell measurement is counteracted by using feedforward techniques (multiplexing).
	\item As mentioned in the main text, the success probability of a standard linear optical BM is limited to $\tfrac{1}{2}$. We use additional photons to boost the success probability to $\tfrac{3}{4}$ \cite{Ewert}. The number of these additional photons is of course included in $N_s$.
\end{itemize}

In the first step of our state generation scheme six unentangled photons are used to heraldedly generate a 3-qubit-GHZ state $\ket{GHZ_3} = \tfrac{1}{\sqrt{2}}(\ket{HHH}+\ket{VVV})$. This is done through a probabilistic process presented in detail in the Supplementary Material of Ref.~\cite{Varnava08}. It succeeds with a probability of $\tfrac{1}{32}$. Next, the GHZ states are combined with Bell measurements to get states of the form
\begin{align}
	\ket{B^{(1,m)}}=\frac{1}{\sqrt{2}}\left[\ket{0}^{(m)}\ket{\phi_{0,0}}+\ket{1}^{(m)}\ket{\phi_{0,1}}\right]. \label{eq:B1m}
\end{align}
A detailed description, how this works, is given below. In the next stage of our state generation scheme, the states $\ket{B^{(1,m)}}$ are connected with BMs to produce
\begin{align}
	\ket{B^{(n,m)}}=\frac{1}{\sqrt{2}}\left[\ket{0}^{(n,m)}\ket{\phi_{0,0}}+\ket{1}^{(n,m)}\ket{\phi_{0,1}}\right]. \label{eq:Bnm}
\end{align}
These basic states, consisting of the codewords $\ket{0}^{(n,m)}$ and $\ket{1}^{(n,m)}$ and two additional photons, can be used to obtain a QPC($n,m$) Bell state by first measuring one of the two additional photons in each state in the $X$-basis and finally performing a Bell measurement on the remaining extra photons.

\begin{figure*}[b]
\includegraphics[height=17.8cm]{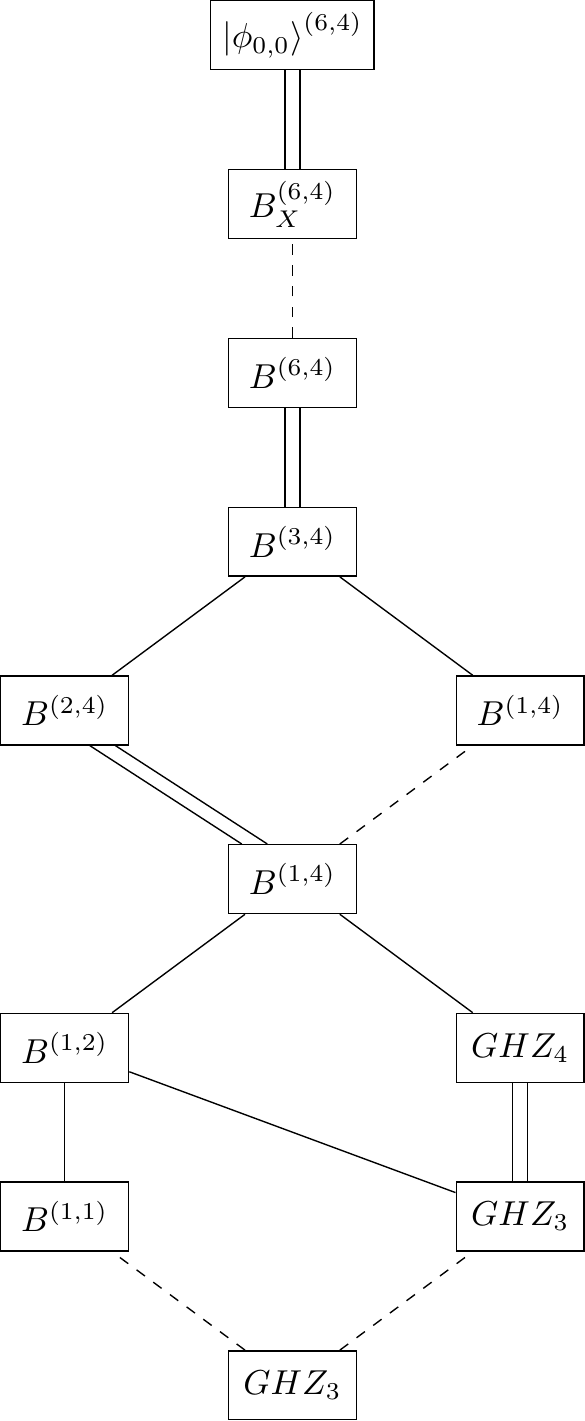} \hfill \includegraphics[height=17.8cm]{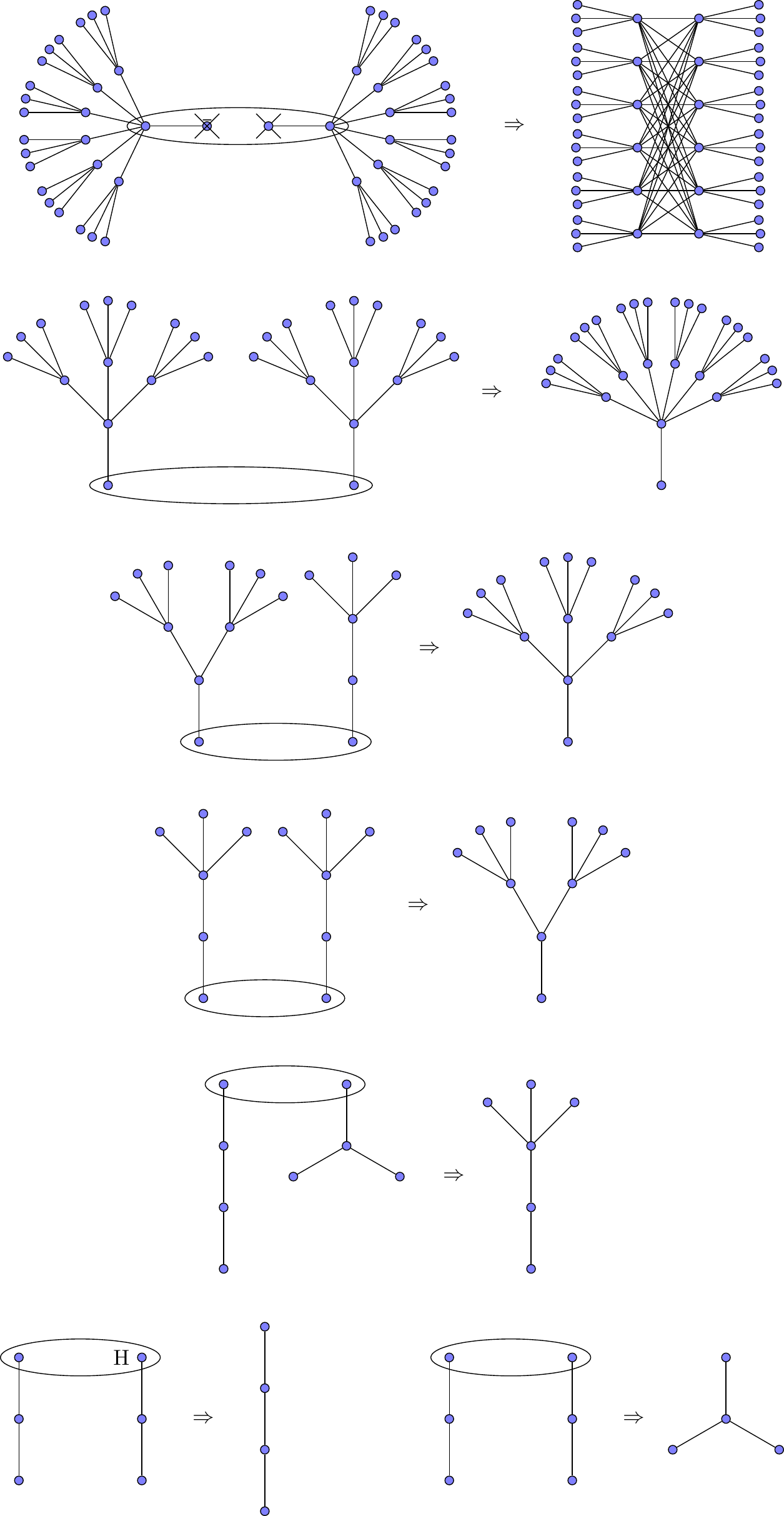}
\caption{State generation for a QPC(6,4) Bell state.\\
Left: This flowchart shows which states are combined on each multiplexing level, to obtain $\ket{\phi_{0,0}}^{(6,4)}$ from $\ket{GHZ_3}$-states. Starting from the bottom the states are either transformed using local, deterministic operations (dashed lines), i.e. Hadamard-gates, X-basis measurements, or simply doing nothing, or two states are combined with a Bell measurement (solid lines). At some points throughout the multiplexing scheme pools, e.g. the $GHZ_3$ pool on the second level, must be split in parts and different actions are performed. The ratio of the parts is chosen to optimize the protocol: since equally many $\ket{B^{(1,2)}}$ and $\ket{GHZ_4}$ states are used in the $\ket{B^{(1,4)}}$ generation, $\tfrac{2}{3}$ of the $\ket{GHZ_3}$-states are used for $\ket{GHZ_4}$ generation and $\tfrac{1}{3}$ for $\ket{B^{(1,2)}}$ generation.\\
Right: Here we show the state combinations as presented in eqs.~\eqref{eq:B12gene}~--~\eqref{eq:Bellgene} in the form of cluster states (note that the states in the equations are not the standard representatives of the cluster states shown here). Ellipses include the photons which are used in a BM, the Hadamard-gate at the bottom left as well as the X-basis measurements at the top are performed before the respective BM. In the final QPC-encoded Bell state at the top right, one can clearly see the $6$ blocks of $4$ photons for each of the logical qubits, as well as the fact, that every block is connected with all blocks of the other logical qubits but not with the blocks of its own logical qubit.}\label{fig:QPC64gen}
\end{figure*}

Let us now turn to the details of creating the states $\ket{B^{(1,m)}}$ from 3-qubit-GHZ states (see also Fig.~\ref{fig:QPC64gen}). The smallest one,
\begin{align}
	\ket{B^{(1,1)}} = \frac{1}{2}\left(\ket{HHH}+\ket{HVV}+\ket{VHH}-\ket{VVV}\right),
\end{align}
can easily be obtained from a 3-qubit-GHZ state by applying a Hadamard gate to the first photon. To create the next one, $\ket{B^{(1,2)}}$, a 3-qubit-GHZ state and the state $\ket{B^{(1,1)}}$ are combined with a Bell measurement. In our notation the Bell measurement is, if not mentioned otherwise, always applied to the first photons of the partaking states. We obtain
\begin{align}
	\ket{GHZ_3} \otimes \ket{B^{(1,1)}} &= \frac{1}{\sqrt{2}}\left(\ket{HHH}+\ket{VVV}\right) \otimes \frac{1}{2}\left(\ket{HHH}+\ket{HVV}+\ket{VHH}-\ket{VVV}\right)\nonumber\\
	&\stackrel{\rm BM}{\longrightarrow} \frac{1}{2} \left(\ket{HHHH}+\ket{HHVV} + \ket{VVHH} - \ket{VVVV} \right) = \ket{B^{(1,2)}}. \label{eq:B12gene}
\end{align}
Here we assumed that the Bell measurement yields the result $\ket{\phi_{0,0}}$. If any other Bell state is identified the above is only true up to some Pauli corrections. Depending on the outcome $\ket{\phi_{k,l}}$ of the BM the correction is given by $(X_1 X_2)^k Z_1^l$. The above result can be generalized to the case of combining any GHZ state with any state $\ket{B^{(1,\nu)}}$:
\begin{align}
	&&&\ket{GHZ_{\mu+2}} \otimes \ket{B^{(1,\nu)}} = \frac{1}{\sqrt{2}}\left(\ket{H}\ket{0}^{(\mu+1)} + \ket{V}\ket{1}^{(\mu+1)} \right) \otimes \frac{1}{\sqrt{2}} \left(\ket{H}\ket{0}^{(\nu-1)}\ket{\phi_{0,0}} + \ket{V}\ket{1}^{(\nu-1)}\ket{\phi_{0,1}}\right)\nonumber\\
	\stackrel{\rm BM}{\longrightarrow}&&& \frac{1}{2} \left[\left(\ket{0}^{(\mu+1)} \ket{0}^{(\nu-1)} + \ket{1}^{(\mu+1)}\ket{1}^{(\nu-1)}\right)\ket{\phi_{0,0}} + \left(\ket{0}^{(\mu+1)} \ket{1}^{(\nu-1)} + \ket{1}^{(\mu+1)}\ket{0}^{(\nu-1)}\right)\ket{\phi_{0,1}} \right]\nonumber\\
	=&&& \frac{1}{\sqrt{2}}\left[\ket{0}^{(\mu+\nu)}\ket{\phi_{0,0}}+\ket{1}^{(\mu+\nu)}\ket{\phi_{0,1}}\right] = \ket{B^{(1,\mu+\nu)}}.\label{eq:BGHZ}
\end{align}
Again, depending on the BM outcome certain Pauli corrections are necessary: $\left(\prod_{j=1}^{\mu-1} X_j\right)^k Z_1^l$.\\
The larger GHZ states, on the other hand, can be obtained by combining two smaller GHZ states:
\begin{align}
	\ket{GHZ_{\mu}} \otimes \ket{GHZ_{\nu}} \stackrel{\rm BM}{\longrightarrow} \ket{GHZ_{\mu+\nu-2}}.\label{eq:GHZGHZ}
\end{align}
Once again, the Bell measurement result implies some Pauli-corrections: $Z_\mu^l \left(\prod_{j=\mu}^{\mu+\nu-2} X_j\right)^k$.\\
Once the states $\ket{B^{(1,m)}}$ are ready, they can be combined into $\ket{B^{(2,m)}}$ with a Bell measurement. However, this time the BM is performed on the last photons of the partaking states, i.e. in each state one of the photons of the Bell-state-part is used. This shall be denoted by BM'.
\begin{align}
	&\vphantom{=}\ket{B^{(1,m)}} \otimes \ket{B^{(1,m)}}\nonumber\\ &= \frac{1}{2}\left[\ket{0}^{(m)}\ket{\phi_{0,0}}\ket{0}^{(m)}\ket{\phi_{0,0}} + \ket{0}^{(m)}\ket{\phi_{0,0}}\ket{1}^{(m)}\ket{\phi_{0,1}}+ \ket{1}^{(m)}\ket{\phi_{0,1}}\ket{0}^{(m)}\ket{\phi_{0,0}} + \ket{1}^{(m)}\ket{\phi_{0,1}}\ket{1}^{(m)}\ket{\phi_{0,1}}\right]\nonumber\\
	&\stackrel{\rm BM'}{\longrightarrow} \frac{1}{2}\left[\left(\ket{0}^{(m)}\ket{0}^{(m)} + \ket{1}^{(m)}\ket{1}^{(m)} \right)\ket{\phi_{0,0}} +\left(\ket{0}^{(m)}\ket{1}^{(m)} + \ket{1}^{(m)}\ket{0}^{(m)} \right)\ket{\phi_{0,0}}\right]\nonumber\\
	&=\frac{1}{\sqrt{2}}\left[\ket{0}^{(2,m)} \ket{\phi_{0,0}} + \ket{1}^{(2,m)}\ket{\phi_{0,0}}\right] = \ket{B^{(2,m)}}\label{eq:B2mgene}.
\end{align}
Here we have collected the remaining photons of the Bell-state-parts at the  end directly after performing the Bell measurement and have furthermore assumed the Bell measurement to have the result $\ket{\phi_{0,0}}$, in any other case the required corrections are: $Z_{m+1}^k Z_{2m+1}^l X_{2m+2}^k$. Additionally we have used the following property of the QPC encoding:
\begin{align}
	\ket{0}^{(n_1,m)}\ket{0}^{(n_2,m)} + \ket{1}^{(n_1,m)}\ket{1}^{(n_2,m)} &= \ket{0}^{(n_1+n_2,m)},\nonumber\\
	\ket{0}^{(n_1,m)}\ket{1}^{(n_2,m)} + \ket{1}^{(n_1,m)}\ket{0}^{(n_2,m)} &= \ket{1}^{(n_1+n_2,m)}.
\end{align}
This also allows us to generalize the above result to
\begin{align}
	\ket{B^{(n_1,m)}} \otimes \ket{B^{(n_2,m)}} \stackrel{\rm BM'}{\longrightarrow} \ket{B^{(n_1+n_2,m)}}, \label{eq:Bn1n2m}
\end{align}
with the correction $\left(\prod_{j=0}^{n_2-1} Z_{(n_1 +j) m + 1}\right)^k Z_{(n_1+n_2)m+1}^l X_{(n_1+n_2)m+2}^k$ depending on the Bell measurement result. With this we can create every $\ket{B^{(n,m)}}$-state from single photons.\\
To finally obtain the QPC($n,m$) Bell states, the $\ket{B^{(n,m)}}$-states are first subjected to an X-basis measurement on the last photon to obtain
\begin{align}
	\ket{B^{(n,m)}_X}=\frac{1}{\sqrt{2}}\left[\ket{0}^{(n,m)}\ket{+}+\ket{1}^{(n,m)}\ket{-}\right]
\end{align}
(assuming the result $\ket{+}$ of the X-basis measurement, otherwise after applying $Z_{n m+1}$), and then a Bell measurement BM' is performed. We get
\begin{align}
	\ket{B^{(n,m)}_X} \otimes \ket{B^{(n,m)}_X} \stackrel{\rm BM'}{\longrightarrow} \frac{1}{\sqrt{2}}\left[\ket{0}^{(n,m)}\ket{0}^{(n,m)} + \ket{1}^{(n,m)}\ket{1}^{(n,m)} \right] = \ket{\phi_{0,0}}^{(n,m)}, \label{eq:Bellgene}
\end{align}
if the BM yields $\ket{\phi_{0,0}}$, otherwise the correction is $\left(\prod_{j_1=1}^{m} X_{j_1}\right)^l \left(\prod_{j_2=0}^{n_2-1} Z_{n_1 m + j_2 m +1}\right)^k$.\\
It should be noted that all the corrections mentioned above do not have to be performed immediately after the respective Bell measurements. It suffices to keep track of all implied corrections and interpret the outcomes of the following Bell measurements accordingly [this includes that the implied corrections by this second Bell measurement depend on the (not executed) corrections of the first BM]. This allows to delay all necessary Pauli corrections to the final step and thus reduces the maximum number of necessary corrections to $2 n m$.

Let us now turn to the resource consumption of this state generation approach. For our communication scheme to work we will require an encoded Bell state at every repeater station. Therefore, the success probability of the state generation $p_{sg}$ must be very close to one, as it will appear in the communication rate as a global factor $p_{sg}^{L_{\rm tot}/L_0}$. For example, if we attempt communicating over $L_{\rm tot}=\unit[1\,000]{km}$ with a repeater spacing of $L_0=\unit[2]{km}$ and demand that the state generation probability reduces the secure key rate by no more than $50\%$, we obtain $1-p_{sg} \leq 10^{-3}$.  To achieve this success probability we attempt to generate a large enough pool of $\ket{B_X^{(n,m)}}$ states, perform the Bell measurement and select one of the pairs where the BM was successful. The size of the pool $n_X$ depends on the required state generation probability and the BM success probability:
\begin{align}
	1-\left(1-p_{\rm BM}\right)^{\lfloor\frac{n_X}{2}\rfloor} \geq p_{sg}. \label{eq:nX}
\end{align}
For the above example of $1-p_{sg} \leq 10^{-3}$ and a BM success probability of $75\%$, as achieved in Ref.~\cite{Ewert}, this yields $n_X\geq10$. It is not extremely important that in every attempt, the pool contains $n_X$ or more $\ket{B_X^{(n,m)}}$-states, since also for $n_X-2$ states the chance of at least one successful BM is relatively high. However, on average, $n_X$ states should be available. This marks an important change in the point of view: from exact probabilities to average values. The latter is less general, but allows for a much easier estimation of the number of photon sources required.

The states $\ket{B_X^{(n,m)}}$ are created from two states $\ket{B^{(n/2,m)}}$ with probability $p_{\rm BM}$ according to \eqref{eq:Bn1n2m} (for simplicity we assume $n=2^\lambda$ here). Thus, to get on average $n_X$ states $\ket{B_X^{(n,m)}}$, we need (on average) $\tfrac{2 n_X}{p_{\rm BM}}$ states $\ket{B^{(n/2,m)}}$. We use feedforward operations to only direct the successfully joined states into the pool of $\ket{B_X^{(n,m)}}$-states. Basically, we are multiplexing. This can be continued down to the $\ket{B^{(1,m)}}$-states: we will need (on average) $n_X \left(\tfrac{2}{p_{\rm BM}}\right)^\lambda$ of them. The states $\ket{B^{(1,m)}}$ themselves can be created from $\ket{B^{(1,m/2)}}$ states and $\ket{GHZ_{m/2+2}}$ states, see Eq.~\eqref{eq:BGHZ} (again we assume $m=2^\kappa$ for simplicity). To obtain the $n_X \left(\tfrac{2}{p_{\rm BM}}\right)^\lambda$ states $\ket{B^{(1,m)}}$, on average, $n_X \left(\tfrac{2}{p_{\rm BM}}\right)^\lambda \tfrac{1}{p_{\rm BM}}$ states of each above type are required. Now this can in turn be continued down to $\ket{B^{(1,1)}}$ and $\ket{GHZ_3}$-states, and since these two differ only by a Hadamard gate, in total, $n_X \left(\tfrac{2}{p_{\rm BM}}\right)^{\lambda+\kappa}$ GHZ states are needed. As was already stated above, GHZ states can be generated from six single photons with a success probability of $\tfrac{1}{32}$. Thus, on average, $192$ photons are required to create a GHZ state and thus, on average,
\begin{align}
	\tilde{N}_s = 192 n_X \left(\frac{2}{p_{\rm BM}}\right)^{\lambda+\kappa} = 192 n_X \left(\frac{2}{p_{\rm BM}}\right)^{\log_2(n)+\log_2(m)} = 192 n_X \left(n m\right)^{\log_2(2/p_{\rm BM})},
\end{align}
single photons are required to obtain one QPC encoded Bell state with a probability larger than $p_{sg}$.

If the code parameters $n$ and $m$ are no powers of 2, the state generation becomes more involved, since unequal states have to be combined at BMs and some states are not used in the next multiplexing-level immediately (see Fig.~\ref{fig:QPC64gen}). However, the estimate above is still quite good. A very conservative estimate would be obtained by using the number of multiplexing levels required, which is $\lceil\log_2(n)\rceil+\lceil\log_2(m)\rceil$ which is bounded by $\log_2(n m)+2$, and furthermore assume that on every level all states are immediately used in a BM. We then get an additional factor $(2/p_{\rm BM})^2$ in the upper bound for the required photon number.

In all of the above calculations the resource cost for the Bell measurement was not yet taken into account. To estimate the number of additional photons injected to boost the BM probability we merely need to count the number of BMs. We will stick to the case of $n=2^\lambda$ and $m=2^\kappa$ for simplicity. In the last step $\tfrac{n_X}{2}$ Bell measurements are performed to attempt joining the $\ket{B_X^{(n,m)}}$ states into QPC-Bell states. On the level below, for each of the $n_X$ states, on average, $\tfrac{1}{p_{\rm BM}}$ BM attempts are made. In the next lower level the number of BMs increases by another factor $\tfrac{2}{p_{\rm BM}}$ and so on. If every BM requires $n_{\rm BM}$ additional photons, we get in total
\begin{align*}
	N_{\rm BM} &= n_{\rm BM} \frac{n_X}{2} \sum_{k=0}^{\lambda+\kappa} \left(\frac{2}{p_{\rm BM}}\right)^k = n_{\rm BM} \frac{n_X}{2} \frac{1}{1-\frac{2}{p_{\rm BM}}} \left[1-\left(\frac{2}{p_{\rm BM}}\right)^{\log_2(n m)+1}\right]\nonumber\\
	&\approx \frac{n_{\rm BM} n_X}{2} \frac{1}{1-\frac{p_{\rm BM}}{2}} \left(n m\right)^{\log_2(2/p_{\rm BM})}.
\end{align*}
This has the same scaling as $\tilde{N}_s$, but the factor $\frac{n_{\rm BM}}{2} \frac{1}{1-\frac{p_{\rm BM}}{2}}$ is much smaller than $192$. When using the BM scheme presented in Ref.~\cite{Ewert}, which uses $4$ additional photons to boost the BM efficiency to $75\%$, we get a factor $\tfrac{16}{5}=3.2$. Thus, the number of photons injected to boost the BM efficiency makes less than $2\%$ of the total number $N_s =\tilde{N}_s+N_{\rm BM}$. We finally obtain the resource consumption of our linear-optics state generation scheme:
\begin{align}
	N_s = \left(192+\tfrac{16}{5}\right) 10 (n m)^{\log_2(8/3)} \approx 1952 (n m)^{1.415}.
\end{align}
We would like to make some additional remarks regarding the error robustness of this state generation scheme. The 3-qubit-GHZ-state generation scheme given in Ref.~\cite{Varnava08} has the appealing property that the GHZ output states have the form of GHZ states on which individual loss channels acted on all three photons. Together with the fact that joining states via Bell measurements does not introduce non-local errors, we find that the final QPC Bell state is afflicted only by local errors that act on individual photons. As we have seen in the main text, our logical Bell measurement and thus the entire communication scheme can handle this type of errors quite well. However, since the errors present in the state preparation (mainly photon loss) will also occur on photons that are measured during the state preparation, the success probability of joining states in a BM and thus the success probability of generating QPC-encoded Bell states is reduced. In a conservative estimate, we assume that all photons that are measured during the state preparation are subject to a loss channel of strength $\eta_{sg}$ (this includes losses from the GHZ generation scheme and losses due to time spent on the state generation chip). The number of photons required for successful state generation is then given by
	\begin{align}
		N_s = \left(\tfrac{192}{\eta_{sg}^3}+\tfrac{16}{5}\right) n_X (n m)^{\log_2\left(\tfrac{8}{3\eta_{sg}^2}\right)}.
	\end{align}
	The pool size $n_X$ for $\ket{B_X^{(n,m)}}$ is increased, since in Eq.~\eqref{eq:nX} $p_{\rm BM}$ is replaced by $p_{\rm BM} \eta_{sg}^4$ (not only the photons of the BM but also those that are measured in the X-basis must be present). If we assume $\eta_{sg}=0.97$ we get $n_X=14$ and obtain the scaling $N_s = 2732.8 (n m)^{1.503}$. This means that for the typical number of photons in a QPC state (10 to 1000), the errors in the state generation scheme imply an increase of the number of required photon sources by a factor of about $2.5$. In Table~\ref{tab:costs} we give the number of required photon sources per repeater station for several QPCs assuming, as above, that over a total distance of $L_{\rm tot}=\unit[1\,000]{km}$ repeater stations are placed with a spacing of $L_0=\unit[2]{km}$ and that the influence of the probabilistic state generation reduces the secure key rate by no more than $50\%$, i.e. $1-p_{sg}\leq10^{-3}$. For comparison, we also give the required number of photons when using the standard linear optical BM instead of the $\tfrac{3}{4}$-efficient one of Ref.~\cite{Ewert}.
	
\begin{table}[b]
\begin{ruledtabular}
\begin{tabular}{crrrr}
$(n,m)$ & $p_{\rm BM}=\tfrac{3}{4}$, $\eta_{sg}=1$ & $p_{\rm BM}=\tfrac{3}{4}$, $\eta_{sg}=0.97$ & $p_{\rm BM}=\tfrac{1}{2}$, $\eta_{sg}=1$ & $p_{\rm BM}=\tfrac{1}{2}$, $\eta_{sg}=0.97$ \\\hline\\[-1em]
(4,2) & $37\cdot 10^3$ & $62\cdot 10^3$ & $246\cdot 10^3$ & $354\cdot 10^3$ \\
(10,3) & $240\cdot 10^3$ & $454\cdot 10^3$ & $3.5\cdot 10^6$ & $5.6\cdot 10^6$ \\
(18,4) & $829\cdot 10^3$ & $1.7\cdot 10^6$ & $19\cdot 10^6$ & $35\cdot 10^6$ \\
(23,5) & $1.6\cdot 10^6$ & $3.4\cdot 10^6$ & $51\cdot 10^6$ & $92\cdot 10^6$ \\
(28,6) & $2.8\cdot 10^6$ & $6.0\cdot 10^6$ & $108\cdot 10^6$ & $204\cdot 10^6$ \\
(38,6) & $4.3\cdot 10^6$ & $9.6\cdot 10^6$ & $200\cdot 10^6$ & $368\cdot 10^6$ \\
(67,11) & $22\cdot 10^6$ & $56\cdot 10^6$ & $2.1\cdot 10^9$ & $4.5\cdot 10^9$
\end{tabular}
\end{ruledtabular}
\caption{Required number of single photon sources to assure that state generation does not reduce the secure key rate by more than $50\%$. Parameters: $L_{\rm tot}=\unit[1\,000]{km}$, $L_0=\unit[2]{km}$.} \label{tab:costs}
\end{table}

\newpage
\FloatBarrier
\newpage
\section{Derivation of the Outcome Probabilities $P^{\rm (dc)}_{u,v}$ in the Presence of Dark Counts}
\label{sec:AppE}

In this section we derive the outcome probabilities $P^{\rm (dc)}_{u,v}$ in the presence of dark counts. It should be noted that we restrict ourselves here to the case of the standard linear optical BM with an efficiency of $\tfrac{1}{2}$, as depicted in Fig.~\ref{fig:BM_setup}~c) of the main text, since this corresponds to the most important case in our communication scenario. Furthermore, we assume the use of detectors that can resolve up to two photons, which is also the standard situation in our work. We have also investigated the combination of dark counts and on-off detectors (see Sec.\ref{subsec: onoff}), but since no conceptually new effects arise in this case, we shall focus here only on the PNRD case.

To calculate the outcome probabilities $P^{\rm (dc)}_{u,v}$ for detectors that exhibit dark counts, we first have to calculate the probabilities $p^{\rm (dc)}_{\mu\rightarrow \nu}$ that a detector recognizes $\nu$ photons, although $\mu$ photons have entered it. In Sec.~\ref{subsec:darkcounts} we have already described the model for a dark-count-afflicted detector: an ideal photon-number-resolving detector is preceded by a beam splitter with transmission coefficient $\eta_d$ whose second input mode is occupied by a thermal state,
\begin{align}
	\rho_{\rm th}(\bar{n}) = \sum_{l=0}^\infty \frac{\bar{n}^l}{(\bar{n}+1)^{l+1}} \ket{l}\bra{l},
\end{align}
with the average photon number $\bar{n}$. If an arbitrary photonic state $\rho_{\rm in} = \sum_{j,k} \rho_{j,k} \ket{j}\bra{k}$ enters the detector, the corresponding two-mode state is given by
\begin{align}
	\rho_{\rm in} \otimes \rho_{\rm th}(\bar{n}) &= \sum_{l} \sum_{j,k} \rho_{j,k} \frac{\bar{n}^l}{(\bar{n}+1)^{l+1}} \ket{j,l}\bra{k,l}\nonumber\\
	&= \sum_{l,j,k} \rho_{j,k} \frac{\bar{n}^l}{(\bar{n}+1)^{l+1}} \frac{1}{\sqrt{j!l!}} \left(a_1^\dagger\right)^j \left(a_2^\dagger\right)^l \ket{0,0}\bra{0,0} a_1^k a_2^l \frac{1}{\sqrt{k! l!}}.
\end{align}
The beam splitter with transmission $\eta_d$ mixes the modes via $a_1^\dagger \rightarrow \sqrt{\eta_d} a_1^\dagger + \sqrt{1-\eta_d} a_2^\dagger$ and $a_2^\dagger \rightarrow \sqrt{\eta_d} a_2^\dagger - \sqrt{1-\eta_d} a_1^\dagger$. With the use of the binomial identity we obtain
\begin{align}
	\rho_{\rm in} \otimes \rho_{\rm th}(\bar{n}) &\stackrel{BS_{\eta_d}}{\longrightarrow} \sum_{l,j,k} \rho_{j,k} \frac{\bar{n}^l}{(\bar{n}+1)^{l+1}} \sum_{n_1,m_1,n_2,m_2} \frac{1}{\sqrt{j!l!}} \binom{j}{n_1}\binom{l}{m_1} \sqrt{\eta_d}^{j-n_1+l-m_1} \sqrt{1-\eta_d}^{n_1+m_1} (-1)^{m_1}\nonumber\\
	&\qquad\quad\times \left(a_1^\dagger\right)^{j-n_1+m_1} \left(a_2^\dagger\right)^{n_1+l-m_1} \ket{0,0}\bra{0,0} a_1^{k-n_2+m_2} a_2^{n_2+l-m_2}\nonumber\\
	&\qquad \quad \times \sqrt{\eta_d}^{k-n_2+l-m_2} \sqrt{1-\eta_d}^{n_2+m_2} (-1)^{m_2} \binom{k}{n_2}\binom{l}{m_2} \frac{1}{\sqrt{k!l!}}.
\end{align}
The state entering the ideal photon detector is now given by tracing over the second mode. We denote the summation index for this partial trace operation by $s$ and get
\begin{align}
	\rho_{\rm in} &\longrightarrow \sum_{l,j,k} \rho_{j,k} \frac{\bar{n}^l}{(\bar{n}+1)^{l+1}} \sum_{n_1,n_2} \sum_s \frac{1}{\sqrt{j!l!}} \binom{j}{n_1}\binom{l}{s-n_1} \sqrt{\eta_d}^{j-2n_1+s} \sqrt{1-\eta_d}^{2n_1+l-s} (-1)^{n_1+l-s}\nonumber\\
	&\qquad\quad\times \sqrt{s!}\sqrt{(j+l-s)!}\ket{j+l-s}\bra{k+l-s}\sqrt{s!}\sqrt{(k+l-s)!}\nonumber\\
	&\qquad \quad \times \sqrt{\eta_d}^{k-2n_2+s} \sqrt{1-\eta_d}^{2n_2+l-s} (-1)^{n_2+l-s} \binom{k}{n_2}\binom{l}{s-n_2} \frac{1}{\sqrt{k!l!}}.
\end{align}
By inserting two unit operations ($\sum_u \delta_{u,j}=\sum_u \braket{u|j}$  and $\sum_v \delta_{k,v}=\sum_v \braket{k|v}$) we are able to isolate the input state and obtain
\begin{align}
	\rho_{\rm in} &\longrightarrow \sum_{l,s} \frac{\bar{n}^l}{(\bar{n}+1)^{l+1}} \left(\sum_{u,n_1} \sqrt{\frac{(u+l-s)!s!}{u!l!}} \binom{u}{n_1}\binom{l}{s-n_1} \sqrt{\eta_d}^{u+s-2n_1} \sqrt{1-\eta_d}^{l-s+2n_1} (-1)^{l-s+n_1}\ket{u+l-s}\bra{u}\right)\nonumber\\
	&\qquad\quad\times \left(\sum_{j,k}\rho_{j,k}\ket{j}\bra{k}\right)\nonumber\\
	&\qquad\quad\times \left(\sum_{v,n_2} \sqrt{\frac{(v+l-s)!s!}{v!l!}} \binom{v}{n_2}\binom{l}{s-n_2} \sqrt{\eta_d}^{v+s-2n_2} \sqrt{1-\eta_d}^{l-s+2n_2} (-1)^{l-s+n_2}\ket{v}\bra{v+l-s}\right)\nonumber\\
	&\quad= \sum_{l,s} \frac{\bar{n}^l}{(\bar{n}+1)^{l+1}}\ K_{s,l}\ \rho_{\rm in}\ K_{s,l}^\dagger,\label{eq:dcerrorchannel}
\end{align}
with the error operators
\begin{align}
	K_{s,l} &= \sum_{u} \sqrt{\frac{(u+l-s)!s!}{u!l!}} \xi_{s,l,u}(\eta_d)\ket{u+l-s}\bra{u} \label{eq:dcerroroperators},\\
	\xi_{s,l,u}(\eta_d)&= \sum_{n_1} \binom{u}{n_1}\binom{l}{s-n_1} \sqrt{\eta_d}^{u+s-2n_1} \sqrt{1-\eta_d}^{l-s+2n_1} (-1)^{l-s+n_1}.
\end{align}
This channel map in terms of Kraus-operator-sum representation describes a Gaussian thermal noise channel with loss parameter $\eta_d$ and thermal mean photon number $\bar{n}$. The probabilities $p^{\rm (dc)}_{\mu\rightarrow \nu}$ are now given by
\begin{align}
	p^{\rm (dc)}_{\mu\rightarrow \nu} &= \braket{\nu|\sum_{l,s} \frac{\bar{n}^l}{(\bar{n}+1)^{l+1}}\ K_{s,l}| \mu} \braket{\mu|\ K_{s,l}^\dagger|\nu} = \sum_{l,s} \frac{\bar{n}^l}{(\bar{n}+1)^{l+1}} \left|\braket{\nu|K_{s,l}|\mu}\right|^2.\label{eq:pdcmunu}
\end{align}
It suffices to investigate the diagonal entries $\ket{\mu}\bra{\mu}$ of the input states, because the result of the ideal photon detector only depends on the diagonal entries in Eq.~\eqref{eq:dcerrorchannel} and the error channel described by the operators in Eq.~\eqref{eq:dcerroroperators} does not mix diagonals of the density matrix.

As already mentioned in Sec.~\ref{subsec:darkcounts} only a few of the probabilities $p^{\rm (dc)}_{\mu\rightarrow \nu}$ are actually needed. With no other source of additional photons besides the dark counts described in this model we have $\mu\leq 2$. Furthermore, we assume that the photon detectors resolve 0,1, and $\geq2$ photons and thus we can restrict the calculation to $\nu\leq1$. In addition to the exact results of Eq.~\eqref{eq:pdcmunu} we also give the first-order terms of the Taylor series with respect to small average photon numbers in the thermal state $\bar{n}\ll1$:
\begin{align*}
	p^{\rm (dc)}_{0\rightarrow 0} &= \sum_{l,s} \frac{\bar{n}^l}{(\bar{n}+1)^{l+1}} \delta_{l,s} \eta_d^l = \frac{1}{1+\bar{n}} \sum_l \left(\frac{\bar{n} \eta_d}{1+\bar{n}}\right)^l = \frac{1}{1+\bar{n}(1-\eta_d)} &&\approx 1-\bar{n}(1-\eta_d)\\
	p^{\rm (dc)}_{0\rightarrow 1} &= \frac{\bar{n}(1-\eta_d)}{(1+\bar{n}(1-\eta_d))^2} &&\approx \bar{n}(1-\eta_d)\\
	p^{\rm (dc)}_{0\rightarrow \geq2} &= 1-p^{\rm (dc)}_{0\rightarrow 0}-p^{\rm (dc)}_{0\rightarrow 1} = \frac{\bar{n}^2(1-\eta_d)^2 }{(1+\bar{n}(1-\eta_d))^2} &&\approx 0
	\\
	p^{\rm (dc)}_{1\rightarrow 0} &= \frac{(1-\eta_d)(1+\bar{n})}{(1+\bar{n}(1-\eta_d))^2}&&\approx (1-\eta_d)+\bar{n}(1-\eta_d)(2\eta_d-1)\\
	p^{\rm (dc)}_{1\rightarrow 1} &= \frac{\eta_d+(1-\eta_d)^2\bar{n}(1+\bar{n})}{(1+\bar{n}(1-\eta_d))^3}&&\approx \eta_d + \bar{n}(1-\eta_d)(1-4\eta_d)\\
	p^{\rm (dc)}_{1\rightarrow \geq2} &= 1-p^{\rm (dc)}_{1\rightarrow 0}-p^{\rm (dc)}_{1\rightarrow 1} = \frac{\bar{n}(1-\eta_d)[2\eta_d+\bar{n}(1-\eta_d)+\bar{n}^2(1-\eta_d)^2]}{(1+\bar{n}(1-\eta_d))^3} &&\approx \bar{n} 2 (1-\eta_d) \eta_d
	\\
	p^{\rm (dc)}_{2\rightarrow 0} &= \frac{(1-\eta_d)^2(1+\bar{n})^2}{(1+\bar{n}(1-\eta_d))^3} &&\approx (1-\eta_d)^2 + \bar{n} (1-\eta_d)^2(3\eta_d-1)\\
	p^{\rm (dc)}_{2\rightarrow 1} &= \frac{(1+\bar{n})(1-\eta_d)[2\eta_d+\bar{n}(1+\bar{n})(1-\eta_d)^2]}{(1+\bar{n}(1-\eta_d))^4} && \approx 2 (1-\eta_d)\eta_d + \bar{n} (1-\eta_d)(1-8\eta_d+9\eta_d^2)\\
	p^{\rm (dc)}_{2\rightarrow \geq2} &= 1-p^{\rm (dc)}_{2\rightarrow 0}-p^{\rm (dc)}_{2\rightarrow 1}% = \frac{\eta_d^2+\bar{n}[2\eta_d(2-3\eta_d+\eta_d^2)+\bar{n}(1-\eta_d)^2(1+4\eta_d)+2\bar{n}^2(1-\eta_d)^3+\bar{n}^3(1-\eta_d)^4]}{(1+\bar{n}(1-\eta_d))^4}
	 && \approx \eta_d^2 +\bar{n} 2 \eta_d (1-\eta_d)(2-3\eta_d)
\end{align*}
In order to obtain the BM outcome probabilities $P^{\rm (dc)}_{u,v}$ from the above detector outcome probabilities, we first have to take a look at the click patterns that an undisturbed Bell state would produce in an ideal detector and then conclude which of the actually occurring click patterns shall be accepted as a certain BM result. The click patterns are given according to the order of the detectors in Fig.~\ref{fig:BM_setup}~c), i.e., $(H_1,V_1,V_2,H_2)$. As already described in Sec.~\ref{sec:QPC BM} both $\ket{\phi_{0,l}}$-states yield the click pattern $(2,0,0,0)$ or permutations hereof. On the other hand, the state $\ket{\phi_{1,0}}$ gives $(1,1,0,0)$ or $(0,0,1,1)$ and the state $\ket{\phi_{1,1}}$ gives patterns $(1,0,1,0)$ or $(0,1,0,1)$.

Let us now discuss the interpretation of actual click patterns obtained from dark-count-afflicted detectors. First, there is the possibility that no photons are detected at all, i.e. the click pattern is $(0,0,0,0)$. Just like in the case without dark counts, this pattern does not contain any information on the original undisturbed Bell states. The same holds for click patterns with only one photon, i.e. $(1,0,0,0)$ and permutations hereof. It is equally probable to obtain these click patterns from all four Bell states. The case of two detected photons corresponds to the accepted Bell measurements in the ideal case. Thus, also in the presence of dark counts, click patterns that match the above ideal ones, should be accepted. However, it is possible now to obtain such click patterns with a combination of photon loss and dark counts. If exactly one of the two initial photons is lost [either during transmission $(1-\eta_t)\eta_d$ or within the detector $\eta_t 2(1-\eta_d)\eta_d$], while at the same time a single dark count occurs, then still two photons will be detected. Depending on the position in which of the four detectors the dark count occurred, the state will be either identified correctly, or as another Bell state, or a pattern indicates failure of the BM [the latter is the case for the patterns $(1,0,0,1)$ and $(0,1,1,0)$]. Since the probability of a dark count is the same for all detectors, this outcome distribution is similar to that of a depolarizing error channel with
\begin{align}
	\epsilon &= 2\epsilon^{(\rm dc)} = 2 \left[(1-\eta_t)\eta_d +\eta_t 2(1-\eta_d)\eta_d \right] \bar{n} (1-\eta_d) \equiv 2 p_{-1} p_{\rm dc} ,
\end{align}
(with the exception of the failing BM). In principle, it is also possible to obtain two detector clicks by losing both initial photons followed by two dark counts. However, this is much more unlikely, since we are mainly concerned with small dark count probabilities. Furthermore, since we assume detectors that can resolve only up to two photons, input states such as $\ket{3000}$ or $\ket{4000}$, and so on, will be recognized as $(2,0,0,0)$ and are thus interpreted as $\ket{\phi_{0,?}}$. This is fine though, because the probability of obtaining such a click pattern from a $\ket{\phi_{1,l}}$-state is smaller, since one more dark count is required there.

Of course, it is also possible to obtain click patterns with three clicks or even more in the presence of dark counts. The most straightforward route, which is also the one we followed for our numerical calculations, is to simply interpret these click patterns as failed physical-level BMs. However, when analyzing especially the patterns with exactly three detector clicks it is possible to obtain some information on the original Bell state. For example, any click pattern with exactly one click in three of the four detectors is most likely the result of a $\ket{\phi_{1,l}}$ state with an additional dark count, since from a $\ket{\phi_{0,l}}$ state, two dark counts and a photon loss would be necessary. Thus these patterns could be interpreted as a $(1,?)$ result. The same thought implies some other identifications:
	\begin{itemize}
		\item	The patterns $(2,1,0,0)$, $(1,2,0,0)$, $(0,0,2,1)$, and $(0,0,1,2)$, which only include detector pairs that would also click simultaneously in the case of ideal $\ket{\phi_{1,0}}$ identification, are most probably results of this state and a single dark count. They could also be the result of a $\ket{\phi_{0,l}}$ state with a dark count, but the probability of that is smaller by a factor $4$.
		\item	Analogously, the patterns $(2,0,1,0)$, $(1,0,2,0)$, $(0,2,0,1)$, and $(0,1,0,2)$ correspond to $\ket{\phi_{1,1}}$.
		\item	Finally, the patterns $(2,0,0,1)$, $(1,0,0,2)$, $(0,2,1,0)$, and $(0,1,2,0)$ correspond to $\ket{\phi_{0,?}}$.
	\end{itemize}
Simply including these patterns into the acceptance sets for the corresponding Bell state, however, can even reduce the obtained secure key rate. This is due to the fact that these additional click patterns favor the identification of Bell states with $k=1$: it is very unlikely that a $\ket{\phi_{1,l}}$ state will be identified as $\ket{\phi_{0,?}}$, because two dark counts are required for that. Because of this asymmetry our standard interpretation rules $f_{u}(\gamma)$ are not well adapted to this scenario. In fact, it might even be necessary to introduce new, additional classes of BM outcomes, such as $(1,0)^*$ which denotes that a (heralded) dark count occurred in the identification as $\ket{\phi_{1,0}}$ state. Because of the associated increase in interpretation complexity and the comparatively little gain in the secure key rate or in understanding of the underlying mechanisms we have decided to focus on the simpler approach of interpreting three-click events as BM failures and to accept the corresponding reduction of the raw transmission rate.

The BM outcome probabilities $P^{\rm (dc)}_{u,v}$ are now obtained by following all possible routes from the undisturbed input state to the click patterns associated with a given BM result:

\begin{align*}
	P^{\rm (dc)}_{1,i} &= P^{\rm (dc)}_{2,i} = P^{\rm (dc)}_{6,i} = 0,\\
	P^{\rm (dc)}_{3,1} &= P^{\rm (dc)}_{3,2} = P^{\rm (dc)}_{4,1} = P^{\rm (dc)}_{4,2}\\
	&= \eta_t \left[p^{\rm (dc)}_{2\rightarrow1} p^{\rm (dc)}_{0\rightarrow1} \left(p^{\rm (dc)}_{0\rightarrow0}\right)^2 + p^{\rm (dc)}_{2\rightarrow0} p^{\rm (dc)}_{0\rightarrow0} \left(p^{\rm (dc)}_{0\rightarrow1}\right)^2 \right] + (1-\eta_t) \left[ p^{\rm (dc)}_{1\rightarrow1} p^{\rm (dc)}_{0\rightarrow1} \left(p^{\rm (dc)}_{0\rightarrow0}\right)^2 + p^{\rm (dc)}_{1\rightarrow0} p^{\rm (dc)}_{0\rightarrow0} \left(p^{\rm (dc)}_{0\rightarrow1}\right)^2 \right]\\
	&\approx p_{-1} p_{\rm dc},\\
	P^{\rm (dc)}_{5,1} &= P^{\rm (dc)}_{5,2}\\
	&= \eta_t \left[\left(p^{\rm (dc)}_{2\rightarrow\geq2}p^{\rm (dc)}_{0\rightarrow0} + 3 p^{\rm (dc)}_{2\rightarrow0} p^{\rm (dc)}_{0\rightarrow\geq2} \right)\left(p^{\rm (dc)}_{0\rightarrow0}\right)^2 \right] + (1-\eta_t) \left[\left(p^{\rm (dc)}_{1\rightarrow\geq2} p^{\rm (dc)}_{0\rightarrow0} + 3 p^{\rm (dc)}_{1\rightarrow0} p^{\rm (dc)}_{0\rightarrow\geq2} \right)\left(p^{\rm (dc)}_{0\rightarrow0}\right)^2 \right]\\
	&\approx \eta_d^2\eta_t \left(1-5p_{\rm dc}\right) + 2 p_{-1} p_{\rm dc},\\
	P^{\rm (dc)}_{7,1} &= P^{\rm (dc)}_{7,2} = 1-\sum_{u=1}^{6} P^{\rm (dc)}_{u,1}\\
	&\approx [1-\eta_d^2\eta_t(1-5p_{\rm dc})] - 4 p_{-1} p_{\rm dc},\\
	P^{\rm (dc)}_{3,3} &= P^{\rm (dc)}_{4,4}\\
	&= \eta_t \left[\left(p^{\rm (dc)}_{1\rightarrow1}\right)^2 \left(p^{\rm (dc)}_{0\rightarrow0}\right)^2 + \left(p^{\rm (dc)}_{0\rightarrow1}\right)^2 \left(p^{\rm (dc)}_{1\rightarrow0}\right)^2 \right] + (1-\eta_t) \left[p^{\rm (dc)}_{1\rightarrow1} p^{\rm (dc)}_{0\rightarrow1} \left(p^{\rm (dc)}_{0\rightarrow0}\right)^2 + p^{\rm (dc)}_{1\rightarrow0} p^{\rm (dc)}_{0\rightarrow0} \left(p^{\rm (dc)}_{0\rightarrow1}\right)^2 \right]\\
	&\approx \eta_d^2\eta_t(1-8p_{\rm dc}) + p_{-1} p_{\rm dc},\\
	P^{\rm (dc)}_{4,3} &= P^{\rm (dc)}_{3,4}\\
	&= \eta_t \left[2 p^{\rm (dc)}_{1\rightarrow1} p^{\rm (dc)}_{1\rightarrow0} p^{\rm (dc)}_{0\rightarrow1} p^{\rm (dc)}_{0\rightarrow0} \right] + (1-\eta_t) \left[p^{\rm (dc)}_{1\rightarrow1} p^{\rm (dc)}_{0\rightarrow1} \left(p^{\rm (dc)}_{0\rightarrow0}\right)^2 + p^{\rm (dc)}_{1\rightarrow0} p^{\rm (dc)}_{0\rightarrow0} \left(p^{\rm (dc)}_{0\rightarrow1}\right)^2 \right]\\
	&\approx p_{-1} p_{\rm dc},\\
	P^{\rm (dc)}_{5,3} &= P^{\rm (dc)}_{5,4}\\
	&= \eta_t \left[\left(p^{\rm (dc)}_{1\rightarrow\geq2} p^{\rm (dc)}_{0\rightarrow0} + p^{\rm (dc)}_{1\rightarrow0} p^{\rm (dc)}_{0\rightarrow\geq2} \right) 2 \left(p^{\rm (dc)}_{1\rightarrow0}p^{\rm (dc)}_{0\rightarrow0}\right) \right] + (1-\eta_t) \left[\left(p^{\rm (dc)}_{1\rightarrow\geq2} p^{\rm (dc)}_{0\rightarrow0} + 3 p^{\rm (dc)}_{1\rightarrow0} p^{\rm (dc)}_{0\rightarrow\geq2} \right) \left(p^{\rm (dc)}_{0\rightarrow0}\right)^2 \right]\\
	&\approx 2 p_{-1} p_{\rm dc},\\
	P^{\rm (dc)}_{7,3} &= P^{\rm (dc)}_{7,4} = 1-\sum_{u=1}^{6} P^{\rm (dc)}_{u,3}\\
	&\approx [1-\eta_d^2\eta_t(1-8p_{\rm dc})] - 4 p_{-1} p_{\rm dc},
\end{align*}
where we have used the abbreviations $p_{\rm dc} = \bar{n} (1-\eta_d)$ for the (approximate) dark count probability and $p_{-1} = \left[(1-\eta_t)\eta_d + \eta_t 2(1-\eta_d)\eta_d\right]$ for the (exact) probability to lose one of the two photons of the BM. By furthermore using $\eta = \eta_d^2 \eta_t$ and $\epsilon^{(\rm dc)} = p_{-1} p_{\rm dc}$ we obtain the matrix $P^{\rm (dc)}$ as given in Table~\ref{tab:pdc}.

\begin{table}[b]
\begin{ruledtabular}
\begin{tabular}{ccccc}
$P_{u,v}^{\rm (dc)}$ & $\ket{\phi_{0,0}}$ & $\ket{\phi_{0,1}}$ & $\ket{\phi_{1,0}}$ & $\ket{\phi_{1,1}}$ \\\hline\\[-1em]
$(1,0)$ & $\epsilon^{(\rm dc)}$ & $\epsilon^{(\rm dc)}$ & $\eta(1-8p_{\rm dc}) + \epsilon^{(\rm dc)}$ & $\epsilon^{(\rm dc)}$ \\
$(1,1)$ & $\epsilon^{(\rm dc)}$ & $\epsilon^{(\rm dc)}$ & $\epsilon^{(\rm dc)}$ & $\eta(1-8p_{\rm dc}) + \epsilon^{(\rm dc)}$ \\
$(0,?)$ & $\eta \left(1-5p_{\rm dc}\right) + 2 \epsilon^{(\rm dc)}$ & $\eta \left(1-5p_{\rm dc}\right) + 2 \epsilon^{(\rm dc)}$ & $2 \epsilon^{(\rm dc)}$ & $2 \epsilon^{(\rm dc)}$ \\
$(?,?)$ & $[1-\eta (1-5p_{\rm dc})] - 4 \epsilon^{(\rm dc)}$ & $[1-\eta (1-5p_{\rm dc})] - 4 \epsilon^{(\rm dc)}$ & $[1-\eta (1-8p_{\rm dc})] - 4 \epsilon^{(\rm dc)}$ & $[1-\eta (1-8p_{\rm dc})] - 4 \epsilon^{(\rm dc)}$
\end{tabular}
\end{ruledtabular}
\caption{BM outcome probabilities $P^{\rm (dc)}$ in the presence of dark counts. Only the leading terms up to linear order are displayed.} \label{tab:pdc}
\end{table}

\end{document}